\documentclass[times,twocolumn,twocolappendix]{aastex701}

\usepackage{tabularray}

\newcommand\msun{\hbox{${M}_{\odot}$}}
\newcommand\mstar{\hbox{${M}_{\star}$}}

\newcommand\Zsun{\hbox{${Z}_{\odot}$}}
\newcommand\ageMW{\hbox{$ \text{age}_{MW}$} }

\graphicspath{{figures/}}

\shorttitle{Age and Metallicity Gradients in MWA progenitors}
\shortauthors{Tan et al.}

\begin{document}

\title{Resolved Ages and Stellar Metallicities in Progenitors of Milky Way Analogs: A Closer Look at their Star Formation Histories since $z=5$}




\author[0000-0002-3503-8899]{Vivian Yun Yan Tan}
\affiliation{Department of Physics and Astronomy, York University, 4700 Keele Street, Toronto, ON, M3J 1P3, Canada}
\email{tanvivia@yorku.ca}

\author[0000-0002-9330-9108]{Adam Muzzin}
\affiliation{Department of Physics and Astronomy, York University, 4700 Keele Street, Toronto, ON, M3J 1P3, Canada}
\email{muzzin@yorku.ca}

\author[0009-0009-9848-3074]{Naadiyah Jagga}
\affiliation{Department of Physics and Astronomy, York University, 4700 Keele Street, Toronto, ON, M3J 1P3, Canada}
\email{jagga@yorku.ca}

\author[0000-0003-0780-9526]{Visal Sok}
\affiliation{Department of Physics and Astronomy, York University, 4700 Keele Street, Toronto, ON, M3J 1P3, Canada}
\email{sokvisal@yorku.ca}

\author[0000-0001-8830-2166]{Ghassan T. E. Sarrouh}
\affiliation{Department of Physics and Astronomy, York University, 4700 Keele Street, Toronto, ON, M3J 1P3, Canada}
\email{gsarrouh@yorku.ca}

\author[0009-0009-4388-898X]{Gregor Rihtar\v{s}i\v{c}}
\affiliation{Faculty of Mathematics and Physics, Jadranska ulica 19, SI-1000 Ljubljana, Slovenia}
\email{gregor.rihtarsic@fmf.uni-lj.si}

\author[0000-0002-4542-921X]{Roberto Abraham}
\affiliation{David A. Dunlap Department of Astronomy and Astrophysics, University of Toronto, 50 St. George Street, Toronto, Ontario, M5S 3H4, Canada}
\affiliation{Dunlap Institute for Astronomy and Astrophysics, 50 St. George Street, Toronto, Ontario, M5S 3H4, Canada}
\email{roberto.abraham@utoronto.ca}

\author[0000-0003-3983-5438]{Yoshihisa Asada}
\affiliation{Dunlap Institute for Astronomy and Astrophysics, 50 St. George Street, Toronto, Ontario, M5S 3H4, Canada}
\email{yoshi.asada@utoronto.ca}

\author[0000-0001-5984-0395]{Maru\v{s}a Brada\v{c}}
\affiliation{Faculty of Mathematics and Physics, Jadranska ulica 19, SI-1000 Ljubljana, Slovenia}
\affiliation{Department of Physics and Astronomy, University of California Davis, 1 Shields Avenue, Davis, CA 95616, USA}
\email{marusa.bradac@fmf.uni-lj.si}

\author[0000-0001-9298-3523]{Kartheik Iyer}
\affiliation{Columbia Astrophysics Laboratory, Columbia University, 550 West 120th Street, New York, NY 10027, USA}
\email{kgi2103@columbia.edu}

\author[0000-0003-3243-9969]{Nicholas S. Martis}
\affiliation{Faculty of Mathematics and Physics, Jadranska ulica 19, SI-1000 Ljubljana, Slovenia}
\email{nicholas.martis@fmf.uni-lj.si}

\author{Ga\"el Noirot}
\affiliation{Space Telescope Science Institute, 3700 San Martin Drive, Baltimore, Maryland 21218, USA}
\email{gnoirot@stsci.edu}

\author[0000-0002-7712-7857]{Marcin Sawicki}
\affiliation{Department of Astronomy and Physics and Institute for Computational Astrophysics, Saint Mary's University, 923 Robie Street, Halifax, Nova Scotia B3H 3C3, Canada}
\email{marcin.sawicki@smu.ca}

\author[0000-0002-4201-7367]{Chris J. Willott}
\affiliation{National Research Council of Canada, Herzberg Astronomy \& Astrophysics Research Centre, 5071 West Saanich Road, Victoria, BC, V9E 2E7, Canada}
\email{chris.willott@nrc.ca}

\author[0009-0000-8716-7695]{Sunna Withers}
\affiliation{Department of Physics and Astronomy, York University, 4700 Keele Street, Toronto, ON, M3J 1P3, Canada}
\email{sunnaw@my.yorku.ca}



\begin{abstract}
We present the evolution of the resolved mass-weighted age, stellar metallicity, and sSFR of 872 Milky Way Analog (MWA) progenitors up to redshift $z=5$ from the Canadian Unbiased Cluster Survey (CANUCS). The metallicity and mass-weighted ages were obtained via spatially resolved SED-fitting with the non-parametric code Dense Basis. We split the sample into mergers versus non-mergers using the merger parameter from the Gini-$M_{20}$ plane obtained through Gini-$M_{20}$ analysis of the morphology of the stellar mass maps with Statmorph. 
Across our redshift range, non-mergers have negative or flat average age gradients from $-0.022$ to 0.005 dex/kpc, and positive or flat sSFR gradients from $-0.089$ to 0.092 dex/kpc, consistent with inside-out assembly. The average $\log(Z/\Zsun)$ gradients for non-mergers range from $-0.029$ to 0.044 dex/kpc, however, positive gradients only appear between $2 < z < 3$. At every redshift epoch, mergers typically have flatter age gradients, more negative sSFR gradients, and similar metallicity gradients compared to non-mergers.
We divide the property maps of ongoing mergers into separate regions based on their component galaxies, and find little to no difference between the components' average ages or metallicities, but the less massive of the merging system is on average $0.1-0.4$ dex higher in sSFR. Our results point to major mergers contributing some momentary disruption to the general trend of inside-out mass assembly, but does not upend the overall picture of MWA disks growing inside-out over cosmic time.
\end{abstract}



\section{Introduction} \label{sec:intro}






The most widely accepted theoretical model of the formation of disk galaxies like the Milky Way is the inside-out growth of the disk (e.g. \citealt{vandenBosch:1998, Abadi:2003a, Munoz-Mateos:2007, Wang:2011, Bird:2013, Licquia:2015b}). 
For the Milky Way, this is reflected in an older central bulge and thick disk and a younger, extended thin disk\citep{Kilic:2017,Frankel:2019}.
From chemical abundance ratios,  such as [Fe/H] measured from Cepheid variables (e.g. \citealt{Lemasle:2007, Lemasle:2008, Luck:2011a, Luck:2011b,Cheng:2012, Genovali:2014,Lemasle:2018}), the Milky Way's radial metallicity gradient is measured as having a shallow declining slope with values ranging from from -0.06 to -0.04 dex/kpc. Other studies using abundances such as [O/H],  or [N/H] in HII regions also give similar results \citep{Maciel:2015, Cunha:2016, Stanghellini:2018, Pena:2019}.  This agrees with high resolution observations of galaxies in the local universe that show that age and metallicity gradients in disk galaxies both generally decline with galactocentric radius for galaxies with stellar masses $9 < \log\mstar[\msun] < 12$ (e.g. \citealt{Sanchez:2014, GonzalezDelgado:2015, Belfiore:2017, Goddard:2017, Zheng:2017, Sanchez-Menguiano:2018, Boardman:2020a} ).

Negative metallicity gradients have usually been interpreted as evidence of ``inside-out" growth, due to the correlation of star formation efficiency with stellar mass (e.g. \citealt{Prantzos:2000,Marcon-Uchida:2010}). However, measurements of elemental abundances of stellar populations in the Milky Way's thick disk show an inverse relation with velocity and with radii (e.g. \citealt{Spagna:2010,Lee:2011,Curir:2012} ).  \cite{Lian:2023} found with APOGEE data that the Milky Way has a ``bow shaped" [Fe/H] gradient, rising from the center, but declining after $R \gtrsim 6$ kpc. Early studies found negative gradients in the $\alpha$-element abundances of the Milky Way (e.g. \citealt{Matteucci:1989, Chiappini:1997}), but this has been challenged by newer results that indicate there are two separate populations of stars in the MW's thick disk: one with a high-$\alpha$ and one with a low-$\alpha$. These two separate populations provide strong evidence for a past merger event known as Gaia-Enceladus with a remnant now embedded in the thick disk (e.g. \citealt{Fuhrmann:2011, Hayes:2018, Helmi:2018}).

Inverse/positive metallicity gradients are rare in the local universe, but occur with more frequency at higher redshift. Studies using gas-phase metallicity such as \cite{Cresci:2010, Jones:2013, Curti:2020, Simons:2021}, and \cite{Li:2022} find star-forming galaxies at $1\lesssim z \lesssim 3$ may have positive metallicity gradients. However, for the most massive galaxies, their metallicity gradients remain negative even at higher redshifts \citep{Wuyts:2016}. While metallicity gradients are more diverse at higher redshift,  recent studies on resolved ages at high redshift such as \cite{Akhshik:2023} for massive galaxies $10.6< \log\mstar [\msun]<12$ and  \cite{Abdurrouf:2023} for a wide range of stellar masses $8.5<\log(\mstar/\msun)<12$ show that radial age gradients remain negative up to $z\sim 6$. However, \cite{Jin:2024} finds evidence of positive color gradients (blue centers, red outskirts) in $z > 4$ galaxies, and color gradients are highly correlated to age. \cite{Jain:2024} compared ages of the central and outskirts of galaxies from $0.5 < z < 2$ and found that galaxies above the star-forming main sequence tend to have younger centers while galaxies below the main sequence have older centers. 
 
There are several explanations for inverse/positive metallicity gradients. For example, \cite{Rupke:2010, Stott:2014} and \cite{Curti:2020} find a correlation between galaxies having higher specific star formation rate (sSFR) and metal-poor centers, and suggest that the cause is the funnelling of pristine or metal-poor gas towards the center during episodes of intense star formation. \cite{Ma:2017} argues feedback is more efficient for low mass systems, which leads to the flattening of metallicity gradients. \cite{Tissera:2022} found in the EAGLE simulations that only negative [O/H] gradients have a mass evolution dependence, while flat or inverted [O/H] gradients are the result of galaxy interactions such as mergers. \cite{Grossi:2020} argues that two dwarf galaxies in the Virgo cluster have positive [O/H] gradients due to a a recent merger triggering inflow of metal-poor gas from the intergalactic medium. \cite{Schonrich:2017} showed that increased SFR can flatten or invert the stellar metallicity gradient, but may steepen negative gas-phase metallicity gradients. 

This has implications for the MW, as the thick disk's metallicity gradient is unusual for the local universe, and its formation was the result of a past merger event. \cite{Ciuca:2024} shows the Gaia-Enceladus merger may have triggered a starburst, and then induced gas to condense into the MW's thin disk afterwards. However, \cite{Kawata:2018} found in an N-body simulation of the thick disk that the most likely formation scenario is still inside-out. Radial mixing transforms the initial mildly positive [Fe/H] radial gradient to a mildly negative one. \cite{Sotillo-Ramos:2022} find in the TNG50 simulation that many disks remain intact after a major merger, and destroyed disks can also reform. \cite{Yu:2023} find in FIRE-2 that a bursty phase of star formation led to the thick disk, and a steady phase of star formation led to the thin disk. These simulations all provide support that mergers do not always disrupt inside-out formation of disk galaxies. 

In our previous paper \cite{Tan:2025}, we showed that the stellar mass and SFR density for MWAs in CANUCS (Canadian NIRISS Unbiased Cluster Survey) are generally in line with inside-out formation up to $z = 5$, although the $z > 3$ universe is burstier and has more merger events. In this paper, we obtain mass-weighted age and stellar metallicity radial gradients for MWA progenitors, and uncover their implications for the mass assembly history of the Milky Way. Furthermore, we separate our samples into non-mergers, ongoing mergers, and late stage mergers to understand the role that merging galaxies play in the age and metallicity distribution of star-forming disk galaxies. We assume a flat $\Lambda$CDM cosmology with $\Omega_{M} = 0.3$, $\Omega_\Lambda = 0.7$, $H_0 = 70$ km s$^{-1}$ Mpc$^{-1}$, and a \cite{Chabrier:2003} IMF.

\section{Data} \label{sec:data}

\subsection{The CANUCS Catalogs}

\begin{figure*}
\gridline{\fig{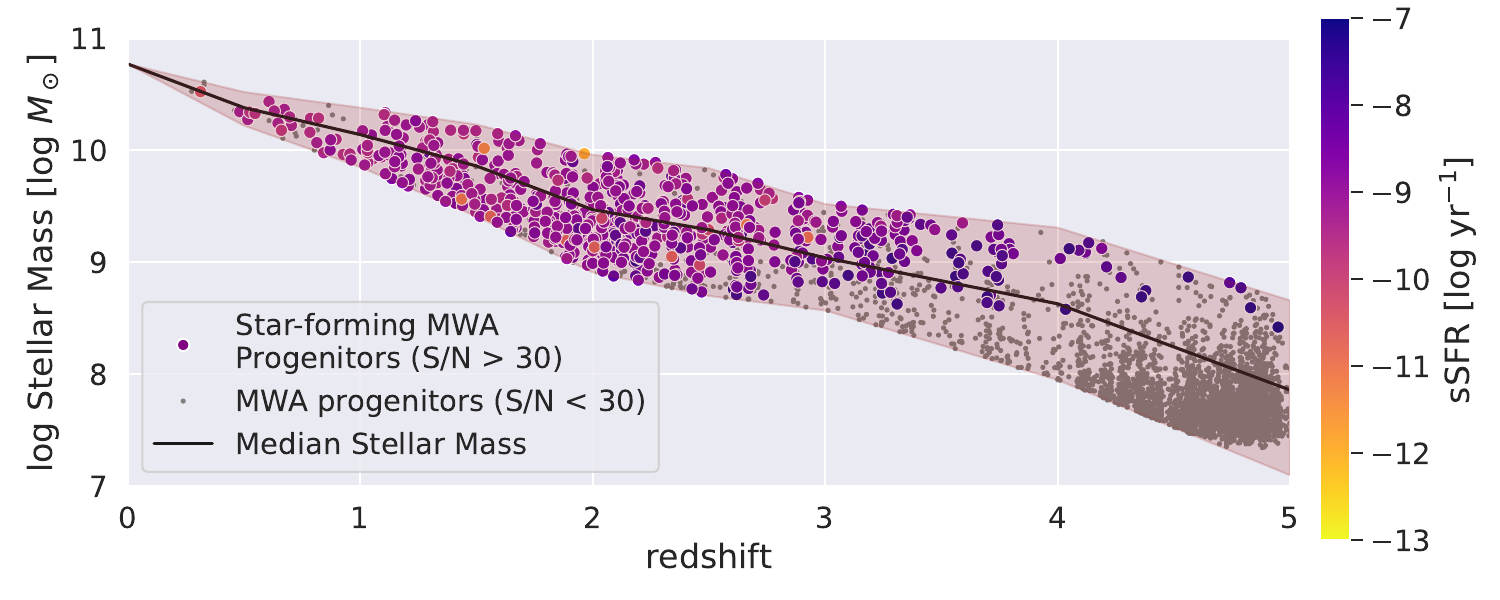}{\textwidth}{}}
\vspace{-1cm}
\gridline{\fig{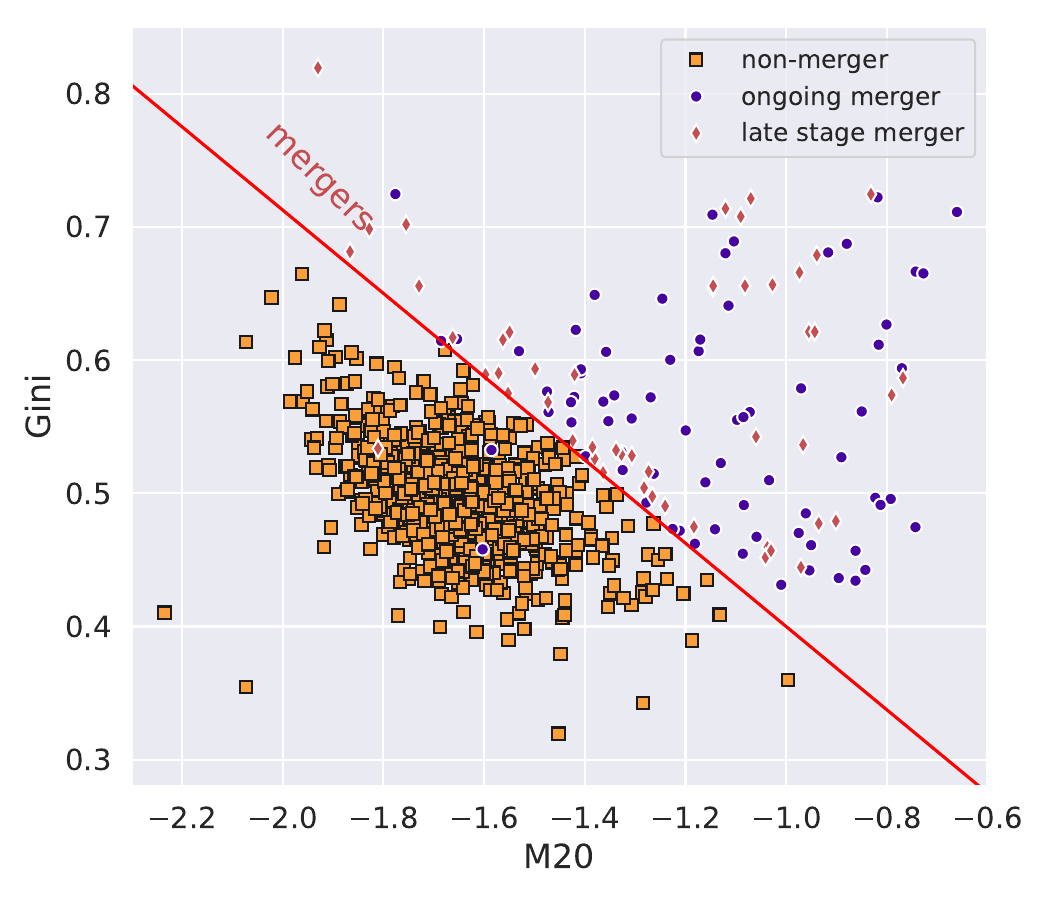}{0.465\textwidth}{}
	\fig{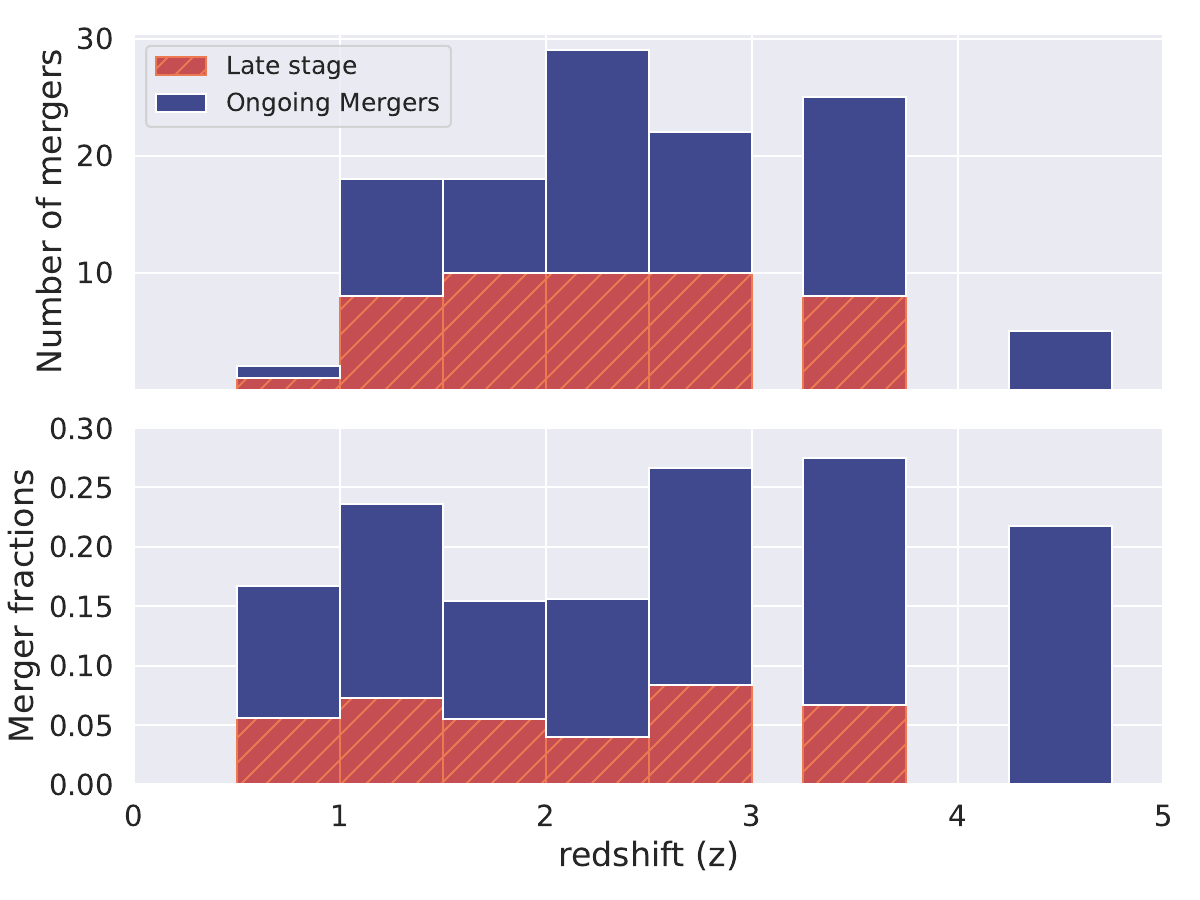}{0.535\textwidth}{}}
	\vspace{-0.5cm}
   \caption{Top panel: Integrated stellar mass of all star-forming MWA progenitors in our sample from the CANUCS DR1 catalogs \citep{Sarrouh:2025}, as a function of redshift. their color indicates specific star formation rate (sSFR).  Bottom left: Gini-$M_{20}$ distribution of our sample. Yellow squares are non-mergers, red diamonds are late stage mergers, blue circles are ongoing mergers. The three non-yellow points below the dividing line were objects at low redshift $z < 1.5$ reclassified as mergers upon visual inspection.  Bottom right: Number of mergers, as well as merger fractions per redshift bin. } \label{fig:this-sample}
\end{figure*}

As with the previous paper \cite{Tan:2025} in this series,  we use the observations from the Canadian NIRISS Unbiased Cluster Survey (CANUCS, see \citealt{Willott:2022}) for its broad wavelength coverage in the infrared and depth of field. The full CANUCS dataset includes NIRCam photometry, with spectroscopy from NIRISS and NIRSpec of five strong lensing clusters Abell 370, MACS J0416.1-2403 (hereafter MACS 0416), MACS J0417.5-1154 (hereafter MACS 0417), MACS J1149.5+2223 (hereafter MACS 1149), and MACS J1423.8+2404 (hereafter MACS 1423).  Lensing models for Abell 370 are provided in \cite{Gledhill:2024}, for MACS0416 in \cite{Rihtarsic:2025}, MACS1149 in Rihtar\v{s}i\v{c} et al. (in prep), and MACS 0417 and MACS 1423 in Desprez et al (in prep).

The work in this paper uses the most up-to-date  photometric catalogs from CANUCS (v2p0.3),  and includes up to 29 different filters in some flanking fields. Details of image processing are given in the CANUCS data release (DR1) paper \cite{Sarrouh:2025} . Details of PSF-matching and catalog creation for CANUCS are described in \cite{Willott:2024} and \cite{Sarrouh:2024}. Source detection and aperture photometry was done using the Photutils package from astropy \citep{Bradley:2016,Bradley:2023} as described in \cite{Asada:2024,Asada:2024b}. For more information on the data reduction process for CANUCS, see \cite{Noirot:2023,Desprez:2024,Asada:2024}, and \citealt{Willott:2024}. 

For the entire cutout of the galaxy, the signal-to-noise cutoff necessary for successful Voronoi binning is $S/N > 30$ in F444W, which is obtained from the measured flux and flux error in the CANUCS photometric catalogs. This threshold is set to ensure that the individual spatial bins have $S/N > 5$. This means that out of all possible MWAs in the CANUCS fields, only the brightest at high-$z$ can be retained in our sample. More information on Voronoi binning is provided in \cite{Tan:2025}. Photometric redshifts in the CANUCS catalogs \citep{Asada:2024} were obtained with an implementation of EAZY in Python known as \texttt{eazy-py} \citep{Brammer:2008}. During fitting, a systematic flux error of $5\%$ was added to the nominal photometric uncertainty. 

\subsection{MWA Progenitor Selection}
For this work, we use the same sample of MWA progenitors as selected in \cite{Tan:2025}. However, as we have updated photometry, we make use of the additional photometric bands in the flanking fields (NCF) for more robust SED-fitting results. 

A summary of the selection criteria for our progenitors is as follows:
\begin{enumerate}
    \item The abundance matching algorithm from \cite{Behroozi:2013a} was applied to a galaxy of stellar mass and cumulative number density as the current Milky Way.
    \item We extrapolate the cumulative number density of possible progenitors up to $z = 5$.
    \item We take stellar mass functions from \cite{Grazian:2015} and \cite{McLeod:2021}, using the number densities and their $1\sigma$ errors to determine the range of stellar masses of MWA progenitors at that redshift range.
    \item We search for galaxies in the CANUCS catalogs with an integrated stellar mass within the bounds obtained from abundance matching to stellar mass functions at that redshift.
    \item Galaxies in the cluster fields with a magnification $ \mu > 2.5$ were removed. The rest of the cluster sample's total stellar mass and SFR were corrected by the magnification factor.  245 cluster galaxies have a magnification $\mu > 1$ and  within that sample, 40 galaxies have $\mu > 1.5$. 
    \item Galaxies which have overall signal-to-noise $S/N < 30$ were also removed, as they cannot be successfully spatially binned such that every bin has sufficient signal-to-noise ($S/N \geq 5$). 
\end{enumerate}

The integrated stellar mass versus catalog photometric redshift is plotted in the top panel of Figure \ref{fig:this-sample}. The galaxies with enough overall signal-to-noise are large colored circles with the color representing their specific star formation rate (sSFR). The grey circles are objects in the catalog which are within the stellar mass range of that redshift epoch, but do not have enough signal-to-noise. Therefore, our sample is not entirely mass complete at $z> 3$. The sample is more biased towards more massive and more star-forming progenitors as redshift increases beyond $z=3$.

\subsection{Selecting the merger sample}

Selecting galaxy mergers is a challenging task, as clumpy galaxies may masquerade as merging systems. Given that a large number of star-forming clumps are low in stellar mass, despite their brightness (e.g. \citealt{Wuyts:2012,Sok:2022}), we therefore use Gini-$M_{20}$ statistics \citep{Lotz:2004,Lotz:2008a} on the resolved stellar mass maps to identify merging galaxies.

Categorizing mergers via the Gini-$M_{20}$ plane was described in detail in \S 5 of \cite{Tan:2025}, which we summarize in this section. We use the code Statmorph \citep{Rodriguez-Gomez:2019} on the stellar mass maps of our galaxies to obtain their Gini coefficients and $M_{20}$ statistics. Since we are working with stellar mass distributions, any time a flux was given in the original equations from \cite{Lotz:2004}, we instead replace that quantity with a stellar mass density $\sigma_{\star}$.

The Gini coefficient is given as:
\begin{equation}
    G = \frac{1}{\bar{\sigma}_{\star} n (n-1)} \sum^{n}_{i = 1} (2i - n - 1)\sigma_{\star,i} ,
\end{equation}
where  $\bar{\sigma}_{\star}$ is the average stellar mass of all the pixels in an image with $n$ pixels, $\sigma_{\star,i}$ is the stellar mass at each individual pixel. A high Gini coefficient indicates higher concentration of stellar mass within a few pixels, with a Gini coefficient of 1 indicating all of the stellar mass is contained in a single pixel. 

Meanwhile, the $M_{20}$ statistic finds the ``brightest pixels" --- the pixels that contain 20\% of all of the image's flux,  and where those pixels are distributed relative to the center of the image. Again, for our purposes, flux is replaced with stellar mass and the $M_{20}$ statistic is tasked with finding the spatial distribution of the ``most massive" pixels. It does so by calculating the second order of the brightest(most massive) moments $\mu_{\star,i}$ compared to the second order of all of the moments  $\mu_{\star,tot}$, which identifies galaxy bulges, clumps, and spiral arms.  $M_{20}$ is defined as:

\begin{equation}
    M_{20} \equiv \log_{10} \left(\frac{\sum_{i} \mu_{\star,i}}{\mu_{star,tot}}\right) , \text{while} \sum_{i} \sigma_{\star,i}< 0.2 \sigma_{\star,tot},
\end{equation}

Potential mergers are identified as galaxies positioned in the top right half of the Gini-M20 plane , with a division line dividing mergers and non-mergers cutting through the diagram from top left to bottom right. Galaxies on the right side of the division line are assigned a positive $S_{merger}$ parameter while galaxies on the left have a negative $S_{merger}$ parameter. We take the merger sample as all galaxies with a positive $S_{merger}$ in their stellar mass distribution. According to \cite{Sazonova:2024}, morphology parameters at various redshifts are affected by angular resolution, and in \cite{Tan:2025}, the merger sample was taken to be the $S_{merger}$ parameter of the stellar mass maps resolution matched to the redshift with the poorest angular resolution ($z \sim 1.6$). However, to take into account as many merging galaxies as possible, the merging population was drawn from galaxies that have either a positive $S_{merger}$ of its original resolution \textit{or} a positive $S_{merger}$ from the reprojected resolution. 

\begin{figure*}
\gridline{\fig{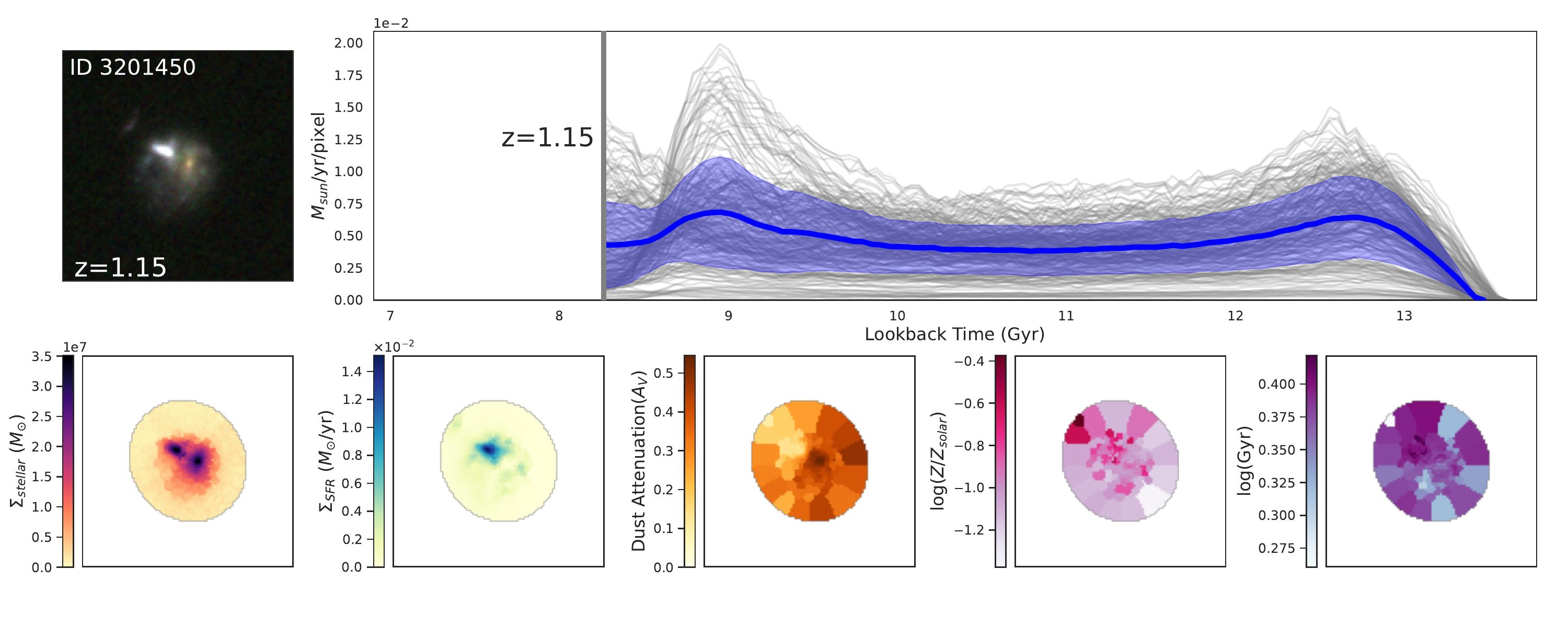}{\textwidth}{}}
\vspace{-1.5cm}
\gridline{\fig{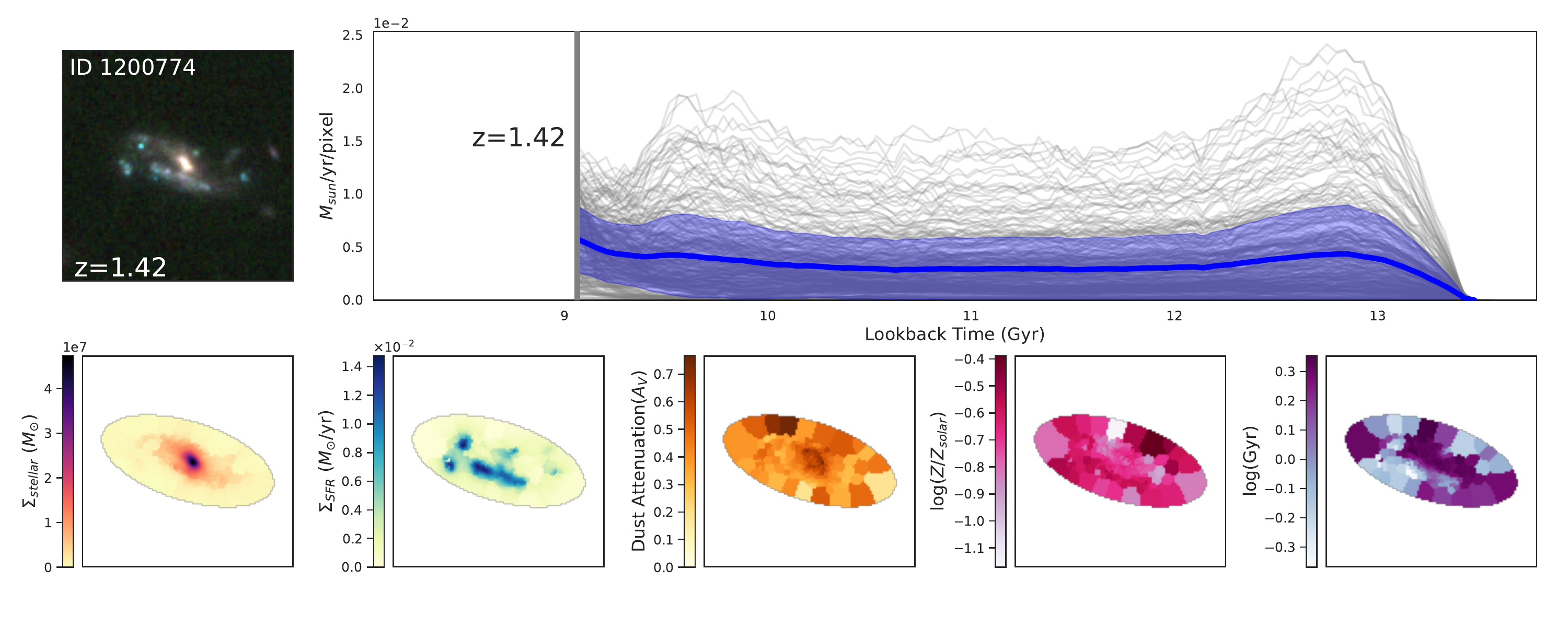}{\textwidth}{}}
\vspace{-1.5cm}
\caption{Color image, SFHs and property maps (left to right, stellar mass density, SFR density, dust attenuation, stellar metallicity, mass-weighted age) for two example galaxies at redshift ranges $1<z<2$, top is a merger, bottom is a non-merger. Every spatial bin's SFH (SFR/yr/pixel) is plotted grey. The blue line is the median SFH, while shaded region is the $1\sigma$ deviation of the same SFHs. The empty region at the beginning of each SFH plot is to indicate that the SFHs are calculate up to the lookback time of when the galaxy is observed.}

    \label{fig:gal-properties} 
\end{figure*}

\begin{figure*}
\gridline{\fig{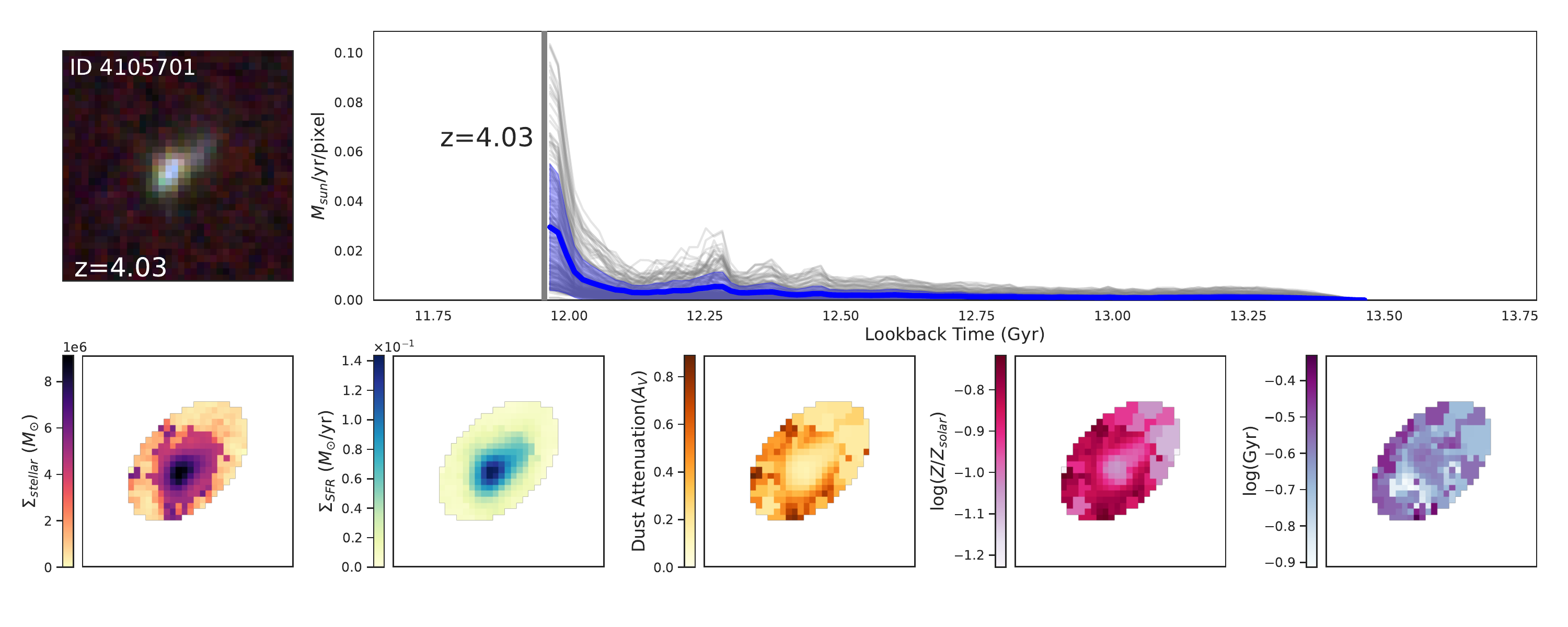}{\textwidth}{}}
\vspace{-1.5cm}
\gridline{\fig{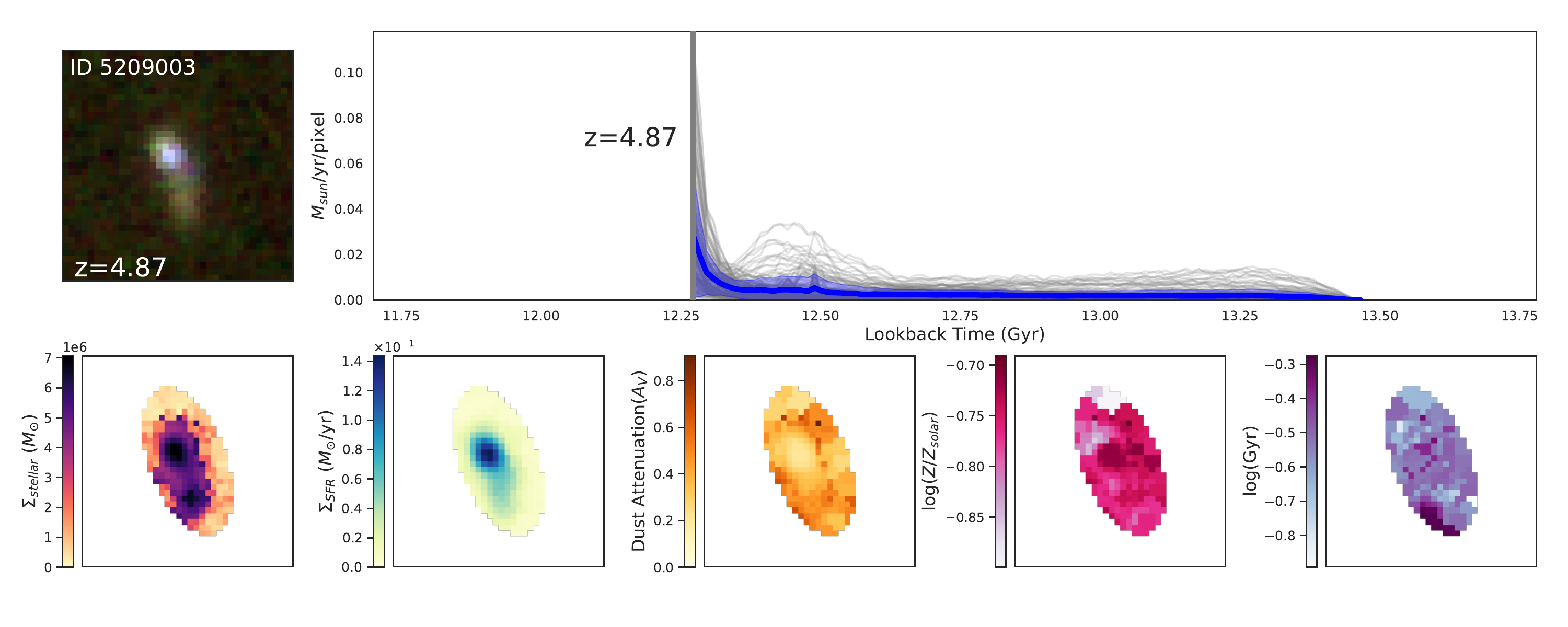}{\textwidth}{}}
\vspace{-1.5cm}
\caption{Color image and property maps for two galaxies at redshift ranges $4<z<5$, top is a non-merger, bottom is a merger. }

    \label{fig:gal-properties2} 
\end{figure*}

The merger sample is then separated by visual inspection into two main categories: ongoing mergers, and potential late stage mergers. Ongoing mergers have two or more distinguishable peaks in the stellar mass distribution, while potential late stage mergers only have a single peak but are asymmetric or otherwise irregular in structure. Furthermore, false positives, where two galaxies are spatially close in the image, but have photometric redshifts greater than $\Delta z = 0.25$ apart were removed from the sample, or reclassified as late-stage mergers if there are enough irregular or asymmetric features. In the bottom left panel of Figure \ref{fig:this-sample}, the Gini-$M_{20}$ plane is plotted for all star-forming MWA progenitors, with the ongoing and late stage mergers plotted in blue and red points respectively. The two blue and red points that are below the merger division line are mergers which had a negative $S_{merger}$ statistic, but upon inspection of their color images and stellar mass maps, are indeed examples of an ongoing or a late stage merger. The bottom right panel of Figure \ref{fig:this-sample} have the number of mergers per redshift bin, as well as the merger fraction per redshift range. Example color images, as well as example property maps of non-merger versus merger galaxies are shown Figure \ref{fig:gal-properties} and \ref{fig:gal-properties2}.  The total number of galaxies per redshift bin, as well as the number of mergers in the same bin is given in Table (table whatever).

\begin{table}
\centering
\begin{tblr}{
  width = \linewidth,
  colspec = {Q[408]Q[154]Q[154]Q[192]},
  hline{1,10} = {-}{0.08em},
  hline{2} = {-}{},
}
{\\Redshift}       & {No. of\\galaxies} & {No. of~\\mergers} & {Ongoing/\\Late stage} \\
$0.0 < z < 0.5$ & 2                  & 0                  & ~0~~~~0                 \\
$0.5 < z < 1.0$ & 16                 & 2                  & ~1~~~~1                 \\
$1.0 < z < 1.5$ & 92                 & 18                 & 10~~~~8                \\
$1.5 < z < 2.0$ & 163                & 18                 & ~8~~~10               \\
$2.0 < z < 2.5$ & 221                & 29                 & 19~~~10                \\
$2.5 < z < 3.0$ & 123                & 22                 & 12~~~10                \\
3.$0 < z < 4.0$ & 95                 & 25                 & 17~~~~8                 \\
$4.0 < z < 5.0$ & 18                 & 5                  & ~5~~~~0                
\end{tblr}
\caption{Number of galaxies selected in total at each redshift range along with the total number of mergers, and mergers divided into ongoing mergers and late stage mergers.}\label{table:samples}
\end{table}

\section{Methods} \label{sec:methods}






\subsection{SED-fitting}\label{sec:sed-fitting}

Our goal is to create spatially resolved property maps from SED-fitting, not only for the stellar mass and star formation rate, but also for metallicity and age. The latter properties, in conjunction mass and SFR, produces a more complete picture of the evolution and growth of MWAs. We use the same spatial bins for our galaxies as \cite{Tan:2025}, which were generated by Voronoi binning \cite{Cappellari:2013},  because we are studying the same sample of galaxies. We use Dense Basis \citep{Iyer:2017, Iyer:2019} to perform non-parametric SED-fitting, with the stellar population synthesis code FSPS \citep{Conroy:2010} loaded into the backend. The default initial mass function (IMF) for Dense Basis is from \cite{Chabrier:2003}, and the default dust law is \cite{Calzetti:2000}.

In addition to using the most updated version of the CANUCS photometry and the photometric catalogs that include the Technicolor filters, we have also updated the prior limits for Dense Basis SED-fitting to obtain more accurate metallicities and ages. This does not affect the derived stellar mass and SFR. We use a flat sSFR prior with $-13 \leq \log(\text{yr}^{-1})\leq -6$, a flat dust $A_V$ prior with a range of 0 to 4, and the range for metallicity in the prior is $-2.5 < \log{(Z/\Zsun)}<0.3$. Note the choice of priors for metallicity and dust attenuation has changed slightly from the ones used in \cite{Tan:2025}. The priors as described in \cite{Tan:2025} were the same priors as for the CANUCS Dense Basis mass catalogs (see Table 4 in \S 6 of \citealt{Sarrouh:2025}). This is reasonable for measuring resolved stellar mass and SFR (although there is also a slight degeneracy between SFR and both metallicity and dust $A_V$, see \citealt{Iyer:2019} for a detailed discussion of SFH uncertainties). However, properties such as metallicity, dust, and age, which are more degenerate with each other \citep{Conroy:2013,Santini:2015,Nagaraj:2022} and thus more prior dependent, may not be as accurate with the catalog priors. For example, the catalog priors had narrower boundaries ($-1.5 < \log{(Z/\Zsun)}<0.25$.) and the dust prior is exponential. This is in order to accommodate for the large number of high redshift galaxies ( $z >> 5$, around $z\sim 10$) which have low signal-to-noise. However, a wider metallicity range with a flat dust prior is more accurate for $z < 5$ galaxies with high enough signal-to-noise for spatial binning. We show in Appendix \ref{appendix-b} the corner plots of the best fit age, metallicity, and dust $A_V$ of every spatial bin in our galaxy sample. The shape of the corner plots demonstrate that age and metallicity are only mildly degenerate, and the degeneracies do not strongly affect our results, since \S \ref{sec:gradients} show that the vast majority of galaxies have radial gradients close to zero. Likewise, degeneracies with dust do not affect the vast majority of output properties for spatial bins.

\begin{figure}
     \centering
     \includegraphics[width=\columnwidth]{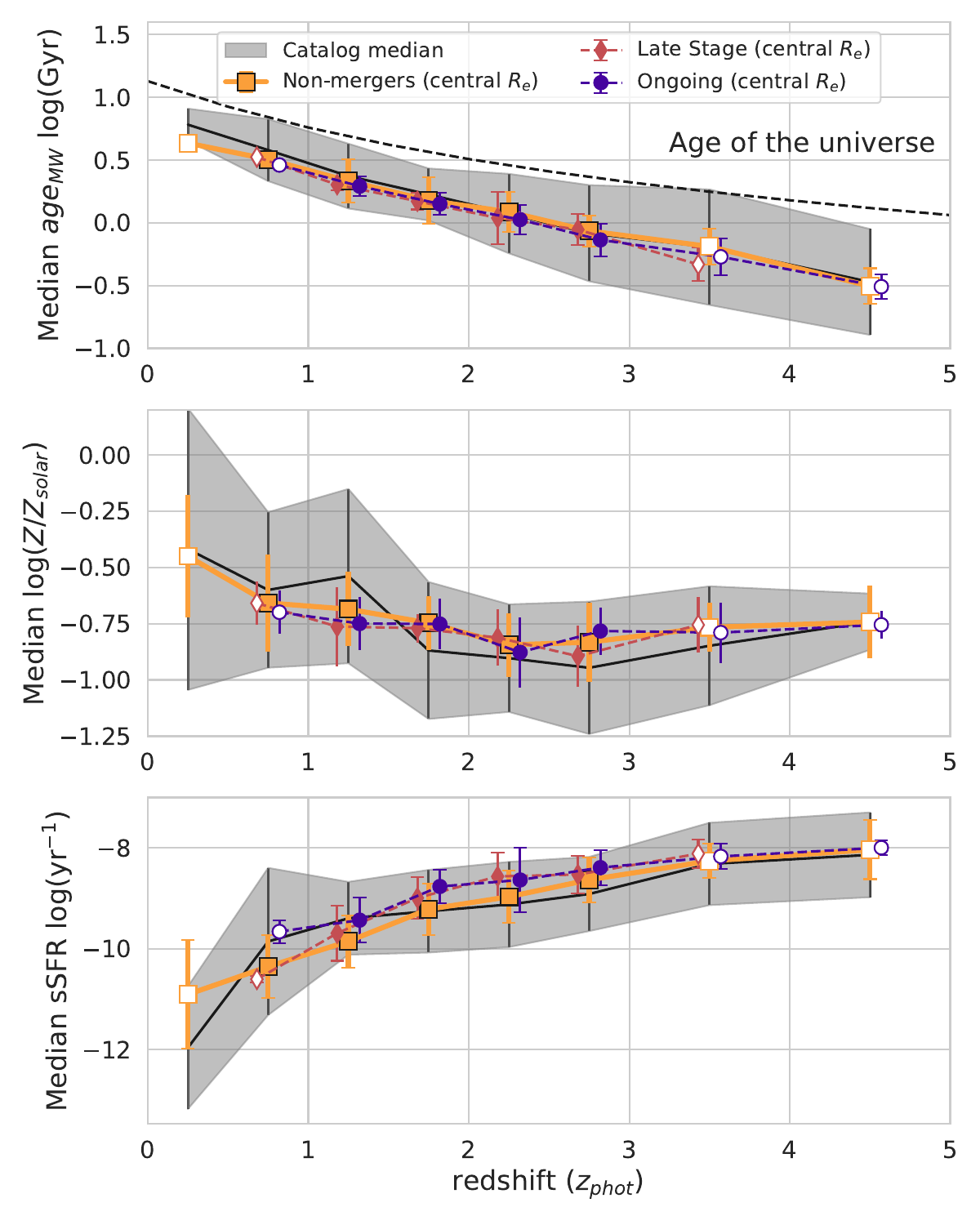}
     \caption{Median mass-weighted age (top panel, stellar metallicity (middle panel), and sSFR (bottom panel) of the inner 1$R_{eff}$ (half-light radius) of the resolved MWA progenitor sample for every redshift bin. Error bars are the 1$\sigma$ scatters of the properties' distributions. Open markers indicate bins affected by mass-incompleteness. The black solid lines and the grey shaded regions represent the median and the 1$\sigma$ scatter respectively, of the \textit{integrated} (unresolved) properties of all possible MWA progenitors from the catalog. The catalog's age is $t_{50}$ (age when 50\% of stellar mass has formed in the galaxy) instead of derived from Equation \ref{eqn:mw_age}.}     \label{fig:avg-props}
 \end{figure}

\subsection{SFHs and Mass-weighted ages}\label{sec:sfhs}
In practice, a mass-weighted age of a galaxy is similar to the average age of all of the stars within that galaxy. Since it is weighed by stellar mass, there is less outshining bias from newly formed stars. Therefore, the mass-weighted age can be a more accurate measurement of the overall amount of star formation that has occurred up to time of observation of the galaxy.

Given an SFH returned by Dense Basis for a particular spatial bin within a galaxy, we get the total mass formed since the birth of the galaxy by integrating the entire SFH over time:
\begin{equation}
    M_{formed}(t) = \int_{t_0}^{t_{age}} \text{SFR}(t) \text{d}t \;\;, \label{eqn:sfh}
\end{equation}
where $t_{age}$ is the age of the galaxy at time of observation, and $t_0$ is when the galaxy first began forming stars. Since Dense Basis returns SFHs in SFR versus lookback time, we must convert it to ``common time" by flipping the time array. Note that $M_{formed}$ is distinct from stellar mass \mstar, because a portion of the formed mass will be returned to the interstellar medium via stellar evolution.
Then we obtain the mass-weighted age $t_{MWage}$ by multiplying the age $t_i$ at each time step by its corresponding $M_{formed}$ at $t_i$, taking the sum, and then dividing by the total mass formed over all time:
\begin{equation}
    t_{MWage} = \frac{\sum_{t_0}^{t_{age}} t_iM_{formed}(t_i) }{\sum M_{formed}(t)} \;\;. \label{eqn:mw_age}
\end{equation}
 The mass-weighted age is more closely associated with the median age of the stars rather than the age of the brightest stellar populations observed for that region of the galaxy. In applying Equation \ref{eqn:mw_age} to each spatial bin, we are using a modified version of the function \texttt{calc\_mw\_age()} from piXedfit \citep{Abdurrouf:2021}.

\begin{figure*}
    \centering
    \includegraphics[height=5.8cm]{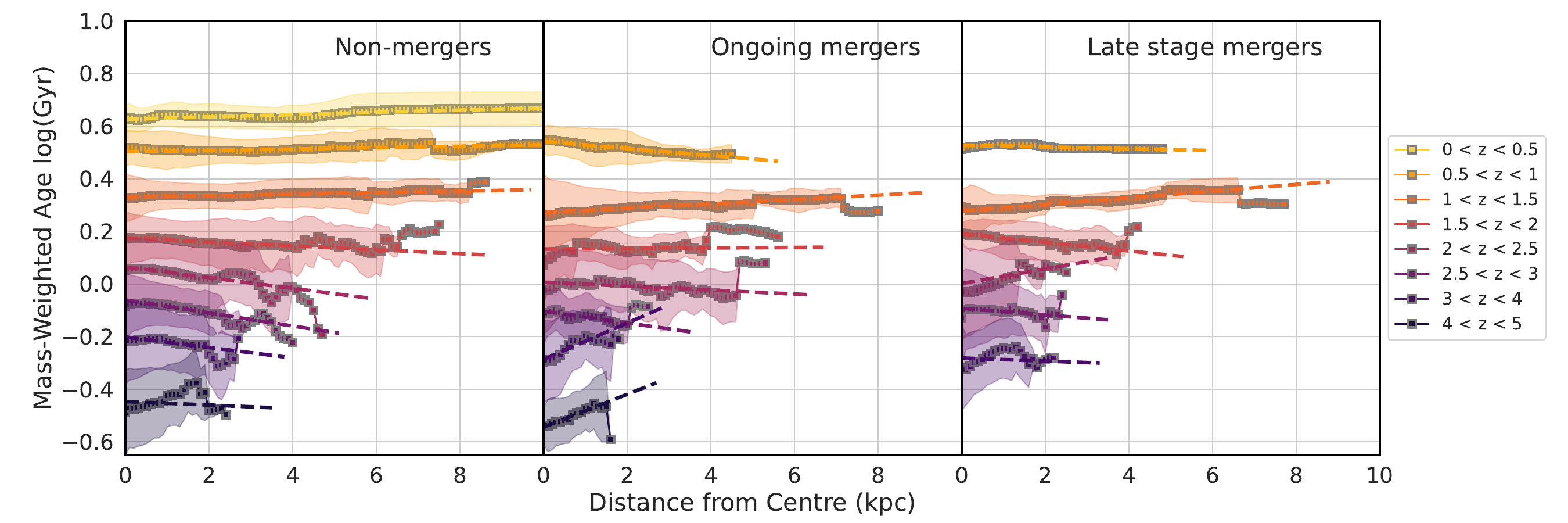}
    \caption{The average mass-weighted age gradients for each redshift epoch created by stacking and normalizing each mass-weighted age profile. The distance from the galactic center is shown in kpc. The shaded regions represent the $1\sigma$ scatter, or the 16th-84th percentile range of the properties. Left panel is the non-merger sample, middle are ongoing mergers, and right panel are late stage mergers. Dashed lines indicate line of best fit to the averaged gradient, which was fit via a weighted linear regression method, with the weights coming from the $1\sigma$ errors of the average profiles (i.e. the shaded regions).}\label{fig:mw_ages_stack}
\end{figure*}
\begin{figure*}
    \centering
    \includegraphics[width=\textwidth]{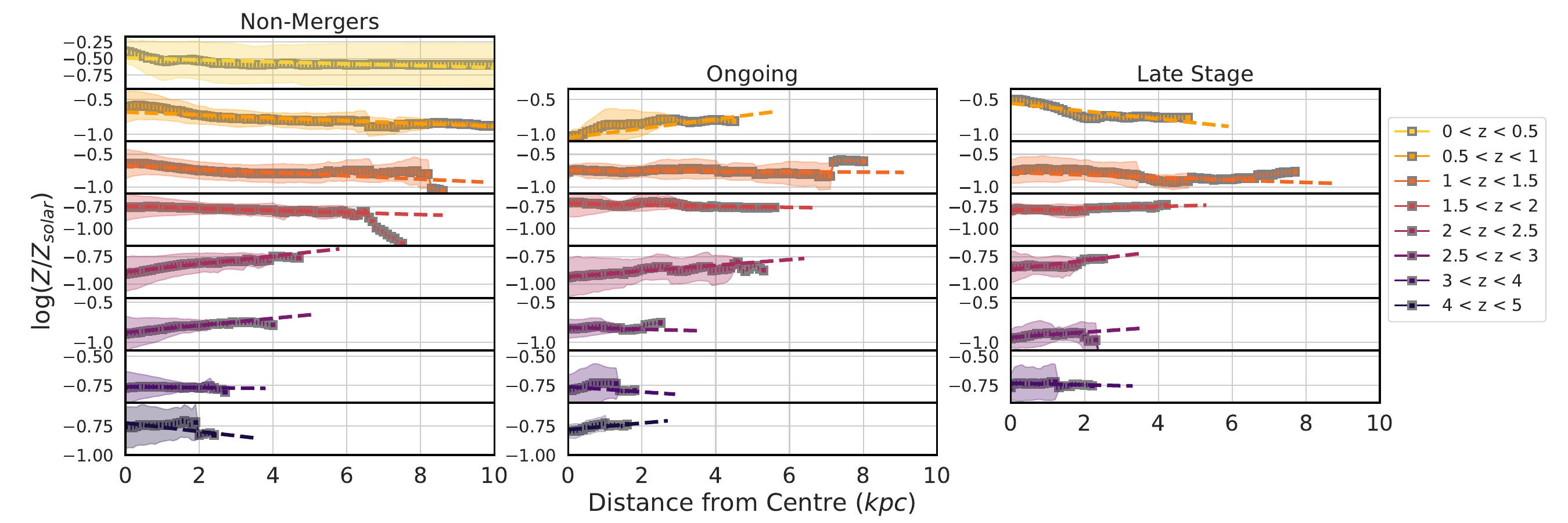}
    \caption{Similar plots as Figure \ref{fig:mw_ages_stack}, but for stellar metallicity $\log(Z/\Zsun)$. Each radial gradient is on its own axis for visual clarity.}\label{fig:met_stack}
\end{figure*}
\begin{figure*}
    \centering
    \includegraphics[width=\textwidth]{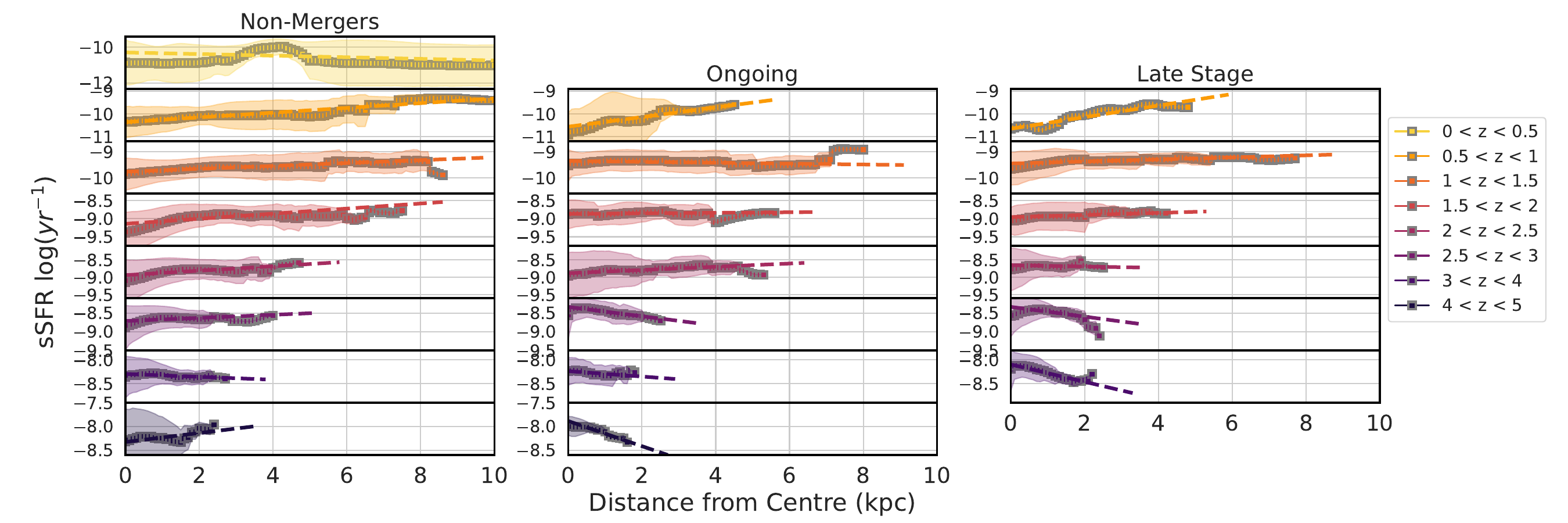}
    \caption{Similar plots as Figures \ref{fig:mw_ages_stack} and \ref{fig:met_stack}, but for sSFR. Each radial gradient is on its own axis for visual clarity.}\label{fig:ssfr_stack}
\end{figure*}
\section{Radial Profiles across the redshift range}\label{sec:gradients}
In this section, we present the results of creating radial property gradients for each of the 872 MWA progenitors, divided into distinct redshift epochs. This was done both with elliptical annuli of 0.1 kpc width, as well as 0.1 effective radii ($R_{eff}$, circularized half-light radii from the CANUCS photometry catalogs \footnote{measured in Photutils from the F150W 20mas resolution image, deconvolved.}) in width, centered on the central pixel of each respective galaxy (or the more massive galaxy for ongoing mergers).  Unlike the stellar mass and SFR maps, the metallicity and age maps are not densities. For the individual age and metallicity gradients, we take the weighted average of the respective property of the pixels of that annulus, with the weight for each pixel determined by its $\Sigma_{\star}$. Radial gradients for sSFR are created by taking $\Sigma_{SFR}/\Sigma_{\star}$ at each annulus. The radial gradients scaled to kpc are to put into context the changes to the spatial distribution of those properties \textit{physically}, as measured between different redshift epochs for MWA progenitors. For this reason, most of our results are presented as radial profiles measured in kpc. However, the majority of radial gradients measured at low-$z$ are presented as a function of effective radius. For the purposes of comparison with local universe resolved studies, the same profiles as a function of $R_{eff}$ and their resultant slopes are shown in Appendix \ref{app:reff}.

Since the radial profiles are affected by lensing magnification, just as in \cite{Tan:2025} we apply a transformation to obtain size measurements as close to the resolution of the source plane as possible. We define a transformation matrix based on $\phi$, the angle of the strongest lensing distortion, and $\theta$, the angle perpendicular to $\phi$, and apply it to $a$, $b$, and $\theta$ (position angle) measured in the observed plane. To preserve resolution and the observed PSF, we scale the angular resolution by the change in the lengths of $a$ and $b$ respectively, as opposed to transforming the images and property maps themselves.

In Figure \ref{fig:avg-props}, we take the median properties: mass-weighted age, stellar metallicity, and sSFR, in of the inner 1 $R_{eff}$ of each galaxy for each subsample (non-merger, ongoing merger, and late stage merger) as a function of redshift. They are also plotted against the median values for the full, unresolved MWA sample, with integrated properties from the CANUCS DR1 \citep{Sarrouh:2025}. Our resolved sample has median properties which agree well at all redshifts with the integrated catalog properties. For median ages, since the catalog does not include a mass-weighted age defined in the same way as in Equation \ref{eqn:mw_age}, we instead plot it against the $t_{50}$ age distribution. The change in mass-weighted age implies the galaxies are ``aging" slower than the universe (indicated by the dashed line on the topmost panel), which is in line with the sample being selected for star-forming galaxies.  

The median metallicity shows almost no evolution with redshift, the values in the middle panel of Figure  \ref{fig:avg-props} are all below $-0.6\log(Z/\Zsun)$ for all three subsamples, except for the non-merger sample at the lowest redshift. The median metallicity of the ongoing mergers also appear to diverge slightly from the non-merger and late stage merger at $z< 1$, being roughly 0.2 dex lower in metallicity. However, the error bars still indicate general agreement between all three populations.

The bottom panel of Figure \ref{fig:avg-props} shows the sSFR evolution with redshift. Although \cite{Tan:2025} shows SFR increases up to $z\sim2$ (cosmic noon) before decreasing once more until $z=0$, the sSFR on the other hand, is consistently decreasing with respect to time.

\begin{figure}
    \centering
    \includegraphics[width=\columnwidth]{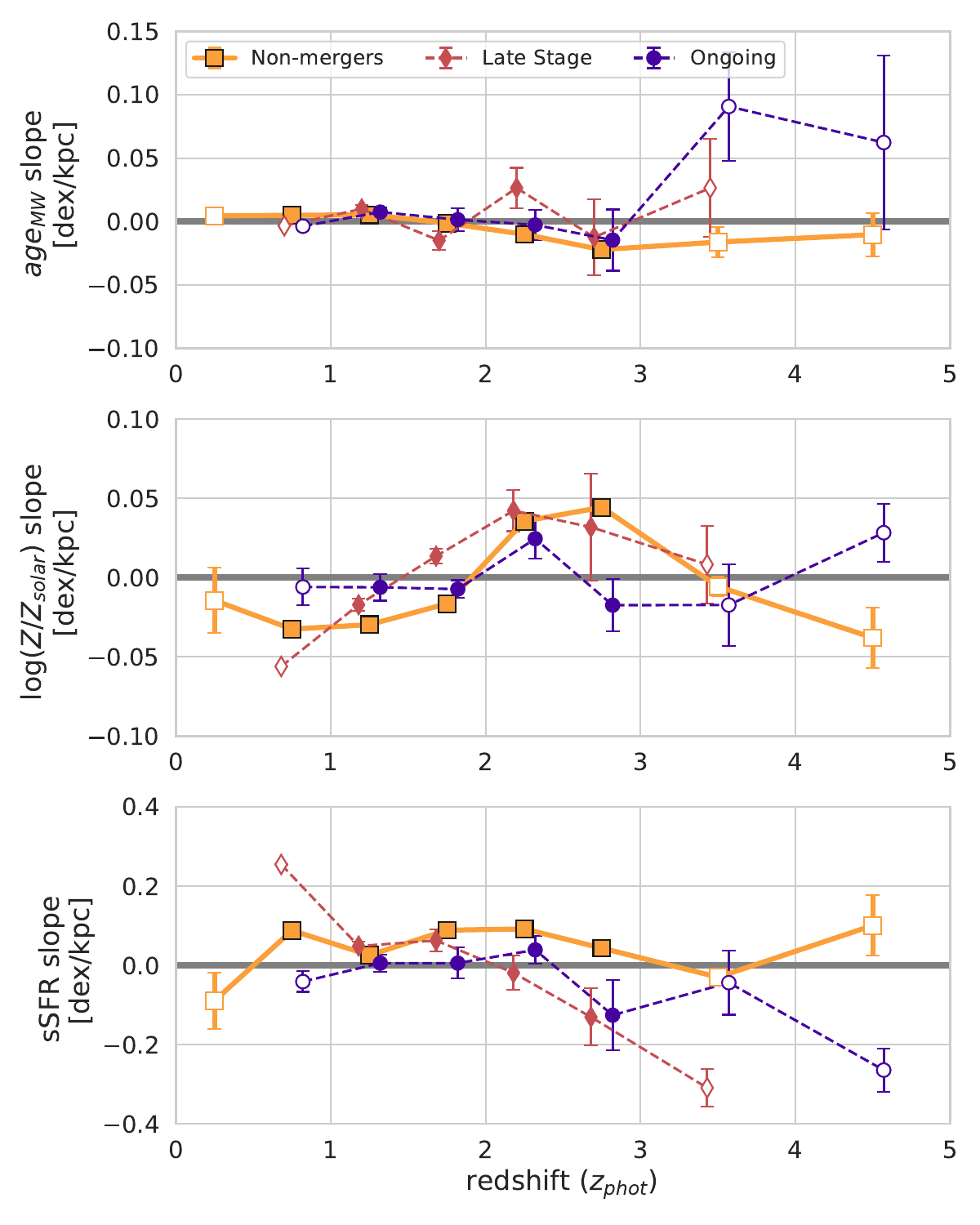}
    \caption{Slopes of the mass-weighted age (top panel), metallicity (middle panel) and sSFR density (bottom panel) gradients for each redshift bin. Yellow squares are non-mergers, red diamonds are late stage mergers, blue circles are ongoing mergers. Error bars represent the fitting error in the slope, weighed by the scatter in the distribution. Open markers indicate bins affected by mass incompleteness.}
    \label{fig:all-slopes}
\end{figure}

\subsection{Average radial gradients for each redshift epoch}
To better understand the relationships between star formation, ages, and metallicity within galaxies, we must examine the behaviour of the radial gradients of their properties. In \cite{Tan:2025}, we have shown using radial gradients of $\Sigma_{\star}$ and sSFR that the MWA progenitors are building mass inside-out. 

We obtain an averaged radial gradient by stacking every profile in each redshift bin, and then normalizing the stacked profile by dividing by the number of galaxies. This is done for the non-merger, ongoing merger, and late stage merger subsamples. These averaged profiles are plotted in Figures \ref{fig:mw_ages_stack}, \ref{fig:met_stack}, and \ref{fig:ssfr_stack}  for mass-weighted age, metallicity, and sSFR respectively, as the colored square points.  The shaded regions represent the $1\sigma$ scatter for the profiles at each redshift. Since the property maps are cut off at 2.5 times the Kron radius of the galaxy from the catalog, each individual gradient may be ``cut off" at a different kpc. Therefore at larger radii, the average gradient is the average of less galaxies, similarly for the error bars. 

To obtain the slope of the radial gradients, we perform a linear fit to each average profile with SciPy's \texttt{curve\_fit()} function. We use the standard error at each $\Delta R$ as uncertainty in the optimization of the parameters, and set \texttt{absolutes\_sigma = True}. Standard error ensures a more accurate linear fit, because as previously stated, different galaxies have different radial cutoffs, in effect biasing the stacking for the average profiles at larger radii as there are fewer galaxies to stack. Standard error fixes the bias because it scales the standard deviation by the square root of the number of objects ($\sigma/\sqrt{N_{galaxies}}$). When there is no standard error available (i.e. on the edges of the gradient where there is only data from one galaxy in the average profile), the uncertainty was set to 1, since the optimized function is $\chi^2 = \sum (r/\sigma)^2$. The line of best fit to each average profile is represented by the dashed lines in Figures \ref{fig:mw_ages_stack}, \ref{fig:met_stack}, and \ref{fig:ssfr_stack} .

\subsection{Mass-weighted age gradients}\label{sec:age-grads}

The trends for the mass-weighted age gradients' slopes versus redshift are plotted in the top panel of  Figure \ref{fig:all-slopes},  as a function of redshift, for more clarity. 
In each each column of Figure \ref{fig:mw_ages_stack}, the averaged non-merger $\log(\ageMW)$ gradients are  well-separated for each redshift epoch. In the left panel, non-merger age gradients are shown, which are the majority of the galaxies in our MWA progenitor sample (730 out of 872). Gradients are flat, but become slightly negative at high-$z$. The average slope of the age gradients at $4<z<5$ is statistically consistent with a negative slope. A negative age gradient indicates that galaxies have younger stars on their outskirts versus their inner regions, which is in line with inside-out mass assembly. A flat age gradient at high-$z$ is expected, because galaxies are still young, so most of their stars will have formed at the same time (at cosmic dawn) and subsequent rounds of star formation are more or less occurring at similar rates and time scales. A flat age gradient at lower-$z$ however, when galaxies have grown to a certain size and have established stellar populations, would require more complex mechanisms, such as radial mixing of stars, or non-preferred location for pristine gas accretion. 

In Figure \ref{fig:all-slopes}, the slopes of the radial age gradients for non-mergers become increasingly negative from $z\sim4.5$ to $z\sim 2.5$, but flattens more with decreasing redshift from $z\sim 2.5$ to $z\sim 0.5$. This matches up with timeline of  MWA progenitors building stellar mass inside-out at $2<z<5$, before switching to a lockstep growth mode at $0.3<z<2$ from our previous work in \cite{Tan:2025}. This may link inside-out mass assembly to negative age gradients, and lockstep assembly with flatter age gradients. Another cause of flattened age gradients are galaxy mergers, and galaxies classified as non-mergers at the moment they are observed could have experienced mergers in the past. However, due to the low merger fractions overall in our sample, mergers would not be the primary mechanism for flattening age gradients, especially since merger fractions are decreasing with decreasing redshift (as seen in the bottom right panel of Figure \ref{fig:this-sample}).

For ongoing mergers, the age gradients are also generally flat at all redshifts, with possible signs of positive gradients at $3<z<5$. We caution that the mildly positive gradients should not be over-interpreted, since they have large error bars. Additionally, small differences in age between merger components (measured in log(Gyr)) are amplified at higher-$z$. Since the middle and right panels of Figure \ref{fig:mw_ages_stack} shows the age gradients for the merger populations are well-separated, the merger population in our sample mainly consists of galaxies that contain stellar populations of similar ages. The overall trend for age gradient evolution in mergers seems to be that there are no large differences in the stellar ages of the merging galaxies.


\subsection{Metallicity gradients} \label{sec:met-grads}

In Figure \ref{fig:met_stack}, for each subsample, we plot the averaged stellar metallicity ($\log(Z/\Zsun)$ gradient of each redshift bin by stacking and normalizing all the individual metallicity profiles of the galaxies within that redshift epoch. At lower redshift ($z < 2$), our non-merger results agree with local universe measurements of an overall negative gradient in star-forming galaxies, with a shallow slope. However at $2 <z < 3$, the metallicity gradient is mildly positive for both redshift bins. At $3<z<5$, the average metallicity gradient is flat. For star-forming galaxies, there exists measurements of a positive \textit{gas phase} metallicity gradients at $2<z<3$ (e.g.  \citealt{Cresci:2010, Li:2022} ). Unlike studies from the local universe ($z < 1$) \citep{GonzalezDelgado:2015,Goddard:2017, Zheng:2017}, we do not find a correlation between negative $\log(\ageMW)$ gradients and negative metallicity gradients, even for non-mergers. This is because at redshifts $2<z<3$, the average age gradients are negative whereas the average metallicity gradients are positive, but at $3<z<5$, average age gradients and average metallicity gradients are both negative. However, we emphasize that our age and metallicity slopes at all redshifts are close to flat. 

While positive metallicity gradients are rare in the local universe, they can appear in low mass systems. If positive gradients are associated with higher sSFR for the galaxy \citep{Stott:2014, Ma:2017, Curti:2020}, it stands to reason that a low mass galaxy would be more likely to have a flat or positive metallicity gradient.  As Figure \ref{fig:avg-props} shows, the global sSFR of our sample of galaxies is increasing as redshift increases, and galaxies are less massive at earlier times. The increased number of low mass galaxies, particularly objects with $\log \mstar [\msun] < 9$, combined with higher sSFR, means there are higher instances of positive metallicity gradients at $2<z<3$. 

For the merger samples in Figure \ref{fig:met_stack}, both the ongoing mergers and the late stage merger gradients appear to follow the trend in redshift as the non-merger sample. If we assume from the flat $\log(\ageMW)$ slopes in Figure \ref{fig:mw_ages_stack} are a result of mergers occurring between galaxies with similar star formation histories, this may explain why their gradients are similar despite galaxy interactions. The underlying stellar populations are not different enough for any distinguishable effect on the average metallicity gradient to differ from the non-merger sample. The only redshift range where the slopes of the merger samples are a drastic departure from the non-merger sample is at $0.5 <z <1$. However, there are only two ongoing mergers and one late stage merger at that redshift. Metallicity gradients and overall stellar metallicity appears to evolve slowly with redshift, and not evolve much between $3<z<5$.

\begin{figure*}
   \centering
\gridline{\fig{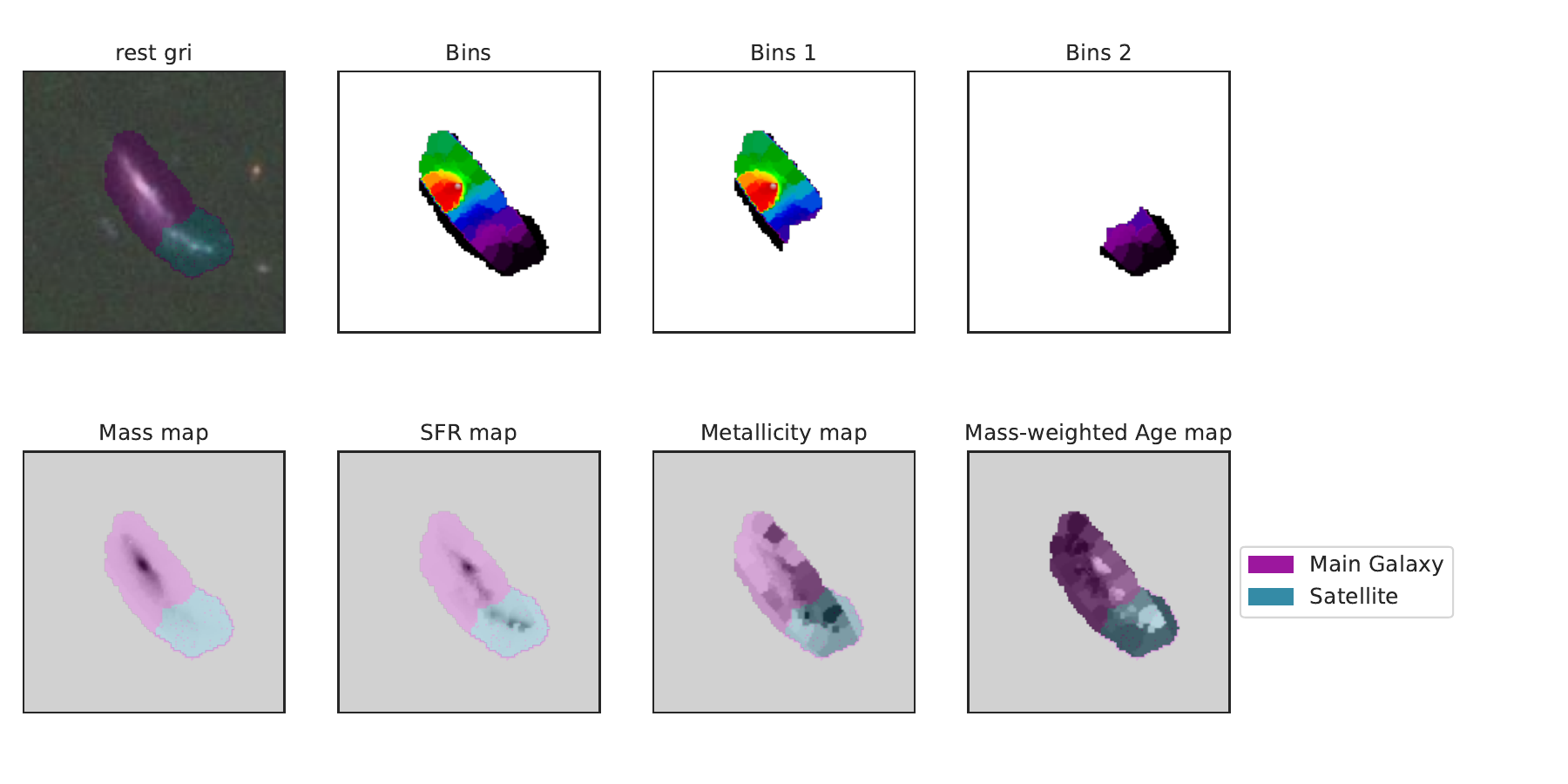}{\textwidth}{}}
\gridline{\fig{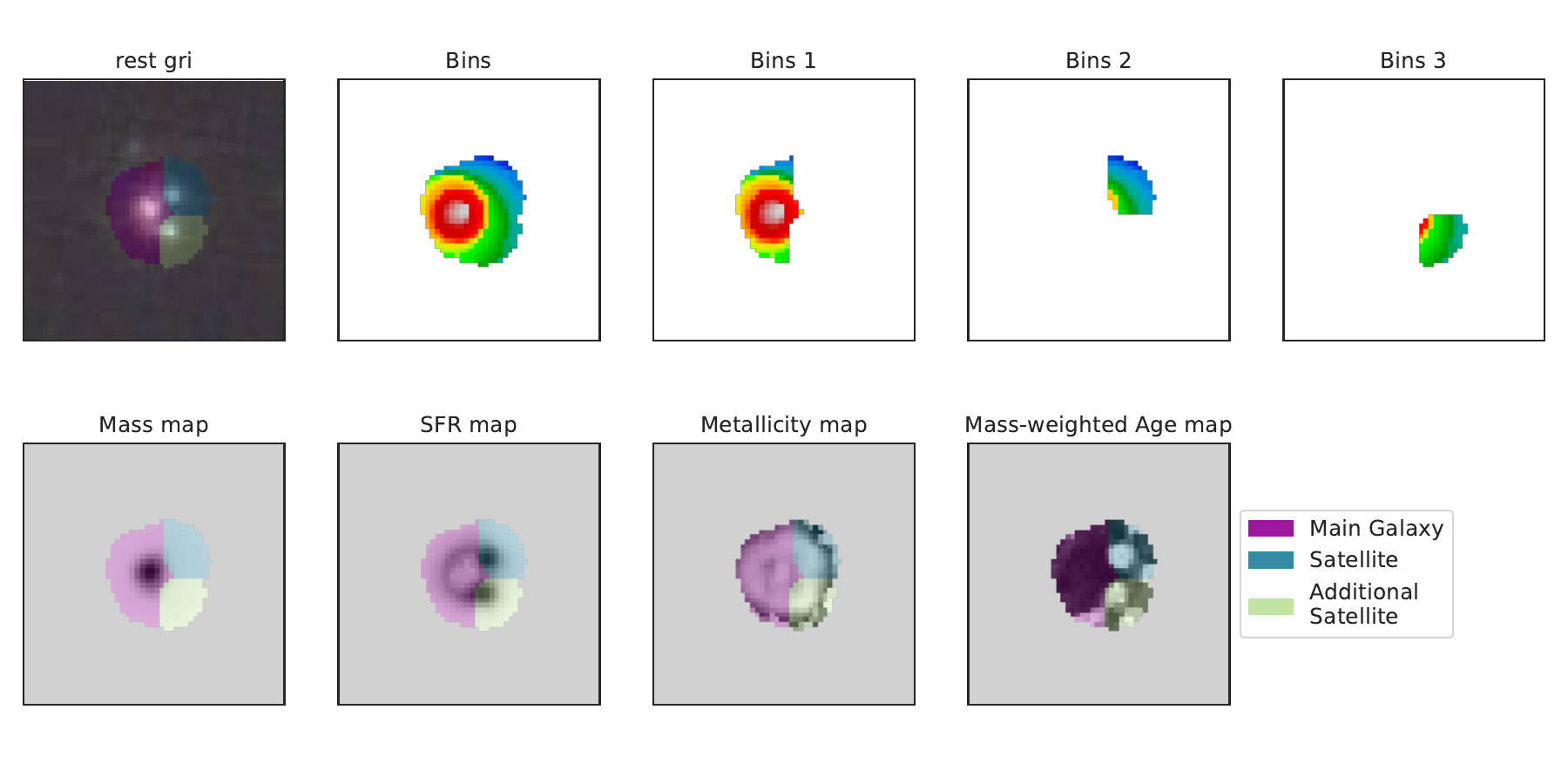}{\textwidth}{}}
   \caption{Two examples of ongoing mergers being divided into separate regions, using the watershed algorithm on a combination of the $\Sigma_\star$ and $\Sigma_{SFR}$ maps. The first cutout is divided into two regions, and the second cutout is divided into three regions. From top left to top right: we display the color image with the regions overlaid on top, the Voronoi bins of all the regions, followed by the Voronoi bins of each separate region separately. From bottom left to bottom right: the watershed segmentation map overlaid on top of the $\Sigma_\star$, $\Sigma_{SFR}$, $\log(Z/\Zsun)$, and $\log(\ageMW)$ maps. The most massive object is defined by the purple region, the second most massive is highlight in teal, and the least massive is highlighted in light green (when there are three objects present).}\label{fig:watershed}
\end{figure*}

\subsection{Specific star formation rate gradients}\label{sec:ssfr-grads}

The link between higher rates of star formation and lower mass-weighted age is clear through the definition of mass-weighted age as outlined in Equation \ref{eqn:mw_age} in \S \ref{sec:methods}. However, it does not appear that age gradients and sSFR gradients have inverse trends, or are even correlated. As Figure \ref{fig:all-slopes} demonstrates, the non-merger sSFR slopes start off flat at $4<z<5$, and then the slopes are on an overall increasing trend with decreasing redshift, while the $\log(\ageMW)$ slopes start off flat at earlier times, become negative, and then flatten out once more closer to the present day. This may appear incongruous with the sSFR slope versus redshift for non-mergers, because the average sSFR gradient slope is positive even at $z\sim0.5$. However, recall that sSFR is the rate of SFR proportional to the stellar mass. Central regions have more stellar mass, so the amount of newly formed stars contribute less to their average age. Once outer regions build up more stellar mass at later times, the age of the younger stars also affect the average age to a lesser degree. In addition, the bottom panel of Figure \ref{fig:avg-props} shows sSFR is decreasing globally with decreasing redshift, and $z < 2$ is after the epoch of cosmic noon. This may also imply that the SFR is decreasing overall, which results in flatter age gradients. Effectively, the outer regions begin to ``catch up" to the inner regions in terms of the age of the stars. 

The sSFR slopes for both ongoing and late stage mergers start off negative, and increase to positive at later times, while their $\log(\ageMW)$ slopes are either mildly positive/negative or flat without any discernible relation to either redshift nor with sSFR slopes. Both merger samples (ongoing and late stage) having a similar sSFR slope trend with redshift that is distinct from the non-merger sample may indicate that galaxy interactions tend to enhance star formation more in regions that already have higher SFR. This may be why the merger samples' sSFR slopes are more steeply negative at $2.5<z<5$, flat at $1 < z < 2.5$, and more steeply positive at $z < 1$.

\section{Properties of component galaxies in ongoing mergers}

Since papers like \cite{Grossi:2020} and \cite{Tissera:2022} find a correlation between mergers and inverted gas-phase metallicity gradients seen in the local universe, we must examine the properties of the merging sample in more detail. As \S \ref{sec:gradients} shows, the general distribution of the slopes of property gradients from mergers are not statistically different from non-mergers. We know that the ongoing merger sample does consist of visually separated objects in both the $\Sigma_\star$ maps and the color images. We  can therefore divide the ongoing merger sample's property maps into regions that are associated with each object, and determine if there are distinct differences in their average properties.

\subsection{Separating galaxies in merger pairs}

The process for separating the ongoing merger maps into different regions based on their component galaxies is as follows:
\begin{enumerate}
\item We combine the normalized stellar mass map and SFR map into one detection image.
\item We create an initial segmentation map (only segmenting foreground and background pixels) from the detection image with the function \texttt{detect\_sources} from Photutils. This segmentation map was then expanded by 5 pixels.
\item The shortest distance for each pixel within the segmentation map to the background is obtained as the scaled distance array.
\item The watershed algorithm, which finds local maxima/peaks in images, was applied to the scaled distance array multiplied by the combined normalized stellar mass and SFR maps. We limit the number of peaks identified in the watershed algorithm to 5, so the initial watershed segmentation maps had up to five regions.
\end{enumerate}

The initial segmentation maps were then inspected for accuracy, and any regions that did not correspond to a peak in the detection image was combined with another region such that no map contained more than three distinct areas for ongoing mergers. These areas were then masked to the property maps of the object's data cube to obtain the averaged properties of each pair of mergers. Any objects separated into three sections would have their third section count as an additional satellite. Two examples are shown in Figure \ref{fig:watershed}.

\begin{figure*}
\centering
    \includegraphics[width=\textwidth]{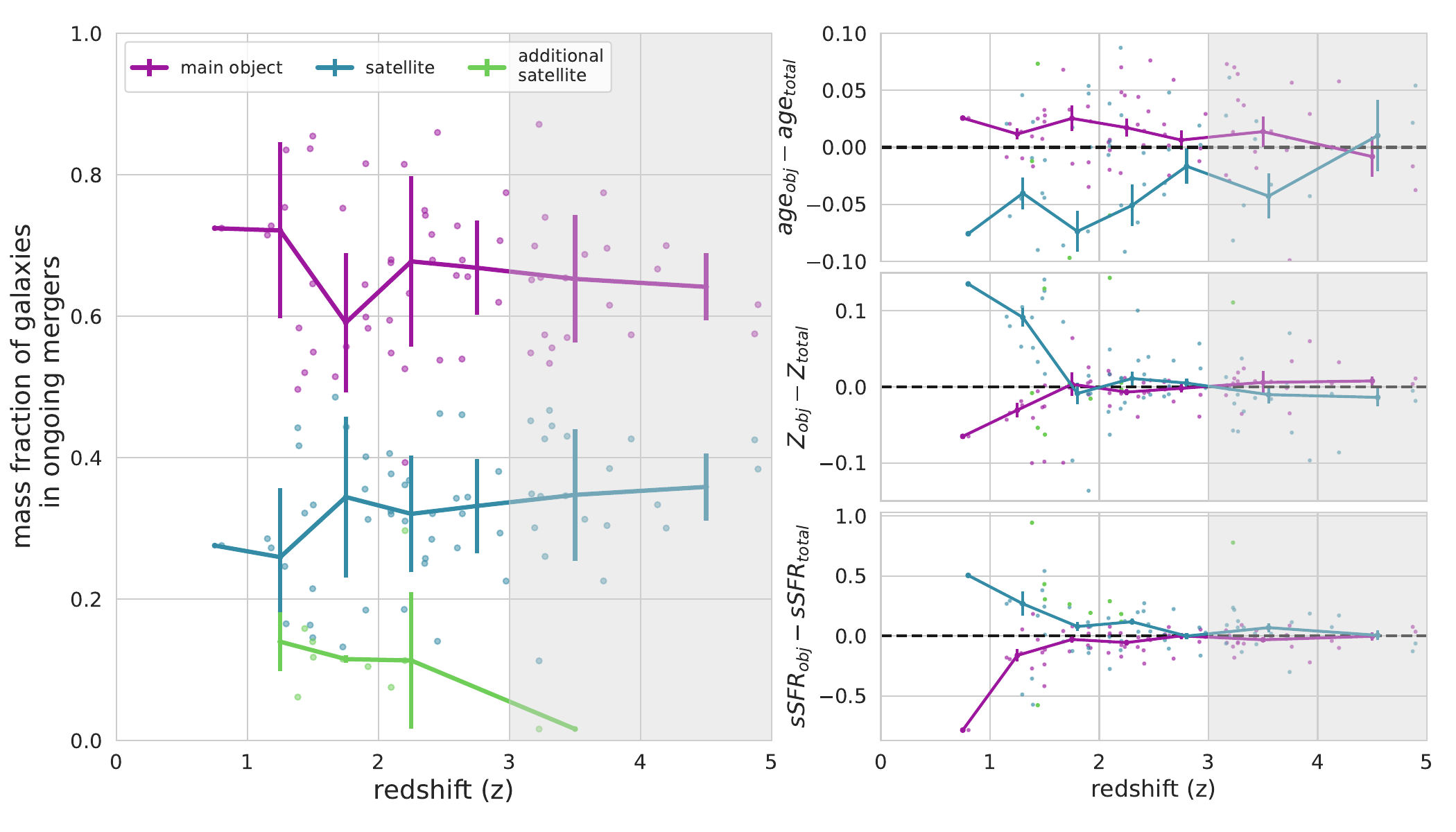}
    \caption{\textit{Left Panel: }Mass fractions of merger components. Error bar is the $\sigma$ scatter. \textit{Right panels:} Difference in average age, average metallicity and average sSFR of the components of the merger versus the average properties of the entire merger system.  Errors are standard error of the mean for more visual clarity of the property differences as a function of redshift. The individual property differences for the main and satellite galaxies are also plotted to demonstrate the true scatter in the relationship. Shaded region at $z > 3$ indicates lack of mass completeness.}\label{fig:merger-pair-props}
\end{figure*}

To identify differences in age and metallicity, we first find which region of the segmentation map contains the most stellar mass. The other region(s) are then counted as ``satellites" of the first galaxy. In Figure \ref{fig:merger-pair-props}, we plot the mass fraction of the main galaxy and its satellite(s). The purple, teal, and green points display individual mass fractions for the three distinct regions outlined in Figure \ref{fig:watershed}, while the purple, teal, and green trend lines are the median mass fractions. The central galaxy of the merging system can be between $\sim40\%$  to $\gtrsim 80\% $of the total mass of the system. The median mass fraction is between 60 and 80 percent, implying roughly a 3:1 or 5:1 ratio for major mergers. There does not appear to be an obvious size evolution with redshift in the mass ratios between the central galaxy and the companion(s), and at all redshifts, there is a large scatter in the distribution of the mass fractions.

One possible explanation for the closeness in mass fraction between the two main components of these merger systems is that the ongoing mergers were selected by visual inspection. Perhaps some of the mergers identified by the Gini-$M_{20}$ plane that were sorted visually into ``late stage" mergers are ongoing mergers that have a satellite galaxy that is not massive enough to be seen by eye on the stellar mass maps. Overall, the more massive galaxy in a merging pair tends to have older stars, which is expected as mass and age are correlated.

\subsection{Difference between merger components and non-mergers}\label{sec:merger-diff}

On the right side of Figure \ref{fig:merger-pair-props}, we take the difference between the average properties for the entire system and the average properties of each galaxy in the merging system. Note that for ages and metallicity, it is a mass-weighted average, whereas for sSFR, the average was not weighed by stellar mass, as sSFR is already dependent on $\Sigma_\star$.  
Any object in our merger sample which was divided into three regions have their two less massive regions count as separate data points for the satellites. The average difference at each redshift bin is plotted as the purple and teal trend lines.  The more massive galaxy (the ``central" galaxy) tends to have older stellar ages, higher metallicity, and lower sSFR. However, only the sSFR difference is statistically significant.

Our sample of mergers are major mergers, as the ratio of masses between the components is similar, and major mergers are more likely to disrupt the inside-out growth that non-mergers exhibit. But the age differences of the merger components are small. Compared to the entire system, the age differences are on average below 0.1 dex. Considering that the sample of mergers is only 15-20\% of the size of the sample of non-mergers, this difference would not contribute to a large change in the age gradient evolution of MWA progenitors.
The lack of difference in metallicity between merger components may also explain why mergers and non-mergers have similar radial profile evolution in Figure \ref{fig:all-slopes}. When merging pairs have notable differences in metallicity, which occurs at $z < 1.5$, the average slope of the metallicity gradient of the mergers also diverge from the non-mergers. {Although, we note there are very small numbers of mergers at $z<1.5$ so this result may not be representative. Overall, the component galaxies of the ongoing mergers do not appear to differ greatly in age at all redshifts, and only outliers in our sample have differences of $> 0.1$ dex in metallicity and $> 0.4$ dex in sSFR at $z< 1.5$. 

We would like to measure how the global properties of the merger galaxies (both ongoing and late stage mergers) differ from the \textit{non-merger properties} at the same redshift. In Figure \ref{fig:merger-nonmerger-compare}, we take the difference between the mass-weighted average $\log(\ageMW)$ and $\log(Z/\Zsun)$ of the ongoing merger components with the non-merger median properties. These are displayed as the purple and teal trend lines. Additionally, we plot the difference between the non-merger and merger sample medians from Figure \ref{fig:avg-props} as a comparison. These are plotted as the red diamonds for the late stage mergers and large blue circles for the ongoing mergers. Recall that the median values in Figure \ref{fig:avg-props} are only the central $1 R_{eff}$ of those galaxies at each redshift, and are more representative of the more massive central galaxy of the merger system.

In Figure \ref{fig:merger-nonmerger-compare}, both the central and the satellite galaxy in the merger exhibit the same mass-weighted age and sSFR trends versus non-mergers. The merger components have lower ages at all redshifts and elevated sSFR with increasing difference at lower $z$. The difference in metallicity between components of a merger and the non-merging sample varies with redshift, but the central and the satellite tend to have similar average metallicity before $z = 1.5$. This implies both merger components undergo the same changes in evolution compared to non-mergers, from $1.5<z < 5$. The changes tend to be more pronounced for the satellite galaxy. Comparing Figure \ref{fig:merger-nonmerger-compare} to the gradient slopes versus redshift in \ref{fig:all-slopes}, there is a greater difference between the median age of mergers and the median age of non-mergers at $1<z<3$ but almost no difference in metallicity. The mergers also exhibit enhanced sSFR at those redshifts. The late stage mergers are generally closer in median age to the non-mergers, but have similar sSFRs to ongoing mergers, implying that merging galaxies on the whole have more active star formation. Since these are the central $1R_{eff}$, this fits with the picture that mergers induce starbursts within galaxies, which would affect the centers more than the outskirts (e.g. \citealt{Ellison:2013, Thorp:2022, Thorp:2024} and references therein).

\begin{figure}
    \centering
    \includegraphics[width=\columnwidth]{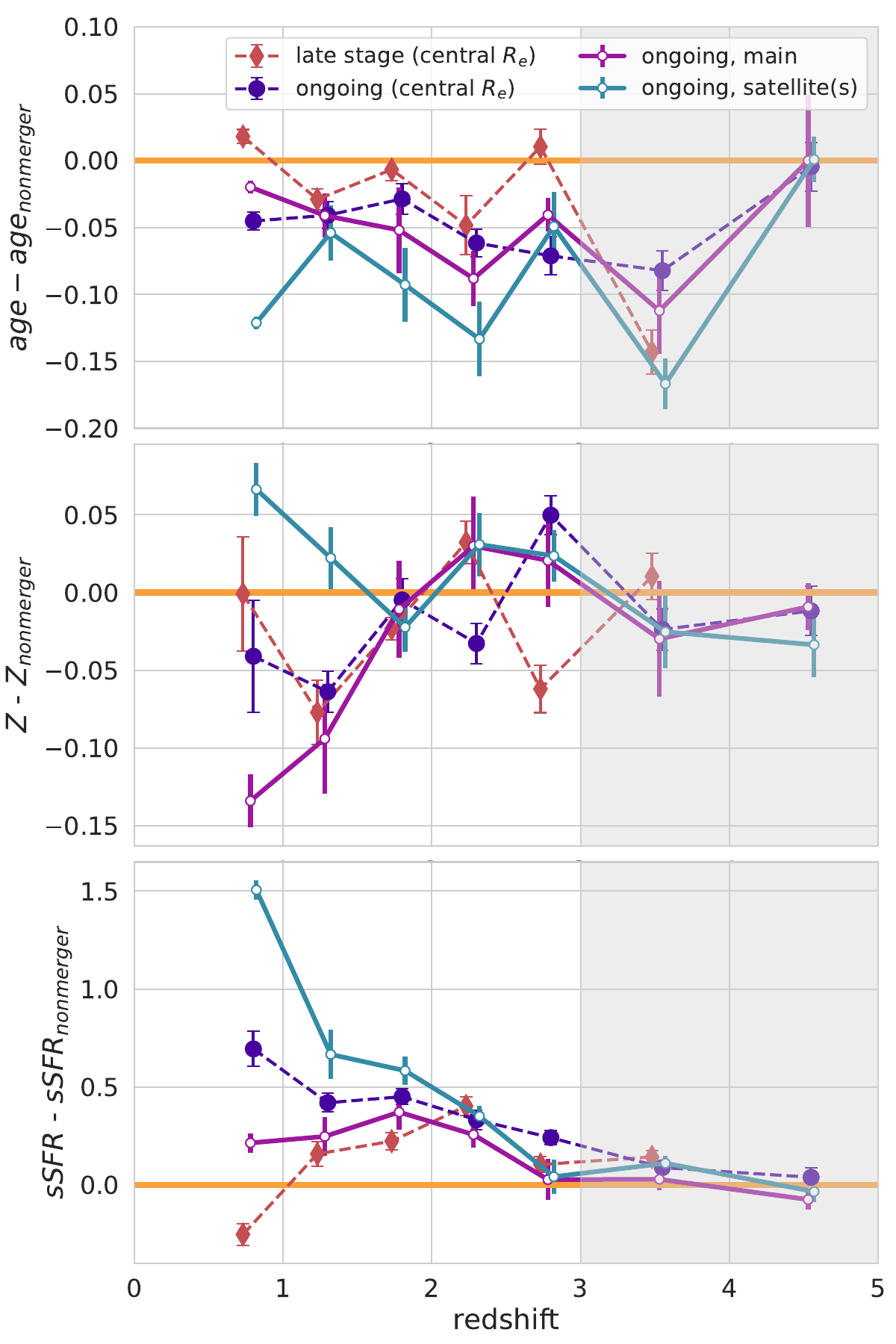}
    \caption{Difference between the median $\log(\ageMW)$, $\log(Z/\Zsun)$, and sSFR of the late stage mergers, ongoing mergers, and the average values of the ongoing merger components. Median values for the radial profile measurements come from Figure \ref{fig:avg-props}, which means that they are the median of the inner 1$R_{eff}$ of the galaxies at each redshift bin. The ongoing merger samples' average values are from the Watershed segmentation maps. Shaded region at $z > 3$ indicates lack of mass completeness.}
    \label{fig:merger-nonmerger-compare}
\end{figure}

\section{Discussion}

\subsection{Metallicity gradient comparison with measurements of the Milky Way}

\begin{figure*}
\centering
\gridline{\fig{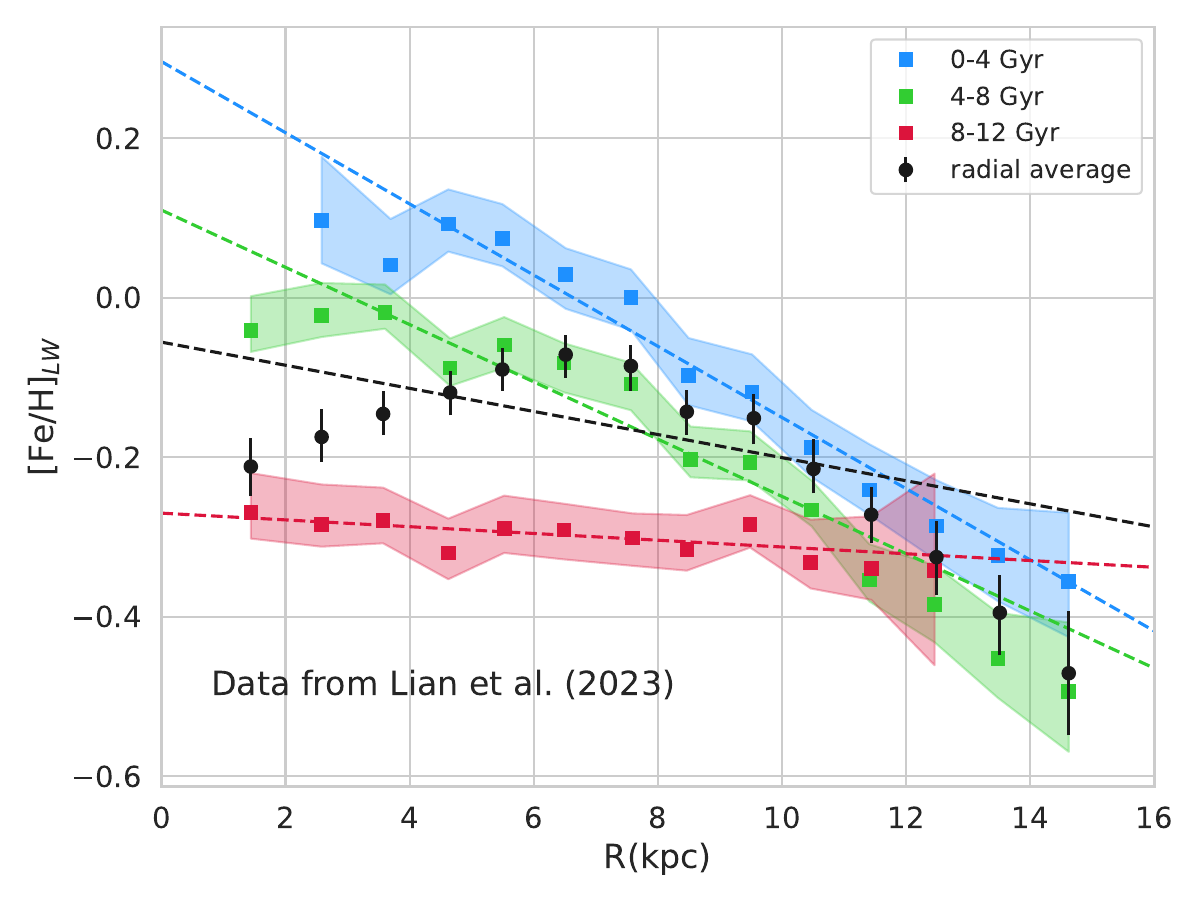}{0.505\textwidth}{} \fig{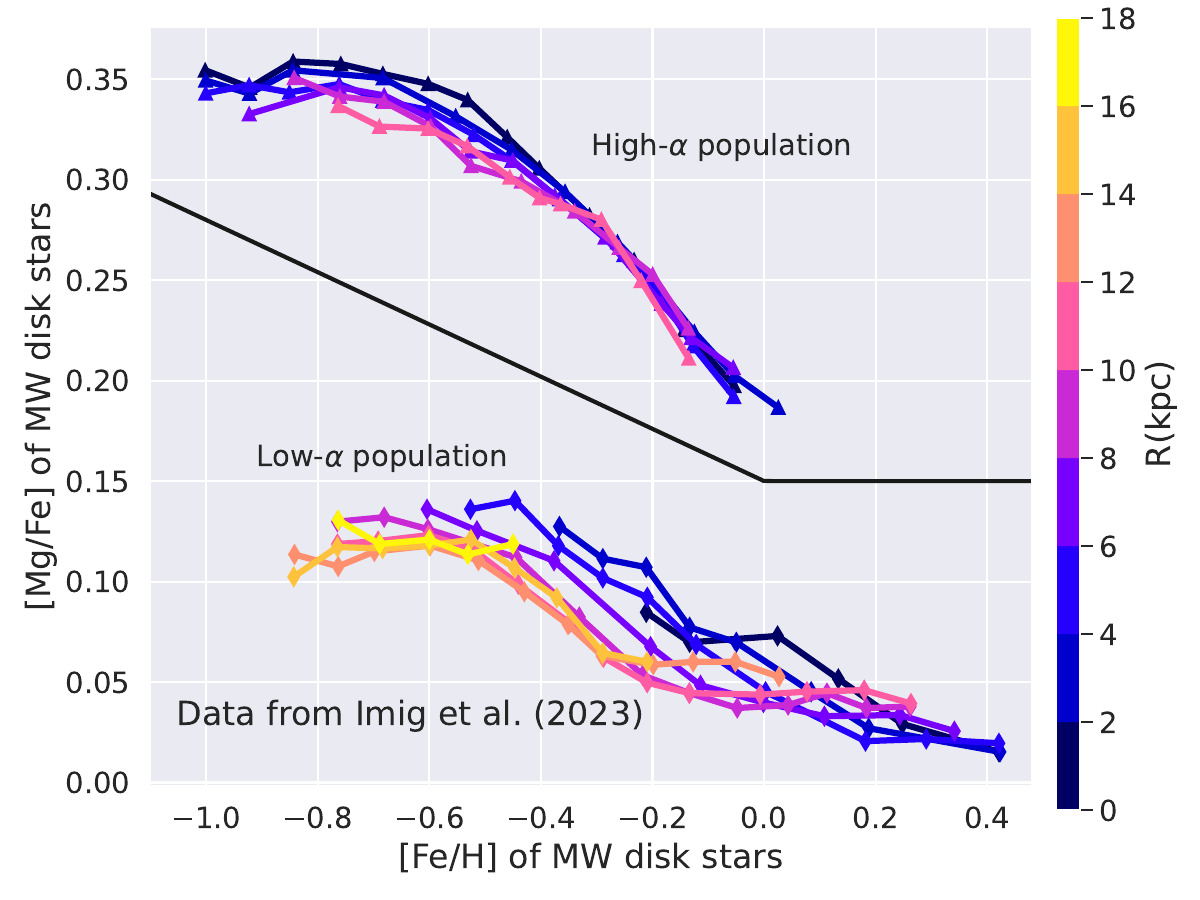}{0.505\textwidth}{}}
\vspace{-1cm}
\gridline{\fig{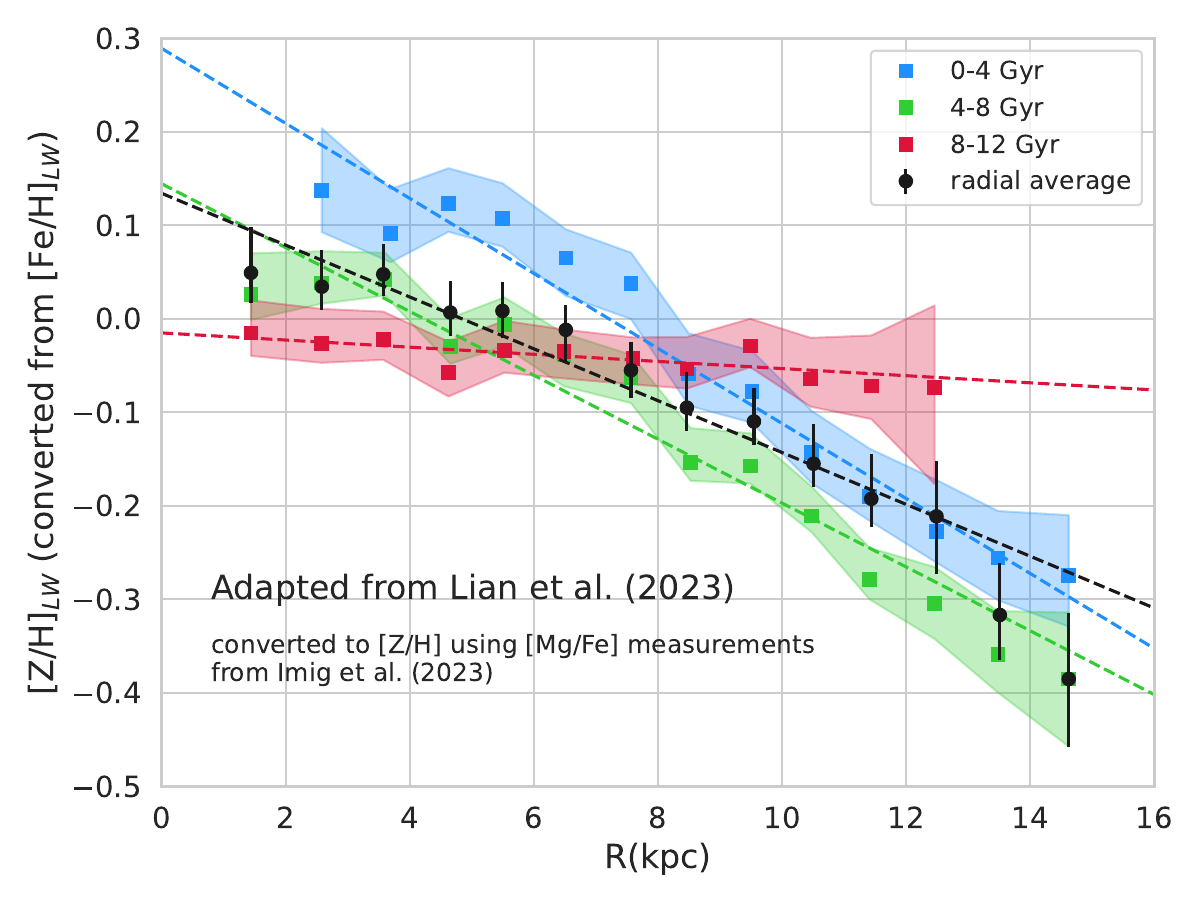}{0.505\textwidth}{}\fig{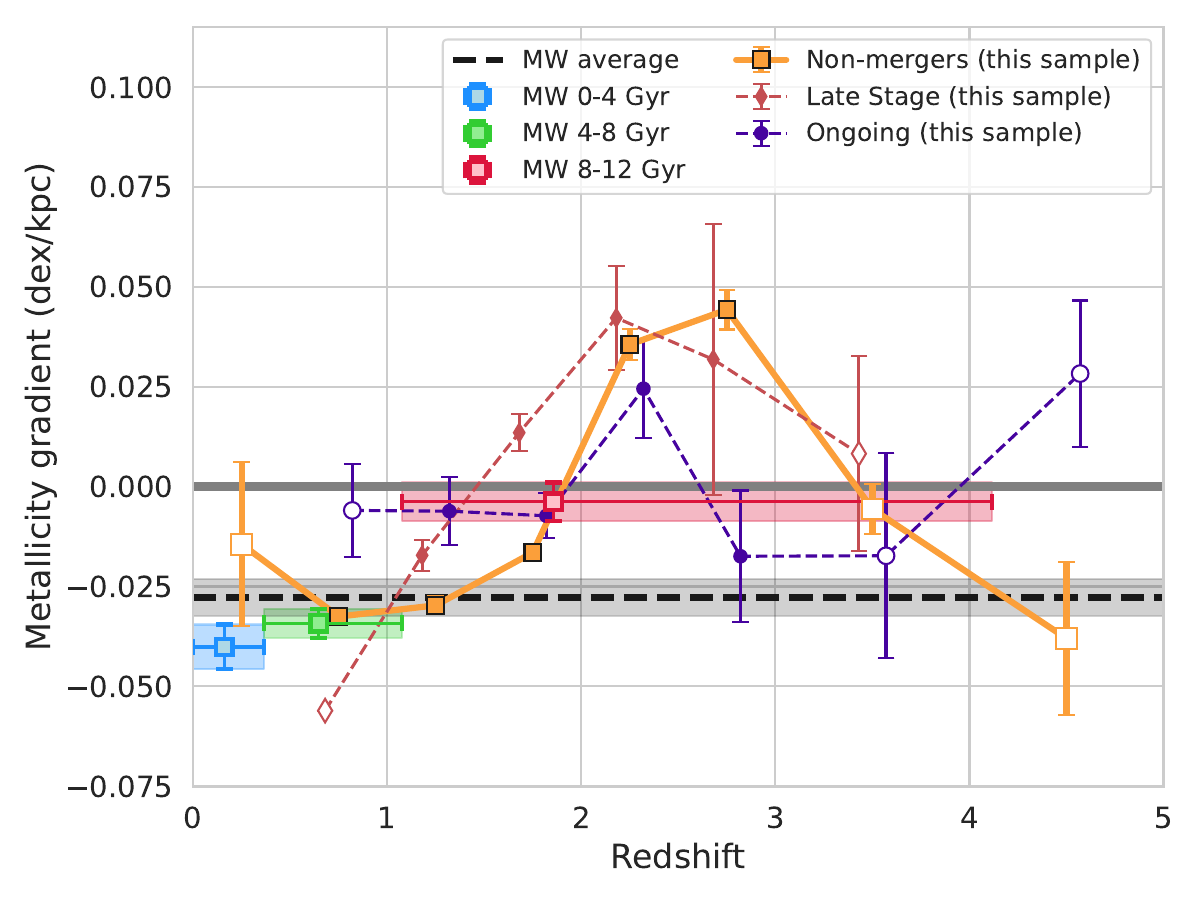}{0.495\textwidth}{}
\vspace{-0.75cm}}
    \caption{\textit{Top Left Panel:} Light-weighted [Fe/H] radial gradients of the MW from \cite{Lian:2023}.  Dashed lines are the slope fitted to the gradients of the respective colors.
    \textit{Top Right Panel:} Median [Mg/Fe] vs [Fe/H] distributions for MW disk stars for high-$\alpha$ and low-$\alpha$ populations, colours represent radial distance in steps of 2 kpc, adapted from Figure 10 of \cite{Imig:2023}.
   \textit{Bottom Left Panel:} [Z/H] radial gradients of the MW from \cite{Lian:2023}, converted from their light-weighted [Fe/H] gradients using Equation \ref{eqn:FeH-to-ZH}  and  corresponding median [Mg/Fe] abundance ratios from \cite{Imig:2023} using the high-$\alpha$ relation for the 8-12 Gyr age bin (red points) and the low-$\alpha$ relation for the 0-4 Gyr and 4-8 Gyr age bins (blue and green points). 
    \textit{Bottom Right Panel:} The average metallicity gradients from this work (same slopes as Figure \ref{fig:all-slopes} ) are compared to the converted [Z/H] slopes from the MW (light blue, light green, and light red squares).  The ages for each population of MW stars are converted to redshift according to our cosmology, and shaded regions in the horizontal represent the age range. Error bars in the vertical represent the weighted error in the slope. \label{fig:lian23-compare}}
\end{figure*}

In order to demonstrate that our results are consistent with the Milky Way itself, we make a comparison with abundance gradients obtained from studies of our own galaxy. \cite{Lian:2023} obtained light-weighted radial [Fe/H] gradients of the Milky Way from the APOGEE survey \citep{Blanton:2017, Ahumada:2020, Jonsson:2020}, split into three different age groups. The original [Fe/H] radial gradients of the MW from \cite{Lian:2023} are plotted in the top-left panel of Figure \ref{fig:lian23-compare}. We convert their radial [Fe/H] gradients to [Z/H] gradients using the equation 
\begin{equation}
    \log(Z/\Zsun) = [\text{Z/H}] = [\text{Fe/H}] + 0.94\times[\text{Mg/Fe}] \;\;, \label{eqn:FeH-to-ZH}
\end{equation}
 which was previously used in \cite{Thomas:2003, Kriek:2019}, and \cite{Carnall:2024} to convert elemental abundances from iron ratios to global stellar metallicity.  

The [Mg/Fe] abundances for stars in the Milky Way disk trace two distinct populations, where one is older with higher $\alpha$-element abundances, and the other is younger with lower $\alpha$-element abundances. This analysis comes from \cite{Imig:2023}, where they plot several median [Mg/Fe] versus [Fe/H] relations for both high-$\alpha$ and low-$\alpha$ populations with respect to radial distance $R$ in kpc in Fig. 10 of their paper. We reproduce their figure in the top-right panel of Figure \ref{fig:lian23-compare}. 

The study by \cite {Imig:2023} also found that the high-$\alpha$ population have stellar ages of $~\sim 10$ Gyr, while the low-$\alpha$ population have an age range of $\sim 2-8$ Gyr. Assuming that the 8-12 Gyr age bin in \cite{Lian:2023} corresponds (roughly) to the high-$\alpha$ population in \cite{Imig:2023}, and the 0-4 Gyr and 4-8 Gyr age bins correspond to the low-$\alpha$ population, we can infer the approximate [Mg/Fe] ratio at each $R$ for each age. Then we can apply Equation \ref{eqn:FeH-to-ZH}  using the [Fe/H] from the original gradients and the estimated [Mg/Fe] ratios. The resulting [Z/H] radial gradients are shown in the bottom-left panel of Figure \ref{fig:lian23-compare}. The slopes fit to the metallicity gradients of the different stellar populations, as well as the MW average are plotted on the bottom-right panel of Figure \ref{fig:lian23-compare}. 

The average metallicity gradients of our MWA progenitors are in fairly good agreement with the metallicity gradient present in the Milky Way itself. The average [Z/H] slope in Figure \ref{fig:lian23-compare} converted from the [Fe/H] slope from \cite{Lian:2023} for the present day Milky Way especially agree well with our sample of non-mergers at $z < 1.5$. The metallicity gradient of the oldest population of stars in the Milky Way (which would primarily be thick disk stars) also show good agreement with the average metallicity slopes at $z\sim 2$ and $z\sim 3.5$ for our sample, for both mergers and non-mergers. Although this does not mean that those stars would have exhibited the same gradient at the time of their formation, since stellar migration and radial mixing could have changed the slope from $z\sim3$ to today. Nevertheless, this broad agreement shows our sample can reasonably represent MWA progenitors in terms of metallicity gradient evolution with redshift.

\subsection{Radial gradients and relation to the star forming main sequence}

In this study, the average age and metallicity gradients for this sample of MWA progenitors are generally flat or slightly negative. However, there are certain cases where the average gradients are positive, although the amplitude of the slopes are small. This is generally consistent with studies of local universe metallicity and age gradients obtained from star-forming galaxies such as \cite{GonzalezDelgado:2015}, \cite{Zheng:2017}, and \cite{Goddard:2017} , which find only very low mass or very high mass galaxies will have positive radial gradients.

In terms of sSFR, our sample of lower-mass non-merger galaxies have flat sSFR radial profiles and this is consistent with previous results at slightly lower-$z$ that also show that lower mass galaxies tend to have flat sSFR gradients. For example, in \cite{Nelson:2016}, the measured sSFR and the H$\alpha$ radial profiles are the flattest for their lowest $\mstar$ sample of $9.0 < \log\mstar [\msun]< 9.5$. Since our sample is an evolutionary sample, the stellar mass is lower at higher $z$, and this implies a scenario where the sSFR profiles for MWA progenitors start off flat in the early universe and gradually become positive at later times (lower $z$). However, the ongoing and late stage mergers' sSFR profiles do not agree with this picture of gradual steepening sSFR profiles towards later times. The merger sSFR radial profile slopes are closest to zero at $1.5<z<3$ and are negative at $z > 3$.

Our merger samples have generally flat age slopes and either flat or negative sSFR slopes, this makes their radial properties similar to the starburst sample in \cite{Bluck:2020}. They found that star-forming and green valley galaxies have a typically negative age slope with a positive $\Sigma$SFR slope while starburst galaxies have an overall \textit{flat} age slope and a negative $\Sigma$SFR slope. This may imply for our MWA progenitors that although mergers are inducing starbursts, they might also be quenching or suppressing star formation in the resulting galaxy afterwards. However, non-mergers follow a steady, inside-out model of disk growth.


Studies from the FIRE and FIRE-2 simulations (e.g. \citealt{Hopkins:2023, Sun:2024, LiFei:2025}) find that a lower galaxy potential is linked to flat and inverted metallicity gradients due to low potentials being more responsive to feedback. Starbursts from galaxy interactions or mergers come from torque due to the inflow of gas into the central galaxy, which strengthens the argument that feedback of gas leads to lower metallicities in mergers. However, \cite{LiFei:2025} do not observe a significant impact of mergers on the SFHs of MW-mass galaxies. This seems to agree with our results of measuring similar radial age gradients in MWA progenitors for both mergers and non-mergers, despite seeing enhancement in sSFR for mergers. In order to fully study the impact of feedback on SFR and metallicity in MWA progenitors, measurements of gas-phase metallicities and kinematics may be necessary.

The measurement of slightly positive metallicity gradients in our $z > 3$ MWA progenitor sample could be consistent with studies of gas-phase metallicity of star-forming galaxies at cosmic noon (i.e. \citealt{Cresci:2010, Jones:2013, Curti:2020, Simons:2021}, and \citealt{Li:2022}), where positive gas-phase metallicity gradients are more prevalent. However, we note that our sample is not mass complete at that redshift, and our metallicity measurements are stellar metallicity, not gas-phase, which hinders a direct comparison. Nevertheless, the process of star formation inevitably turns gas into stars, and thus gas-phase metallicity contributes to stellar metallicity. If positive radial gradients are measured for both stellar and gas-phase metallicities, this would be stronger evidence that the enhanced star formation at cosmic noon causes starbursts that affect the central regions more than the outskirts. Therefore, the positive metallicity gradients of this study show that the inflow of metal-poor gas into galaxy centers may result in a different population of younger but more metal poor stars.

\subsection{Galaxy interactions and their effects on the mass assembly history of MWAs}

We find that ongoing and late stage mergers in our sample show elevated total sSFR compared to non-mergers at all redshifts. This is in line with previous studies that show mergers and close pairs have SFR enhancement at all radii in the local universe (e.g. \citealt{Ellison:2013}  and \citealt{Patton:2013}) We also find that the sSFR enhancement in mergers decreases as redshift increases, which seems to agree with other studies of high-$z$ mergers such as \cite{Shah:2022} and \cite{Duan:2025}. However, other studies (e.g. \citealt{Silva:2018, RodriguezMontero:2019, Hani:2020} ) find either no SFR enhancement or the SFR enhancement decreases with increasing redshift. This is in disagreement with our result, since stellar mass increases towards lower redshift. Note that our data is not mass complete at $z > 3$, and there are also few data points at $z < 1$ redshift bins. More resolved studies of high-redshift galaxies are needed for a complete picture of  how galaxy interactions impact the mass assembly histories of MWAs and of galaxies in general.

Studies such as \cite{Cibinel:2019} show the merger fraction increases to $\geq 70\%$ above the main sequence, which shows that starbursts are more likely to be merger-induced at $z > 1$. However, \cite{Pearson:2019} who performed a similar analysis as \cite{Cibinel:2019} did not find that mergers have a much higher SFR than non-mergers, only an average change of 0.1 dex. But our sample does show consistent enhancement in sSFR for mergers compared to non-mergers, with a minimum  difference of $\sim 0.04$ dex at $4<z<5$ up to $\sim0.69$ dex at $0.5<z<1$.

In addition to the sSFR enhancement, our results also show lower stellar metallicities in both ongoing and late stage mergers at $z < 2$ and $3<z <5$ (although the latter is not mass complete for lower stellar masses). This is roughly consistent with the results presented in \cite{Bustamante:2020}  mergers in the local universe and \cite{Horstman:2021} for mergers at $z\sim 2$, which could indicate that mergers are offset from the fundamental metallicity relation (FMR). However, the metallicities of mergers in this study are not consistently below the non-mergers, especially at $2<z<3$. Thus we cannot draw any conclusions that mergers in the MWA progenitor sample are offset from the FMR.

\cite{Qi:2025} compared both resolved star-forming main sequence (rSFMS) and resolved mass-metallicity relations (rMZR) within galaxies from the TNG50 and EAGLE simulations against resolved low-$z$ surveys ($z\lesssim 0.1)$. They found that resolved SFR has a dependence on the host galaxy's mass, but the rMZR is a \textit{localized} property independent of host galaxy mass. In our sample, the sSFR gradients evolves with redshift - and thus evolves with stellar mass, but our metallicity gradients do not evolve with redshift. This seems to agree with their result that resolved metallicity is independent of global stellar mass.

In general, our results are in agreement with studies that demonstrate that major mergers alter the radial gradients as they occur, especially causing central starbursts. Ultimately, the average trends across redshift show mergers do not disrupt the general process of inside-out mass assembly of disk galaxies. It is possible that since a merger-induced starburst quenches the galaxy, its quiescent ``descendants" will be excluded from the selection of our progenitor sample, since it is no longer a MWA progenitor. This can also explain the shift to negative metallicity gradients at lower redshifts. There may be a ``survivorship bias" in the case of lower redshift progenitors needing to still be actively star-forming to remain a MWA progenitor.

\section{Summary}
In this study we used spatially resolved SED-fitting on NIRCam, NIRISS, and HST photometry to produce specific star formation rate, mass-weighted age, and stellar metallicity maps for MWA progenitors up to $z=5$. Using a combination of statistical morphology classification tools (the Gini-$M_{20}$ plane) and visual inspection, we separated the sample of 872 galaxies into ongoing mergers, late stage mergers, and non-mergers. We analyzed the radial gradient of the property maps of the galaxies as a function of redshift. Our main conclusions are:
\begin{itemize}
    \item The age gradient slopes of non-mergers are consistently flat across the redshift range. The most negative age gradient slope is at $z\sim3$, and age gradients at $z\sim 5$ and $z < 0.5$ are both relatively flat. This corresponds with a mostly positive sSFR gradient for the non-merger sample, save for the lowest and highest redshift bins. Since the sSFR gradients are still positive, there are two main interpretations: MWAs may still be growing inside-out at $z < 2$, but there is good radial mixing of stars, or MWAs switch from inside-out growth to lockstep growth at all radii, and due to the sSFR decreasing globally at $z < 2$, the higher sSFR at the outer regions relative to the inner regions do not translate to a noticeable age difference in the age gradients.
    \item Metallicity gradients of non-mergers are typically flat or negative, but the average metallicity gradient is mildly positive at $2 < z < 3.5$. However, the average age gradients at those redshifts are still negative, and the most negative at all times, and the sSFR gradients are positive, which means the positive metallicity gradients do not contradict inside-out growth.
    \item For merging galaxies, age gradients are typically flat for ongoing and late stage mergers, but some redshift bins have average age gradients which are positive, and correspondingly negative sSFR gradients. This potentially points to enhanced star formation in the central regions during the merging process. 
    \item For ongoing mergers, the average age difference (in log Gyr) between the more massive and the less massive object is $< 0.1$ dex at every redshift bin, but there is a large scatter. There is almost no difference in the average metallicity between the more massive and less massive object in the ongoing mergers until $z \sim 1.5$, although again, there is a scatter of $\sim 0.07$ dex in the individual metallicity differences. This shows that typically, MWA mergers are usually between galaxies with similar star formation histories, especially at $z > 1.5$.
    \item Although the merger components have relatively similar properties (except perhaps in sSFR), they are quite different in sSFR and age than the non-merger sample. Ongoing mergers are overall younger, have more star formation enhancement, and slightly more metal-poor than non-mergers. Late stage mergers have properties in between ongoing mergers and non-mergers. Differences in metallicity and sSFR become more pronounced at lower redshift, but differences in age lessen. 
\end{itemize}

Mergers may comprise $10-20\%$ of the MWA progenitor sample at any given redshift, which means that barring instances where mergers quench the galaxy, almost all MWAs in the local universe have experienced at least one merger. Our results show that mergers, especially before $z \sim 2$ do not appear to disrupt the inside-out picture of mass assembly for disk galaxies.

\section{Acknowledgments}
We would like to thank the referee for their suggestions which greatly improved this paper. This research was supported by grant 18JWST-GTO1 and 23JWGO2A13 from the Canadian Space Agency (CSA), and funding from the Natural Sciences and Engineering Research Council of Canada (NSERC). The MAST DOI for CANUCS is \href{https://doi.org/10.17909/ph4n-6n76}{doi:10.17909/ph4n-6n76}.  The JWST in Technicolor DOI is \href{https://doi.org/10.17909/cyh7-mm53}{doi:10.17909/cyh7-mm53}. The data release HLSP DOI is \href{https://doi.org/10.17909/18nv-np70}{doi:10.17909/18nv-np70}. AM acknowledges generous support from the Yavin Family Fund. MB acknowledges support from the ERC Grant FIRSTLIGHT and Slovenian national research agency ARIS through grants N1-0238 and P1-0188.
This research used the Canadian Advanced Network For Astronomy Research (CANFAR) operated in partnership by the Canadian Astronomy Data Centre and The Digital Research Alliance of Canada with support from the National Research Council of Canada the Canadian Space Agency, CANARIE, and the Canadian Foundation for Innovation. 

\bibliography{paper}

@ARTICLE{Sarrouh:2025,
       author = {{Sarrouh}, Ghassan T.~E. and {Asada}, Yoshihisa and {Martis}, Nicholas S. and {Willott}, Chris J. and {Iyer}, Kartheik G. and {Noirot}, Ga{\"e}l and {Muzzin}, Adam and {Sawicki}, Marcin and {Brammer}, Gabriel and {Desprez}, Guillaume and {Rihtar{\v{s}}i{\v{c}}}, Gregor and {Zabl}, Johannes and {Abraham}, Roberto and {Brada{\v{c}}}, Maru{\v{s}}a and {Doyon}, Ren{\'e} and {Antwi-Danso}, Jacqueline and {Berek}, Samantha and {Brown}, Westley and {Estrada-Carpenter}, Vince and {Favaro}, Jeremy and {Felicioni}, Giordano and {Forrest}, Ben and {Gaspar}, Gaia and {Gould}, Katriona M.~L. and {Gledhill}, Rachel and {Harshan}, Anishya and {Jahan}, Nusrath and {Jagga}, Naadiyah and {Jude{\v{z}}}, Jon and {Marchesini}, Danilo and {Markov}, Vladan and {Matharu}, Jasleen and {MacFarland}, Shannon and {Merchant}, Maya and {M{\'e}rida}, Rosa M. and {Mowla}, Lamiya and {Myers}, Katherine and {Omori}, Kiyoaki C. and {Pacifici}, Camilla and {Ravindranath}, Swara and {Robbins}, Luke and {Strait}, Victoria and {Sok}, Visal and {Tan}, Vivian Yun Yan and {Tripodi}, Roberta and {Wilson}, Gillian and {Withers}, Sunna},
        title = "{CANUCS/Technicolor Data Release 1: Imaging, Photometry, Slit Spectroscopy, and Stellar Population Parameters}",
      journal = {arXiv e-prints},
         year = 2025,
        month = jun,
          eid = {arXiv:2506.21685},
        pages = {arXiv:2506.21685},
          doi = {10.48550/arXiv.2506.21685},
archivePrefix = {arXiv},
       eprint = {2506.21685},
 primaryClass = {astro-ph.GA},
       adsurl = {https://ui.adsabs.harvard.edu/abs/2025arXiv250621685S}
}

@ARTICLE{Bell:2000,
       author = {{Bell}, Eric F. and {de Jong}, Roelof S.},
        title = "{The stellar populations of spiral galaxies}",
      journal = {\mnras},
         year = 2000,
        month = mar,
       volume = {312},
       number = {3},
        pages = {497-520},
          doi = {10.1046/j.1365-8711.2000.03138.x},
archivePrefix = {arXiv},
       eprint = {astro-ph/9909402},
 primaryClass = {astro-ph},
       adsurl = {https://ui.adsabs.harvard.edu/abs/2000MNRAS.312..497B}
}

@ARTICLE{MacArthur:2004,
       author = {{MacArthur}, Lauren A. and {Courteau}, St{\'e}phane and {Bell}, Eric and {Holtzman}, Jon A.},
        title = "{Structure of Disk-dominated Galaxies. II. Color Gradients and Stellar Population Models}",
      journal = {\apjs},
         year = 2004,
        month = jun,
       volume = {152},
       number = {2},
        pages = {175-199},
          doi = {10.1086/383525},
archivePrefix = {arXiv},
       eprint = {astro-ph/0401437},
 primaryClass = {astro-ph},
       adsurl = {https://ui.adsabs.harvard.edu/abs/2004ApJS..152..175M}
}

@ARTICLE{Lee:2007,
       author = {{Lee}, Hyun-chul and {Worthey}, Guy and {Trager}, Scott C. and {Faber}, S.~M.},
        title = "{On the Age and Metallicity Estimation of Spiral Galaxies Using Optical and Near-Infrared Photometry}",
      journal = {\apj},
         year = 2007,
        month = jul,
       volume = {664},
       number = {1},
        pages = {215-225},
          doi = {10.1086/518855},
archivePrefix = {arXiv},
       eprint = {astro-ph/0605425},
 primaryClass = {astro-ph},
       adsurl = {https://ui.adsabs.harvard.edu/abs/2007ApJ...664..215L}
}

@ARTICLE{Eminian:2008,
       author = {{Eminian}, C. and {Kauffmann}, G. and {Charlot}, S. and {Wild}, V. and {Bruzual}, G. and {Rettura}, A. and {Loveday}, J.},
        title = "{Physical interpretation of the near-infrared colours of low-redshift galaxies}",
      journal = {\mnras},
         year = 2008,
        month = mar,
       volume = {384},
       number = {3},
        pages = {930-942},
          doi = {10.1111/j.1365-2966.2007.12742.x},
archivePrefix = {arXiv},
       eprint = {0709.1147},
 primaryClass = {astro-ph},
       adsurl = {https://ui.adsabs.harvard.edu/abs/2008MNRAS.384..930E}
}

@ARTICLE{Wuyts:2007,
       author = {{Wuyts}, Stijn and {Labb{\'e}}, Ivo and {Franx}, Marijn and {Rudnick}, Gregory and {van Dokkum}, Pieter G. and {Fazio}, Giovanni G. and {F{\"o}rster Schreiber}, Natascha M. and {Huang}, Jiasheng and {Moorwood}, Alan F.~M. and {Rix}, Hans-Walter and {R{\"o}ttgering}, Huub and {van der Werf}, Paul},
        title = "{What Do We Learn from IRAC Observations of Galaxies at 2 < z < 3.5?}",
      journal = {\apj},
         year = 2007,
        month = jan,
       volume = {655},
       number = {1},
        pages = {51-65},
          doi = {10.1086/509708},
archivePrefix = {arXiv},
       eprint = {astro-ph/0609548},
 primaryClass = {astro-ph},
       adsurl = {https://ui.adsabs.harvard.edu/abs/2007ApJ...655...51W}
}

@ARTICLE{Williams:2009,
       author = {{Williams}, Rik J. and {Quadri}, Ryan F. and {Franx}, Marijn and {van Dokkum}, Pieter and {Labb{\'e}}, Ivo},
        title = "{Detection of Quiescent Galaxies in a Bicolor Sequence from Z = 0-2}",
      journal = {\apj},
         year = 2009,
        month = feb,
       volume = {691},
       number = {2},
        pages = {1879-1895},
          doi = {10.1088/0004-637X/691/2/1879},
archivePrefix = {arXiv},
       eprint = {0806.0625},
 primaryClass = {astro-ph},
       adsurl = {https://ui.adsabs.harvard.edu/abs/2009ApJ...691.1879W}
}

@ARTICLE{Burgarella:2005,
       author = {{Burgarella}, D. and {Buat}, V. and {Iglesias-P{\'a}ramo}, J.},
        title = "{Star formation and dust attenuation properties in galaxies from a statistical ultraviolet-to-far-infrared analysis}",
      journal = {\mnras},
         year = 2005,
        month = jul,
       volume = {360},
       number = {4},
        pages = {1413-1425},
          doi = {10.1111/j.1365-2966.2005.09131.x},
archivePrefix = {arXiv},
       eprint = {astro-ph/0504434},
 primaryClass = {astro-ph},
       adsurl = {https://ui.adsabs.harvard.edu/abs/2005MNRAS.360.1413B}
}

@ARTICLE{daCunha:2015,
       author = {{da Cunha}, E. and {Walter}, F. and {Smail}, I.~R. and {Swinbank}, A.~M. and {Simpson}, J.~M. and {Decarli}, R. and {Hodge}, J.~A. and {Weiss}, A. and {van der Werf}, P.~P. and {Bertoldi}, F. and {Chapman}, S.~C. and {Cox}, P. and {Danielson}, A.~L.~R. and {Dannerbauer}, H. and {Greve}, T.~R. and {Ivison}, R.~J. and {Karim}, A. and {Thomson}, A.},
        title = "{An ALMA Survey of Sub-millimeter Galaxies in the Extended Chandra Deep Field South: Physical Properties Derived from Ultraviolet-to-radio Modeling}",
      journal = {\apj},
         year = 2015,
        month = jun,
       volume = {806},
       number = {1},
          eid = {110},
        pages = {110},
          doi = {10.1088/0004-637X/806/1/110},
archivePrefix = {arXiv},
       eprint = {1504.04376},
 primaryClass = {astro-ph.GA},
       adsurl = {https://ui.adsabs.harvard.edu/abs/2015ApJ...806..110D}
}

@ARTICLE{Salim:2018,
       author = {{Salim}, Samir and {Boquien}, M{\'e}d{\'e}ric and {Lee}, Janice C.},
        title = "{Dust Attenuation Curves in the Local Universe: Demographics and New Laws for Star-forming Galaxies and High-redshift Analogs}",
      journal = {\apj},
         year = 2018,
        month = may,
       volume = {859},
       number = {1},
          eid = {11},
        pages = {11},
          doi = {10.3847/1538-4357/aabf3c},
archivePrefix = {arXiv},
       eprint = {1804.05850},
 primaryClass = {astro-ph.GA},
       adsurl = {https://ui.adsabs.harvard.edu/abs/2018ApJ...859...11S}
}

@ARTICLE{Jones:2023,
       author = {{Jones}, Gareth T. and {Stanway}, Elizabeth R.},
        title = "{Exploring the evolution of dust temperature using spectral energy distribution fitting in a large photometric survey}",
      journal = {\mnras},
         year = 2023,
        month = nov,
       volume = {525},
       number = {4},
        pages = {5720-5736},
          doi = {10.1093/mnras/stad2683},
archivePrefix = {arXiv},
       eprint = {2309.02502},
 primaryClass = {astro-ph.GA},
       adsurl = {https://ui.adsabs.harvard.edu/abs/2023MNRAS.525.5720J}
}

@ARTICLE{Iyer:2025,
       author = {{Iyer}, Kartheik G. and {Pacifici}, Camilla and {Calistro-Rivera}, Gabriela and {Lovell}, Christopher C.},
        title = "{The Spectral Energy Distributions of Galaxies}",
      journal = {arXiv e-prints},
         year = 2025,
        month = feb,
          eid = {arXiv:2502.17680},
        pages = {arXiv:2502.17680},
          doi = {10.48550/arXiv.2502.17680},
archivePrefix = {arXiv},
       eprint = {2502.17680},
 primaryClass = {astro-ph.GA},
       adsurl = {https://ui.adsabs.harvard.edu/abs/2025arXiv250217680I}
}

@ARTICLE{Abdurrouf:2021,
       author = {{Abdurro'uf} and {Lin}, Yen-Ting and {Wu}, Po-Feng and {Akiyama}, Masayuki},
        title = "{Introducing piXedfit: A Spectral Energy Distribution Fitting Code Designed for Resolved Sources}",
      journal = {\apjs},
         year = 2021,
        month = may,
       volume = {254},
       number = {1},
          eid = {15},
        pages = {15},
          doi = {10.3847/1538-4365/abebe2},
archivePrefix = {arXiv},
       eprint = {2101.09717},
 primaryClass = {astro-ph.GA},
       adsurl = {https://ui.adsabs.harvard.edu/abs/2021ApJS..254...15A}
}

@software{Bradley:2023,
       author = {{Bradley}, Larry and {Sip{\H{o}}cz}, Brigitta and {Robitaille}, Thomas and {Tollerud}, Erik and {Vin{\'\i}cius}, Z{\'e} and {Deil}, Christoph and {Barbary}, Kyle and {Wilson}, Tom J and {Busko}, Ivo and {Donath}, Axel and {G{\"u}nther}, Hans Moritz and {Cara}, Mihai and {Lim}, P.~L. and {Me{\ss}linger}, Sebastian and {Conseil}, Simon and {Burnett}, Zach and {Bostroem}, Azalee and {Droettboom}, Michael and {Bray}, E.~M. and {Andersen Bratholm}, Lars and {Jamieson}, William and {Ginsburg}, Adam and {Barentsen}, Geert and {Craig}, Matt and {Morris}, Brett M. and {Perrin}, Marshall and {Rathi}, Shivangee and {Pascual}, Sergio and {Perren}, Gabriel and {Georgiev}, Iskren Y.},
        title = "{astropy/photutils: 1.10.0}",
         year = 2023,
        month = nov,
          eid = {10.5281/zenodo.1035865},
          doi = {10.5281/zenodo.1035865},
      version = {1.10.0},
    publisher = {Zenodo},
       adsurl = {https://ui.adsabs.harvard.edu/abs/2023zndo...1035865B}
}

@ARTICLE{Rodriguez-Gomez:2019,
       author = {{Rodriguez-Gomez}, Vicente and {Snyder}, Gregory F. and {Lotz}, Jennifer M. and {Nelson}, Dylan and {Pillepich}, Annalisa and {Springel}, Volker and {Genel}, Shy and {Weinberger}, Rainer and {Tacchella}, Sandro and {Pakmor}, R{\"u}diger and {Torrey}, Paul and {Marinacci}, Federico and {Vogelsberger}, Mark and {Hernquist}, Lars and {Thilker}, David A.},
        title = "{The optical morphologies of galaxies in the IllustrisTNG simulation: a comparison to Pan-STARRS observations}",
      journal = {\mnras},
         year = 2019,
        month = mar,
       volume = {483},
       number = {3},
        pages = {4140-4159},
          doi = {10.1093/mnras/sty3345},
archivePrefix = {arXiv},
       eprint = {1809.08239},
 primaryClass = {astro-ph.GA},
       adsurl = {https://ui.adsabs.harvard.edu/abs/2019MNRAS.483.4140R}
}

@software{Bradley:2016,
       author = {{Bradley}, Larry and {Sipocz}, Brigitta and {Robitaille}, Thomas and {Tollerud}, Erik and {Deil}, Christoph and {Vin{\'\i}cius}, Z{\`e} and {Barbary}, Kyle and {G{\"u}nther}, Hans Moritz and {Bostroem}, Azalee and {Droettboom}, Michael and {Bray}, Erik and {Bratholm}, Lars Andersen and {Pickering}, T.~E. and {Craig}, Matt and {Pascual}, Sergio and {Greco}, Johnny and {Donath}, Axel and {Kerzendorf}, Wolfgang and {Littlefair}, Stuart and {Barentsen}, Geert and {D'Eugenio}, Francesco and {Weaver}, Benjamin Alan},
        title = "{Photutils: Photometry tools}",
 howpublished = {Astrophysics Source Code Library, record ascl:1609.011},
         year = 2016,
        month = sep,
          eid = {ascl:1609.011},
       adsurl = {https://ui.adsabs.harvard.edu/abs/2016ascl.soft09011B}
}

@ARTICLE{Brammer:2008,
       author = {{Brammer}, Gabriel B. and {van Dokkum}, Pieter G. and {Coppi}, Paolo},
        title = "{EAZY: A Fast, Public Photometric Redshift Code}",
      journal = {\apj},
         year = 2008,
        month = oct,
       volume = {686},
       number = {2},
        pages = {1503-1513},
          doi = {10.1086/591786},
archivePrefix = {arXiv},
       eprint = {0807.1533},
 primaryClass = {astro-ph},
       adsurl = {https://ui.adsabs.harvard.edu/abs/2008ApJ...686.1503B}
}

@ARTICLE{Asada:2024b,
       author = {{Asada}, Yoshihisa and {Desprez}, Guillaume and {Willott}, Chris J. and {Sawicki}, Marcin and {Brada{\v{c}}}, Maru{\v{s}}a and {Brammer}, Gabriel and {Dubath}, Florian and {Iyer}, Kartheik G. and {Martis}, Nicholas S. and {Muzzin}, Adam and {Noirot}, Ga{\"e}l and {Paltani}, St{\'e}phane and {Sarrouh}, Ghassan T.~E. and {Harshan}, Anishya and {Markov}, Vladan},
        title = "{Improving photometric redshifts of Epoch of Reionization galaxies: a new transmission curve with the neutral hydrogen damped Ly$\alpha$ absorption}",
      journal = {arXiv e-prints},
         year = 2024,
        month = oct,
          eid = {arXiv:2410.21543},
        pages = {arXiv:2410.21543},
          doi = {10.48550/arXiv.2410.21543},
archivePrefix = {arXiv},
       eprint = {2410.21543},
 primaryClass = {astro-ph.GA},
       adsurl = {https://ui.adsabs.harvard.edu/abs/2024arXiv241021543A}
}

@ARTICLE{Rihtarsic:2025,
       author = {{Rihtar{\v{s}}i{\v{c}}}, G. and {Brada{\v{c}}}, M. and {Desprez}, G. and {Harshan}, A. and {Noirot}, G. and {Estrada-Carpenter}, V. and {Martis}, N.~S. and {Abraham}, R.~G. and {Asada}, Y. and {Brammer}, G. and {Iyer}, K.~G. and {Matharu}, J. and {Mowla}, L. and {Muzzin}, A. and {Sarrouh}, G.~T.~E. and {Sawicki}, M. and {Strait}, V. and {Willott}, C.~J. and {Gledhill}, R. and {Markov}, V. and {Tripodi}, R.},
        title = "{CANUCS: Constraining the MACS J0416.1-2403 strong lensing model with JWST NIRISS, NIRSpec, and NIRCam}",
      journal = {\aap},
         year = 2025,
        month = apr,
       volume = {696},
          eid = {A15},
        pages = {A15},
          doi = {10.1051/0004-6361/202451117},
archivePrefix = {arXiv},
       eprint = {2406.10332},
 primaryClass = {astro-ph.CO},
       adsurl = {https://ui.adsabs.harvard.edu/abs/2025A&A...696A..15R}
}

@ARTICLE{Asada:2024,
       author = {{Asada}, Yoshihisa and {Sawicki}, Marcin and {Abraham}, Roberto and {Brada{\v{c}}}, Maru{\v{s}}a and {Brammer}, Gabriel and {Desprez}, Guillaume and {Estrada-Carpenter}, Vince and {Iyer}, Kartheik and {Martis}, Nicholas and {Matharu}, Jasleen and {Mowla}, Lamiya and {Muzzin}, Adam and {Noirot}, Ga{\"e}l and {Sarrouh}, Ghassan T.~E. and {Strait}, Victoria and {Willott}, Chris J. and {Harshan}, Anishya},
        title = "{Bursty star formation and galaxy-galaxy interactions in low-mass galaxies 1 Gyr after the Big Bang}",
      journal = {\mnras},
         year = 2024,
        month = feb,
       volume = {527},
       number = {4},
        pages = {11372-11392},
          doi = {10.1093/mnras/stad3902},
archivePrefix = {arXiv},
       eprint = {2310.02314},
 primaryClass = {astro-ph.GA},
       adsurl = {https://ui.adsabs.harvard.edu/abs/2024MNRAS.52711372A}
}

@ARTICLE{Desprez:2024,
       author = {{Desprez}, Guillaume and {Martis}, Nicholas S. and {Asada}, Yoshihisa and {Sawicki}, Marcin and {Willott}, Chris J. and {Muzzin}, Adam and {Abraham}, Roberto G. and {Brada{\v{c}}}, Maru{\v{s}}a and {Brammer}, Gabe and {Estrada-Carpenter}, Vicente and {Iyer}, Kartheik G. and {Matharu}, Jasleen and {Mowla}, Lamiya and {Noirot}, Ga{\"e}l and {Sarrouh}, Ghassan T.~E. and {Strait}, Victoria and {Gledhill}, Rachel and {Rihtar{\v{s}}i{\v{c}}}, Gregor},
        title = "{{\ensuremath{\Lambda}}CDM not dead yet: massive high-z Balmer break galaxies are less common than previously reported}",
      journal = {\mnras},
         year = 2024,
        month = may,
       volume = {530},
       number = {3},
        pages = {2935-2952},
          doi = {10.1093/mnras/stae1084},
archivePrefix = {arXiv},
       eprint = {2310.03063},
 primaryClass = {astro-ph.GA},
       adsurl = {https://ui.adsabs.harvard.edu/abs/2024MNRAS.530.2935D}
}

@ARTICLE{Gledhill:2024,
       author = {{Gledhill}, Rachel and {Strait}, Victoria and {Desprez}, Guillaume and {Rihtar{\v{s}}i{\v{c}}}, Gregor and {Brada{\v{c}}}, Maru{\v{s}}a and {Brammer}, Gabriel and {Willott}, Chris J. and {Martis}, Nicholas and {Sawicki}, Marcin and {Noirot}, Ga{\"e}l and {Sarrouh}, Ghassan T.~E. and {Muzzin}, Adam},
        title = "{CANUCS: An Updated Mass and Magnification Model of Abell 370 with JWST}",
      journal = {arXiv e-prints},
         year = 2024,
        month = mar,
          eid = {arXiv:2403.07062},
        pages = {arXiv:2403.07062},
          doi = {10.48550/arXiv.2403.07062},
archivePrefix = {arXiv},
       eprint = {2403.07062},
 primaryClass = {astro-ph.GA},
       adsurl = {https://ui.adsabs.harvard.edu/abs/2024arXiv240307062G}
}

@ARTICLE{Sarrouh:2024,
       author = {{Sarrouh}, Ghassan T. and {Muzzin}, Adam and {Iyer}, Kartheik G. and {Mowla}, Lamiya and {Abraham}, Roberto G. and {Asada}, Yoshihisa and {Bradac}, Marusa and {Brammer}, Gabriel B. and {Desprez}, Guillaume and {Martis}, Nicholas S. and {Matharu}, Jasleen and {Noirot}, Ga{\"e}l and {Sawicki}, Marcin and {Strait}, Victoria and {Willott}, Chris and {Zabl}, Johannes},
        title = "{Exposing Line Emission: A First Look At The Systematic Differences of Measuring Stellar Masses With JWST NIRCam Medium Versus Wide Band Photometry}",
      journal = {arXiv e-prints},
         year = 2024,
        month = jan,
          eid = {arXiv:2401.08781},
        pages = {arXiv:2401.08781},
          doi = {10.48550/arXiv.2401.08781},
archivePrefix = {arXiv},
       eprint = {2401.08781},
 primaryClass = {astro-ph.GA},
       adsurl = {https://ui.adsabs.harvard.edu/abs/2024arXiv240108781S}
}

@ARTICLE{Willott:2024,
       author = {{Willott}, Chris J. and {Desprez}, Guillaume and {Asada}, Yoshihisa and {Sarrouh}, Ghassan T.~E. and {Abraham}, Roberto and {Brada{\v{c}}}, Maru{\v{s}}a and {Brammer}, Gabe and {Estrada-Carpenter}, Vince and {Iyer}, Kartheik G. and {Martis}, Nicholas S. and {Matharu}, Jasleen and {Mowla}, Lamiya and {Muzzin}, Adam and {Noirot}, Ga{\"e}l and {Sawicki}, Marcin and {Strait}, Victoria and {Rihtar{\v{s}}i{\v{c}}}, Gregor and {Withers}, Sunna},
        title = "{A Steep Decline in the Galaxy Space Density beyond Redshift 9 in the CANUCS UV Luminosity Function}",
      journal = {\apj},
         year = 2024,
        month = may,
       volume = {966},
       number = {1},
          eid = {74},
        pages = {74},
          doi = {10.3847/1538-4357/ad35bc},
archivePrefix = {arXiv},
       eprint = {2311.12234},
 primaryClass = {astro-ph.GA},
       adsurl = {https://ui.adsabs.harvard.edu/abs/2024ApJ...966...74W}
}

@ARTICLE{Noirot:2023,
       author = {{Noirot}, Ga{\"e}l and {Desprez}, Guillaume and {Asada}, Yoshihisa and {Sawicki}, Marcin and {Estrada-Carpenter}, Vicente and {Martis}, Nicholas and {Sarrouh}, Ghassan and {Strait}, Victoria and {Abraham}, Roberto and {Brada{\v{c}}}, Maru{\v{s}}a and {Brammer}, Gabriel and {Iyer}, Kartheik and {MacFarland}, Shannon and {Matharu}, Jasleen and {Mowla}, Lamiya and {Muzzin}, Adam and {Pacifici}, Camilla and {Ravindranath}, Swara and {Willott}, Chris J. and {Albert}, Lo{\"\i}c and {Doyon}, Ren{\'e} and {Hutchings}, John B. and {Rowlands}, Neil},
        title = "{The first large catalogue of spectroscopic redshifts in Webb's first deep field, SMACS J0723.3-7327}",
      journal = {\mnras},
         year = 2023,
        month = oct,
       volume = {525},
       number = {2},
        pages = {1867-1884},
          doi = {10.1093/mnras/stad1019},
archivePrefix = {arXiv},
       eprint = {2212.07366},
 primaryClass = {astro-ph.GA},
       adsurl = {https://ui.adsabs.harvard.edu/abs/2023MNRAS.525.1867N}
}

@ARTICLE{Withers:2023,
       author = {{Withers}, Sunna and {Muzzin}, Adam and {Ravindranath}, Swara and {Sarrouh}, Ghassan T.~E. and {Abraham}, Roberto and {Asada}, Yoshihisa and {Brada{\v{c}}}, Maru{\v{s}}a and {Brammer}, Gabriel and {Desprez}, Guillaume and {Iyer}, Kartheik and {Martis}, Nicholas and {Mowla}, Lamiya and {Noirot}, Ga{\"e}l and {Sawicki}, Marcin and {Strait}, Victoria and {Willott}, Chris J.},
        title = "{Spectroscopy from Photometry: A Population of Extreme Emission Line Galaxies at 1.7 {\ensuremath{\lesssim}} z {\ensuremath{\lesssim}} 6.7 Selected with JWST Medium Band Filters}",
      journal = {\apjl},
         year = 2023,
        month = nov,
       volume = {958},
       number = {1},
          eid = {L14},
        pages = {L14},
          doi = {10.3847/2041-8213/ad01c0},
archivePrefix = {arXiv},
       eprint = {2304.11181},
 primaryClass = {astro-ph.GA},
       adsurl = {https://ui.adsabs.harvard.edu/abs/2023ApJ...958L..14W}
}

@ARTICLE{Willott:2022,
       author = {{Willott}, Chris J. and {Doyon}, Ren{\'e} and {Albert}, Loic and {Brammer}, Gabriel B. and {Dixon}, William V. and {Muzic}, Koraljka and {Ravindranath}, Swara and {Scholz}, Aleks and {Abraham}, Roberto and {Artigau}, {\'E}tienne and {Brada{\v{c}}}, Maru{\v{s}}a and {Goudfrooij}, Paul and {Hutchings}, John B. and {Iyer}, Kartheik G. and {Jayawardhana}, Ray and {LaMassa}, Stephanie and {Martis}, Nicholas and {Meyer}, Michael R. and {Morishita}, Takahiro and {Mowla}, Lamiya and {Muzzin}, Adam and {Noirot}, Ga{\"e}l and {Pacifici}, Camilla and {Rowlands}, Neil and {Sarrouh}, Ghassan and {Sawicki}, Marcin and {Taylor}, Joanna M. and {Volk}, Kevin and {Zabl}, Johannes},
        title = "{The Near-infrared Imager and Slitless Spectrograph for the James Webb Space Telescope. II. Wide Field Slitless Spectroscopy}",
      journal = {\pasp},
         year = 2022,
        month = feb,
       volume = {134},
       number = {1032},
          eid = {025002},
        pages = {025002},
          doi = {10.1088/1538-3873/ac5158},
archivePrefix = {arXiv},
       eprint = {2202.01714},
 primaryClass = {astro-ph.IM},
       adsurl = {https://ui.adsabs.harvard.edu/abs/2022PASP..134b5002W}
}

@ARTICLE{Qi:2025,
       author = {{Qi}, Jia and {Garcia}, Alex M. and {Torrey}, Paul and {Moreno}, Jorge and {Green}, Kara N. and {Evans}, Aaron S. and {Hemler}, Z.~S. and {Hernquist}, Lars and {Ellison}, Sara L.},
        title = "{Star Formation Rates, Metallicities, and Stellar Masses on kpc-scales in TNG50}",
      journal = {arXiv e-prints},
         year = 2025,
        month = jan,
          eid = {arXiv:2501.18687},
        pages = {arXiv:2501.18687},
          doi = {10.48550/arXiv.2501.18687},
archivePrefix = {arXiv},
       eprint = {2501.18687},
 primaryClass = {astro-ph.GA},
       adsurl = {https://ui.adsabs.harvard.edu/abs/2025arXiv250118687Q}
}

@ARTICLE{Sotillo-Ramos:2022,
       author = {{Sotillo-Ramos}, Diego and {Pillepich}, Annalisa and {Donnari}, Martina and {Nelson}, Dylan and {Eisert}, Lukas and {Rodriguez-Gomez}, Vicente and {Joshi}, Gandhali and {Vogelsberger}, Mark and {Hernquist}, Lars},
        title = "{The merger and assembly histories of Milky Way- and M31-like galaxies with TNG50: disc survival through mergers}",
      journal = {\mnras},
         year = 2022,
        month = nov,
       volume = {516},
       number = {4},
        pages = {5404-5427},
          doi = {10.1093/mnras/stac2586},
archivePrefix = {arXiv},
       eprint = {2211.00036},
 primaryClass = {astro-ph.GA},
       adsurl = {https://ui.adsabs.harvard.edu/abs/2022MNRAS.516.5404S}
}

@ARTICLE{LiFei:2025,
       author = {{Li}, Fei and {Rahman}, Mubdi and {Murray}, Norman and {Kere{\v{s}}}, Du{\v{s}}an and {Wetzel}, Andrew and {Faucher-Gigu{\`e}re}, Claude-Andr{\'e} and {Hopkins}, Philip F. and {Moreno}, Jorge},
        title = "{The Effect of Galaxy Interactions on Starbursts in Milky Way-mass Galaxies in FIRE Simulations}",
      journal = {\apj},
         year = 2025,
        month = jan,
       volume = {979},
       number = {1},
          eid = {7},
        pages = {7},
          doi = {10.3847/1538-4357/ad94ef},
archivePrefix = {arXiv},
       eprint = {2411.10373},
 primaryClass = {astro-ph.GA},
       adsurl = {https://ui.adsabs.harvard.edu/abs/2025ApJ...979....7L}
}

@ARTICLE{Yu:2023,
       author = {{Yu}, Sijie and {Bullock}, James S. and {Gurvich}, Alexander B. and {Hafen}, Zachary and {Stern}, Jonathan and {Boylan-Kolchin}, Michael and {Faucher-Gigu{\`e}re}, Claude-Andr{\'e} and {Wetzel}, Andrew and {Hopkins}, Philip F. and {Moreno}, Jorge},
        title = "{Born this way: thin disc, thick disc, and isotropic spheroid formation in FIRE-2 Milky Way-mass galaxy simulations}",
      journal = {\mnras},
         year = 2023,
        month = aug,
       volume = {523},
       number = {4},
        pages = {6220-6238},
          doi = {10.1093/mnras/stad1806},
archivePrefix = {arXiv},
       eprint = {2210.03845},
 primaryClass = {astro-ph.GA},
       adsurl = {https://ui.adsabs.harvard.edu/abs/2023MNRAS.523.6220Y}
}

@ARTICLE{Ahumada:2020,
       author = {{Ahumada}, Romina and {Allende Prieto}, Carlos and {Almeida}, Andr{\'e}s and {Anders}, Friedrich and {Anderson}, Scott F. and {Andrews}, Brett H. and {Anguiano}, Borja and {Arcodia}, Riccardo and {Armengaud}, Eric and {Aubert}, Marie and {Avila}, Santiago and {Avila-Reese}, Vladimir and {Badenes}, Carles and {Balland}, Christophe and {Barger}, Kat and {Barrera-Ballesteros}, Jorge K. and {Basu}, Sarbani and {Bautista}, Julian and {Beaton}, Rachael L. and {Beers}, Timothy C. and {Benavides}, B. Izamar T. and {Bender}, Chad F. and {Bernardi}, Mariangela and {Bershady}, Matthew and {Beutler}, Florian and {Bidin}, Christian Moni and {Bird}, Jonathan and {Bizyaev}, Dmitry and {Blanc}, Guillermo A. and {Blanton}, Michael R. and {Boquien}, M{\'e}d{\'e}ric and {Borissova}, Jura and {Bovy}, Jo and {Brandt}, W.~N. and {Brinkmann}, Jonathan and {Brownstein}, Joel R. and {Bundy}, Kevin and {Bureau}, Martin and {Burgasser}, Adam and {Burtin}, Etienne and {Cano-D{\'\i}az}, Mariana and {Capasso}, Raffaella and {Cappellari}, Michele and {Carrera}, Ricardo and {Chabanier}, Sol{\`e}ne and {Chaplin}, William and {Chapman}, Michael and {Cherinka}, Brian and {Chiappini}, Cristina and {Doohyun Choi}, Peter and {Chojnowski}, S. Drew and {Chung}, Haeun and {Clerc}, Nicolas and {Coffey}, Damien and {Comerford}, Julia M. and {Comparat}, Johan and {da Costa}, Luiz and {Cousinou}, Marie-Claude and {Covey}, Kevin and {Crane}, Jeffrey D. and {Cunha}, Katia and {Ilha}, Gabriele da Silva and {Dai}, Yu Sophia and {Damsted}, Sanna B. and {Darling}, Jeremy and {Davidson}, Jr., James W. and {Davies}, Roger and {Dawson}, Kyle and {De}, Nikhil and {de la Macorra}, Axel and {De Lee}, Nathan and {Queiroz}, Anna B{\'a}rbara de Andrade and {Deconto Machado}, Alice and {de la Torre}, Sylvain and {Dell'Agli}, Flavia and {du Mas des Bourboux}, H{\'e}lion and {Diamond-Stanic}, Aleksandar M. and {Dillon}, Sean and {Donor}, John and {Drory}, Niv and {Duckworth}, Chris and {Dwelly}, Tom and {Ebelke}, Garrett and {Eftekharzadeh}, Sarah and {Davis Eigenbrot}, Arthur and {Elsworth}, Yvonne P. and {Eracleous}, Mike and {Erfanianfar}, Ghazaleh and {Escoffier}, Stephanie and {Fan}, Xiaohui and {Farr}, Emily and {Fern{\'a}ndez-Trincado}, Jos{\'e} G. and {Feuillet}, Diane and {Finoguenov}, Alexis and {Fofie}, Patricia and {Fraser-McKelvie}, Amelia and {Frinchaboy}, Peter M. and {Fromenteau}, Sebastien and {Fu}, Hai and {Galbany}, Llu{\'\i}s and {Garcia}, Rafael A. and {Garc{\'\i}a-Hern{\'a}ndez}, D.~A. and {Garma Oehmichen}, Luis Alberto and {Ge}, Junqiang and {Geimba Maia}, Marcio Antonio and {Geisler}, Doug and {Gelfand}, Joseph and {Goddy}, Julian and {Gonzalez-Perez}, Violeta and {Grabowski}, Kathleen and {Green}, Paul and {Grier}, Catherine J. and {Guo}, Hong and {Guy}, Julien and {Harding}, Paul and {Hasselquist}, Sten and {Hawken}, Adam James and {Hayes}, Christian R. and {Hearty}, Fred and {Hekker}, S. and {Hogg}, David W. and {Holtzman}, Jon A. and {Horta}, Danny and {Hou}, Jiamin and {Hsieh}, Bau-Ching and {Huber}, Daniel and {Hunt}, Jason A.~S. and {Ider Chitham}, J. and {Imig}, Julie and {Jaber}, Mariana and {Jimenez Angel}, Camilo Eduardo and {Johnson}, Jennifer A. and {Jones}, Amy M. and {J{\"o}nsson}, Henrik and {Jullo}, Eric and {Kim}, Yerim and {Kinemuchi}, Karen and {Kirkpatrick}, IV, Charles C. and {Kite}, George W. and {Klaene}, Mark and {Kneib}, Jean-Paul and {Kollmeier}, Juna A. and {Kong}, Hui and {Kounkel}, Marina and {Krishnarao}, Dhanesh and {Lacerna}, Ivan and {Lan}, Ting-Wen and {Lane}, Richard R. and {Law}, David R. and {Le Goff}, Jean-Marc and {Leung}, Henry W. and {Lewis}, Hannah and {Li}, Cheng and {Lian}, Jianhui and {Lin}, Lihwai and {Long}, Dan and {Longa-Pe{\~n}a}, Pen{\'e}lope and {Lundgren}, Britt and {Lyke}, Brad W. and {Mackereth}, J. Ted and {MacLeod}, Chelsea L. and {Majewski}, Steven R. and {Manchado}, Arturo and {Maraston}, Claudia and {Martini}, Paul and {Masseron}, Thomas and {Masters}, Karen L. and {Mathur}, Savita and {McDermid}, Richard M. and {Merloni}, Andrea and {Merrifield}, Michael and {M{\'e}sz{\'a}ros}, Szabolcs and {Miglio}, Andrea and {Minniti}, Dante and {Minsley}, Rebecca and {Miyaji}, Takamitsu and {Mohammad}, Faizan Gohar and {Mosser}, Benoit and {Mueller}, Eva-Maria and {Muna}, Demitri and {Mu{\~n}oz-Guti{\'e}rrez}, Andrea and {Myers}, Adam D. and {Nadathur}, Seshadri and {Nair}, Preethi and {Nandra}, Kirpal and {Correa do Nascimento}, Janaina and {Nevin}, Rebecca Jean and {Newman}, Jeffrey A. and {Nidever}, David L. and {Nitschelm}, Christian and {Noterdaeme}, Pasquier and {O'Connell}, Julia E. and {Olmstead}, Matthew D. and {Oravetz}, Daniel and {Oravetz}, Audrey and {Osorio}, Yeisson and {Pace}, Zachary J. and {Padilla}, Nelson and {Palanque-Delabrouille}, Nathalie and {Palicio}, Pedro A.},
        title = "{The 16th Data Release of the Sloan Digital Sky Surveys: First Release from the APOGEE-2 Southern Survey and Full Release of eBOSS Spectra}",
      journal = {\apjs},
         year = 2020,
        month = jul,
       volume = {249},
       number = {1},
          eid = {3},
        pages = {3},
          doi = {10.3847/1538-4365/ab929e},
archivePrefix = {arXiv},
       eprint = {1912.02905},
 primaryClass = {astro-ph.GA},
       adsurl = {https://ui.adsabs.harvard.edu/abs/2020ApJS..249....3A}
}

@ARTICLE{Jonsson:2020,
       author = {{J{\"o}nsson}, Henrik and {Holtzman}, Jon A. and {Allende Prieto}, Carlos and {Cunha}, Katia and {Garc{\'\i}a-Hern{\'a}ndez}, D.~A. and {Hasselquist}, Sten and {Masseron}, Thomas and {Osorio}, Yeisson and {Shetrone}, Matthew and {Smith}, Verne and {Stringfellow}, Guy S. and {Bizyaev}, Dmitry and {Edvardsson}, Bengt and {Majewski}, Steven R. and {M{\'e}sz{\'a}ros}, Szabolcs and {Souto}, Diogo and {Zamora}, Olga and {Beaton}, Rachael L. and {Bovy}, Jo and {Donor}, John and {Pinsonneault}, Marc H. and {Poovelil}, Vijith Jacob and {Sobeck}, Jennifer},
        title = "{APOGEE Data and Spectral Analysis from SDSS Data Release 16: Seven Years of Observations Including First Results from APOGEE-South}",
      journal = {\aj},
         year = 2020,
        month = sep,
       volume = {160},
       number = {3},
          eid = {120},
        pages = {120},
          doi = {10.3847/1538-3881/aba592},
archivePrefix = {arXiv},
       eprint = {2007.05537},
 primaryClass = {astro-ph.GA},
       adsurl = {https://ui.adsabs.harvard.edu/abs/2020AJ....160..120J}
}

@ARTICLE{Blanton:2017,
       author = {{Blanton}, Michael R. and {Bershady}, Matthew A. and {Abolfathi}, Bela and {Albareti}, Franco D. and {Allende Prieto}, Carlos and {Almeida}, Andres and {Alonso-Garc{\'\i}a}, Javier and {Anders}, Friedrich and {Anderson}, Scott F. and {Andrews}, Brett and {Aquino-Ort{\'\i}z}, Erik and {Arag{\'o}n-Salamanca}, Alfonso and {Argudo-Fern{\'a}ndez}, Maria and {Armengaud}, Eric and {Aubourg}, Eric and {Avila-Reese}, Vladimir and {Badenes}, Carles and {Bailey}, Stephen and {Barger}, Kathleen A. and {Barrera-Ballesteros}, Jorge and {Bartosz}, Curtis and {Bates}, Dominic and {Baumgarten}, Falk and {Bautista}, Julian and {Beaton}, Rachael and {Beers}, Timothy C. and {Belfiore}, Francesco and {Bender}, Chad F. and {Berlind}, Andreas A. and {Bernardi}, Mariangela and {Beutler}, Florian and {Bird}, Jonathan C. and {Bizyaev}, Dmitry and {Blanc}, Guillermo A. and {Blomqvist}, Michael and {Bolton}, Adam S. and {Boquien}, M{\'e}d{\'e}ric and {Borissova}, Jura and {van den Bosch}, Remco and {Bovy}, Jo and {Brandt}, William N. and {Brinkmann}, Jonathan and {Brownstein}, Joel R. and {Bundy}, Kevin and {Burgasser}, Adam J. and {Burtin}, Etienne and {Busca}, Nicol{\'a}s G. and {Cappellari}, Michele and {Delgado Carigi}, Maria Leticia and {Carlberg}, Joleen K. and {Carnero Rosell}, Aurelio and {Carrera}, Ricardo and {Chanover}, Nancy J. and {Cherinka}, Brian and {Cheung}, Edmond and {G{\'o}mez Maqueo Chew}, Yilen and {Chiappini}, Cristina and {Choi}, Peter Doohyun and {Chojnowski}, Drew and {Chuang}, Chia-Hsun and {Chung}, Haeun and {Cirolini}, Rafael Fernando and {Clerc}, Nicolas and {Cohen}, Roger E. and {Comparat}, Johan and {da Costa}, Luiz and {Cousinou}, Marie-Claude and {Covey}, Kevin and {Crane}, Jeffrey D. and {Croft}, Rupert A.~C. and {Cruz-Gonzalez}, Irene and {Garrido Cuadra}, Daniel and {Cunha}, Katia and {Damke}, Guillermo J. and {Darling}, Jeremy and {Davies}, Roger and {Dawson}, Kyle and {de la Macorra}, Axel and {Dell'Agli}, Flavia and {De Lee}, Nathan and {Delubac}, Timoth{\'e}e and {Di Mille}, Francesco and {Diamond-Stanic}, Aleks and {Cano-D{\'\i}az}, Mariana and {Donor}, John and {Downes}, Juan Jos{\'e} and {Drory}, Niv and {du Mas des Bourboux}, H{\'e}lion and {Duckworth}, Christopher J. and {Dwelly}, Tom and {Dyer}, Jamie and {Ebelke}, Garrett and {Eigenbrot}, Arthur D. and {Eisenstein}, Daniel J. and {Emsellem}, Eric and {Eracleous}, Mike and {Escoffier}, Stephanie and {Evans}, Michael L. and {Fan}, Xiaohui and {Fern{\'a}ndez-Alvar}, Emma and {Fernandez-Trincado}, J.~G. and {Feuillet}, Diane K. and {Finoguenov}, Alexis and {Fleming}, Scott W. and {Font-Ribera}, Andreu and {Fredrickson}, Alexander and {Freischlad}, Gordon and {Frinchaboy}, Peter M. and {Fuentes}, Carla E. and {Galbany}, Llu{\'\i}s and {Garcia-Dias}, R. and {Garc{\'\i}a-Hern{\'a}ndez}, D.~A. and {Gaulme}, Patrick and {Geisler}, Doug and {Gelfand}, Joseph D. and {Gil-Mar{\'\i}n}, H{\'e}ctor and {Gillespie}, Bruce A. and {Goddard}, Daniel and {Gonzalez-Perez}, Violeta and {Grabowski}, Kathleen and {Green}, Paul J. and {Grier}, Catherine J. and {Gunn}, James E. and {Guo}, Hong and {Guy}, Julien and {Hagen}, Alex and {Hahn}, ChangHoon and {Hall}, Matthew and {Harding}, Paul and {Hasselquist}, Sten and {Hawley}, Suzanne L. and {Hearty}, Fred and {Gonzalez Hern{\'a}ndez}, Jonay I. and {Ho}, Shirley and {Hogg}, David W. and {Holley-Bockelmann}, Kelly and {Holtzman}, Jon A. and {Holzer}, Parker H. and {Huehnerhoff}, Joseph and {Hutchinson}, Timothy A. and {Hwang}, Ho Seong and {Ibarra-Medel}, H{\'e}ctor J. and {da Silva Ilha}, Gabriele and {Ivans}, Inese I. and {Ivory}, KeShawn and {Jackson}, Kelly and {Jensen}, Trey W. and {Johnson}, Jennifer A. and {Jones}, Amy and {J{\"o}nsson}, Henrik and {Jullo}, Eric and {Kamble}, Vikrant and {Kinemuchi}, Karen and {Kirkby}, David and {Kitaura}, Francisco-Shu and {Klaene}, Mark and {Knapp}, Gillian R. and {Kneib}, Jean-Paul and {Kollmeier}, Juna A. and {Lacerna}, Ivan and {Lane}, Richard R. and {Lang}, Dustin and {Law}, David R. and {Lazarz}, Daniel and {Lee}, Youngbae and {Le Goff}, Jean-Marc and {Liang}, Fu-Heng and {Li}, Cheng and {Li}, Hongyu and {Lian}, Jianhui and {Lima}, Marcos and {Lin}, Lihwai and {Lin}, Yen-Ting and {Bertran de Lis}, Sara and {Liu}, Chao and {de Icaza Lizaola}, Miguel Angel C. and {Long}, Dan and {Lucatello}, Sara and {Lundgren}, Britt and {MacDonald}, Nicholas K. and {Deconto Machado}, Alice and {MacLeod}, Chelsea L. and {Mahadevan}, Suvrath and {Geimba Maia}, Marcio Antonio and {Maiolino}, Roberto and {Majewski}, Steven R. and {Malanushenko}, Elena and {Malanushenko}, Viktor and {Manchado}, Arturo and {Mao}, Shude and {Maraston}, Claudia and {Marques-Chaves}, Rui and {Masseron}, Thomas and {Masters}, Karen L. and {McBride}, Cameron K. and {McDermid}, Richard M. and {McGrath}, Brianne and {McGreer}, Ian D. and {Medina Pe{\~n}a}, Nicol{\'a}s and {Melendez}, Matthew},
        title = "{Sloan Digital Sky Survey IV: Mapping the Milky Way, Nearby Galaxies, and the Distant Universe}",
      journal = {\aj},
         year = 2017,
        month = jul,
       volume = {154},
       number = {1},
          eid = {28},
        pages = {28},
          doi = {10.3847/1538-3881/aa7567},
archivePrefix = {arXiv},
       eprint = {1703.00052},
 primaryClass = {astro-ph.GA},
       adsurl = {https://ui.adsabs.harvard.edu/abs/2017AJ....154...28B}
}

@ARTICLE{Duan:2025,
       author = {{Duan}, Qiao and {Conselice}, Christopher J. and {Li}, Qiong and {Austin}, Duncan and {Harvey}, Thomas and {Adams}, Nathan J. and {Duncan}, Kenneth J. and {Trussler}, James and {Ferreira}, Leonardo and {Westcott}, Lewi and {Harris}, Honor and {Windhorst}, Rogier A. and {Holwerda}, Benne W. and {Broadhurst}, Thomas J. and {Coe}, Dan and {Cohen}, Seth H. and {Du}, Xiaojing and {Driver}, Simon P. and {Frye}, Brenda and {Grogin}, Norman A. and {Hathi}, Nimish P. and {Jansen}, Rolf A. and {Koekemoer}, Anton M. and {Marshall}, Madeline A. and {Nonino}, Mario and {Ortiz}, III, Rafael and {Pirzkal}, Nor and {Robotham}, Aaron and {Ryan}, Russell E. and {Summers}, Jake and {D'Silva}, Jordan C.~J. and {Willmer}, Christopher N.~A. and {Yan}, Haojing},
        title = "{Galaxy mergers in the epoch of reionization ─ I. A JWST study of pair fractions, merger rates, and stellar mass accretion rates at z = 4.5─11.5}",
      journal = {\mnras},
         year = 2025,
        month = jun,
       volume = {540},
       number = {1},
        pages = {774-805},
          doi = {10.1093/mnras/staf638},
archivePrefix = {arXiv},
       eprint = {2407.09472},
 primaryClass = {astro-ph.GA},
       adsurl = {https://ui.adsabs.harvard.edu/abs/2025MNRAS.540..774D}
}

@ARTICLE{Carnall:2024,
       author = {{Carnall}, A.~C. and {Cullen}, F. and {McLure}, R.~J. and {McLeod}, D.~J. and {Begley}, R. and {Donnan}, C.~T. and {Dunlop}, J.~S. and {Shapley}, A.~E. and {Rowlands}, K. and {Almaini}, O. and {Arellano-C{\'o}rdova}, K.~Z. and {Barrufet}, L. and {Cimatti}, A. and {El:2025s}, R.~S. and {Grogin}, N.~A. and {Hamadouche}, M.~L. and {Illingworth}, G.~D. and {Koekemoer}, A.~M. and {Leung}, H. -H. and {Lovell}, C.~C. and {P{\'e}rez-Gonz{\'a}lez}, P.~G. and {Santini}, P. and {Stanton}, T.~M. and {Wild}, V.},
        title = "{The JWST EXCELS survey: too much, too young, too fast? Ultra-massive quiescent galaxies at 3 < z < 5}",
      journal = {\mnras},
         year = 2024,
        month = oct,
       volume = {534},
       number = {1},
        pages = {325-348},
          doi = {10.1093/mnras/stae2092},
archivePrefix = {arXiv},
       eprint = {2405.02242},
 primaryClass = {astro-ph.GA},
       adsurl = {https://ui.adsabs.harvard.edu/abs/2024MNRAS.534..325C}
}

@ARTICLE{Ciuca:2024,
       author = {{Ciuc{\u{a}}}, Ioana and {Kawata}, Daisuke and {Ting}, Yuan-Sen and {Grand}, Robert J.~J. and {Miglio}, Andrea and {Hayden}, Michael and {Baba}, Junichi and {Fragkoudi}, Francesca and {Monty}, Stephanie and {Buder}, Sven and {Freeman}, Ken},
        title = "{Chasing the impact of the Gaia-Sausage-Enceladus merger on the formation of the Milky Way thick disc}",
      journal = {\mnras},
         year = 2024,
        month = feb,
       volume = {528},
       number = {1},
        pages = {L122-L126},
          doi = {10.1093/mnrasl/slad033},
archivePrefix = {arXiv},
       eprint = {2211.01006},
 primaryClass = {astro-ph.GA},
       adsurl = {https://ui.adsabs.harvard.edu/abs/2024MNRAS.528L.122C}
}

@ARTICLE{Jain:2024,
       author = {{Jain}, Shweta and {Tacchella}, Sandro and {Mosleh}, Moein},
        title = "{Self-regulated growth of galaxy sizes along the star-forming main sequence}",
      journal = {The Open Journal of Astrophysics},
         year = 2024,
        month = dec,
       volume = {7},
          eid = {113},
        pages = {113},
          doi = {10.33232/001c.126775},
archivePrefix = {arXiv},
       eprint = {2412.00599},
 primaryClass = {astro-ph.GA},
       adsurl = {https://ui.adsabs.harvard.edu/abs/2024OJAp....7E.113J}
}

@ARTICLE{Jin:2024,
       author = {{Jin}, Bingcheng and {Ho}, Luis C. and {Sun}, Wen},
        title = "{A High Incidence of Central Star Formation Inferred from the Color Gradients of Galaxies at $z>4$}",
      journal = {arXiv e-prints},
         year = 2024,
        month = dec,
          eid = {arXiv:2412.03455},
        pages = {arXiv:2412.03455},
          doi = {10.48550/arXiv.2412.03455},
archivePrefix = {arXiv},
       eprint = {2412.03455},
 primaryClass = {astro-ph.GA},
       adsurl = {https://ui.adsabs.harvard.edu/abs/2024arXiv241203455J}
}

@ARTICLE{Sazonova:2024,
       author = {{Sazonova}, Elizaveta and {Morgan}, Cameron and {Balogh}, Michael and {Alatalo}, Katherine and {Benavides}, Jose A. and {Bluck}, Asa and {Brough}, Sarah and {Busa}, Innocenza and {Demarco}, Ricardo and {Donevski}, Darko and {Figueira}, Miguel and {Martin}, Garreth and {Rodriguez-Gomez}, Vicente and {Rom{\'a}n}, Javier and {Rowlands}, Kate},
        title = "{RMS asymmetry: a robust metric of galaxy shapes in images with varied depth and resolution}",
      journal = {arXiv e-prints},
         year = 2024,
        month = apr,
          eid = {arXiv:2404.05792},
        pages = {arXiv:2404.05792},
          doi = {10.48550/arXiv.2404.05792},
archivePrefix = {arXiv},
       eprint = {2404.05792},
 primaryClass = {astro-ph.GA},
       adsurl = {https://ui.adsabs.harvard.edu/abs/2024arXiv240405792S}
}

@ARTICLE{Sun:2024,
       author = {{Sun}, Xunda and {Wang}, Xin and {Ma}, Xiangcheng and {Wang}, Kai and {Wetzel}, Andrew and {Faucher-Gigu{\`e}re}, Claude-Andr{\'e} and {Hopkins}, Philip F. and {Kere{\v{s}}}, Du{\v{s}}an and {Graf}, Russell L. and {Marszewski}, Andrew and {Stern}, Jonathan and {Sun}, Guochao and {Sun}, Lei and {Thyme}, Keyer},
        title = "{The physical origin of positive metallicity radial gradients in high-redshift galaxies: insights from the FIRE-2 cosmological hydrodynamic simulations}",
      journal = {arXiv e-prints},
         year = 2024,
        month = sep,
          eid = {arXiv:2409.09290},
        pages = {arXiv:2409.09290},
          doi = {10.48550/arXiv.2409.09290},
archivePrefix = {arXiv},
       eprint = {2409.09290},
 primaryClass = {astro-ph.GA},
       adsurl = {https://ui.adsabs.harvard.edu/abs/2024arXiv240909290S}
}

@ARTICLE{Tan:2025,
       author = {{Tan}, Vivian Yun Yan and {Muzzin}, Adam and {Sarrouh}, Ghassan T.~E. and {Antwi-Danso}, Jacqueline and {Sok}, Visal and {Jagga}, Naadiyah and {Rihtar{\v{s}}i{\v{c}}}, Gregor and {Brown}, Westley and {Abraham}, Roberto and {Asada}, Yoshihisa and {Desprez}, Guillaume and {Iyer}, Kartheik and {Martis}, Nicholas S. and {M{\'e}rida}, Rosa M. and {Mowla}, Lamiya A. and {Noirot}, Ga{\"e}l and {Omori}, Kiyoaki Christopher and {Sawicki}, Marcin and {Tripodi}, Roberta and {Willott}, Chris J.},
        title = "{Resolved Mass Assembly and Star Formation in Milky Way Progenitors since z = 5 from JWST/CANUCS: From Clumps and Mergers to Well-ordered Disks}",
      journal = {\apj},
         year = 2025,
        month = nov,
       volume = {994},
       number = {1},
          eid = {94},
        pages = {94},
          doi = {10.3847/1538-4357/ae0ffe},
archivePrefix = {arXiv},
       eprint = {2412.07829},
 primaryClass = {astro-ph.GA},
       adsurl = {https://ui.adsabs.harvard.edu/abs/2025ApJ...994...94T}
}

@ARTICLE{Antwi-Danso:2023,
       author = {{Antwi-Danso}, Jacqueline and {Papovich}, Casey and {Leja}, Joel and {Marchesini}, Danilo and {Marsan}, Z. Cemile and {Martis}, Nicholas S. and {Labb{\'e}}, Ivo and {Muzzin}, Adam and {Glazebrook}, Karl and {Straatman}, Caroline M.~S. and {Tran}, Kim-Vy H.},
        title = "{Beyond UVJ: Color Selection of Galaxies in the JWST Era}",
      journal = {\apj},
         year = 2023,
        month = feb,
       volume = {943},
       number = {2},
          eid = {166},
        pages = {166},
          doi = {10.3847/1538-4357/aca294},
archivePrefix = {arXiv},
       eprint = {2207.07170},
 primaryClass = {astro-ph.GA},
       adsurl = {https://ui.adsabs.harvard.edu/abs/2023ApJ...943..166A}
}

@ARTICLE{Imig:2023,
       author = {{Imig}, Julie and {Price}, Cathryn and {Holtzman}, Jon A. and {Stone-Martinez}, Alexander and {Majewski}, Steven R. and {Weinberg}, David H. and {Johnson}, Jennifer A. and {Allende Prieto}, Carlos and {Beaton}, Rachael L. and {Beers}, Timothy C. and {Bizyaev}, Dmitry and {Blanton}, Michael R. and {Brownstein}, Joel R. and {Cunha}, Katia and {Fern{\'a}ndez-Trincado}, Jos{\'e} G. and {Feuillet}, Diane K. and {Hasselquist}, Sten and {Hayes}, Christian R. and {J{\"o}nsson}, Henrik and {Lane}, Richard R. and {Lian}, Jianhui and {M{\'e}sz{\'a}ros}, Szabolcs and {Nidever}, David L. and {Robin}, Annie C. and {Shetrone}, Matthew and {Smith}, Verne and {Wilson}, John C.},
        title = "{A Tale of Two Disks: Mapping the Milky Way with the Final Data Release of APOGEE}",
      journal = {\apj},
         year = 2023,
        month = sep,
       volume = {954},
       number = {2},
          eid = {124},
        pages = {124},
          doi = {10.3847/1538-4357/ace9b8},
archivePrefix = {arXiv},
       eprint = {2307.13887},
 primaryClass = {astro-ph.GA},
       adsurl = {https://ui.adsabs.harvard.edu/abs/2023ApJ...954..124I}
}

@ARTICLE{Abdurrouf:2023,
       author = {{Abdurro'uf} and {Coe}, Dan and {Jung}, Intae and {Ferguson}, Henry C. and {Brammer}, Gabriel and {Iyer}, Kartheik G. and {Bradley}, Larry D. and {Dayal}, Pratika and {Windhorst}, Rogier A. and {Zitrin}, Adi and {Meena}, Ashish Kumar and {Oguri}, Masamune and {Diego}, Jose M. and {Kokorev}, Vasily and {Dimauro}, Paola and {Adamo}, Angela and {Conselice}, Christopher J. and {Welch}, Brian and {Vanzella}, Eros and {Hsiao}, Tiger Yu-Yang and {Xu}, Xinfeng and {Roy}, Namrata and {Mulcahey}, Celia R.},
        title = "{Spatially Resolved Stellar Populations of 0.3 < z < 6.0 Galaxies in WHL 0137-08 and MACS 0647+70 Clusters as Revealed by JWST: How Do Galaxies Grow and Quench over Cosmic Time?}",
      journal = {\apj},
         year = 2023,
        month = mar,
       volume = {945},
       number = {2},
          eid = {117},
        pages = {117},
          doi = {10.3847/1538-4357/acba06},
archivePrefix = {arXiv},
       eprint = {2301.02209},
 primaryClass = {astro-ph.GA},
       adsurl = {https://ui.adsabs.harvard.edu/abs/2023ApJ...945..117A}
}

@ARTICLE{Akhshik:2023,
       author = {{Akhshik}, Mohammad and {Whitaker}, Katherine E. and {Leja}, Joel and {Richard}, Johan and {Spilker}, Justin S. and {Song}, Mimi and {Brammer}, Gabriel and {Bezanson}, Rachel and {Ebeling}, Harald and {Gallazzi}, Anna R. and {Mahler}, Guillaume and {Mowla}, Lamiya A. and {Nelson}, Erica J. and {Pacifici}, Camilla and {Sharon}, Keren and {Toft}, Sune and {Williams}, Christina C. and {Wright}, Lillian and {Zabl}, Johannes},
        title = "{REQUIEM-2D: A Diversity of Formation Pathways in a Sample of Spatially Resolved Massive Quiescent Galaxies at z   2}",
      journal = {\apj},
         year = 2023,
        month = feb,
       volume = {943},
       number = {2},
          eid = {179},
        pages = {179},
          doi = {10.3847/1538-4357/aca677},
archivePrefix = {arXiv},
       eprint = {2203.04979},
 primaryClass = {astro-ph.GA},
       adsurl = {https://ui.adsabs.harvard.edu/abs/2023ApJ...943..179A}
}

@ARTICLE{Hopkins:2023,
       author = {{Hopkins}, Philip F. and {Gurvich}, Alexander B. and {Shen}, Xuejian and {Hafen}, Zachary and {Grudi{\'c}}, Michael Y. and {Kurinchi-Vendhan}, Shalini and {Hayward}, Christopher C. and {Jiang}, Fangzhou and {Orr}, Matthew E. and {Wetzel}, Andrew and {Kere{\v{s}}}, Du{\v{s}}an and {Stern}, Jonathan and {Faucher-Gigu{\`e}re}, Claude-Andr{\'e} and {Bullock}, James and {Wheeler}, Coral and {El-Badry}, Kareem and {Loebman}, Sarah R. and {Moreno}, Jorge and {Boylan-Kolchin}, Michael and {Quataert}, Eliot},
        title = "{What causes the formation of discs and end of bursty star formation?}",
      journal = {\mnras},
         year = 2023,
        month = oct,
       volume = {525},
       number = {2},
        pages = {2241-2286},
          doi = {10.1093/mnras/stad1902},
archivePrefix = {arXiv},
       eprint = {2301.08263},
 primaryClass = {astro-ph.GA},
       adsurl = {https://ui.adsabs.harvard.edu/abs/2023MNRAS.525.2241H}
}

@ARTICLE{Lian:2023,
       author = {{Lian}, Jianhui and {Bergemann}, Maria and {Pillepich}, Annalisa and {Zasowski}, Gail and {Lane}, Richard R.},
        title = "{The integrated metallicity profile of the Milky Way}",
      journal = {Nature Astronomy},
         year = 2023,
        month = aug,
       volume = {7},
        pages = {951-958},
          doi = {10.1038/s41550-023-01977-z},
archivePrefix = {arXiv},
       eprint = {2306.14100},
 primaryClass = {astro-ph.GA},
       adsurl = {https://ui.adsabs.harvard.edu/abs/2023NatAs...7..951L}
}

@ARTICLE{Li:2022,
       author = {{Li}, Zihao and {Wang}, Xin and {Cai}, Zheng and {Shi}, Dong Dong and {Fan}, Xiaohui and {Zheng}, Xian Zhong and {Malkan}, Matthew A. and {Teplitz}, Harry I. and {Henry}, Alaina L. and {Bian}, Fuyan and {Colbert}, James},
        title = "{First Census of Gas-phase Metallicity Gradients of Star-forming Galaxies in Overdense Environments at Cosmic Noon}",
      journal = {\apjl},
         year = 2022,
        month = apr,
       volume = {929},
       number = {1},
          eid = {L8},
        pages = {L8},
          doi = {10.3847/2041-8213/ac626f},
archivePrefix = {arXiv},
       eprint = {2204.03008},
 primaryClass = {astro-ph.GA},
       adsurl = {https://ui.adsabs.harvard.edu/abs/2022ApJ...929L...8L}
}

@ARTICLE{Nagaraj:2022,
       author = {{Nagaraj}, Gautam and {Forbes}, John C. and {Leja}, Joel and {Foreman-Mackey}, Daniel and {Hayward}, Christopher C.},
        title = "{A Bayesian Population Model for the Observed Dust Attenuation in Galaxies}",
      journal = {\apj},
         year = 2022,
        month = jun,
       volume = {932},
       number = {1},
          eid = {54},
        pages = {54},
          doi = {10.3847/1538-4357/ac6c80},
archivePrefix = {arXiv},
       eprint = {2202.05102},
 primaryClass = {astro-ph.GA},
       adsurl = {https://ui.adsabs.harvard.edu/abs/2022ApJ...932...54N}
}

@ARTICLE{Shah:2022,
       author = {{Shah}, Ekta A. and {Kartaltepe}, Jeyhan S. and {Magagnoli}, Christina T. and {Cox}, Isabella G. and {Wetherell}, Caleb T. and {Vanderhoof}, Brittany N. and {Cooke}, Kevin C. and {Calabro}, Antonello and {Chartab}, Nima and {Conselice}, Christopher J. and {Croton}, Darren J. and {de la Vega}, Alexander and {Hathi}, Nimish P. and {Ilbert}, Olivier and {Inami}, Hanae and {Kocevski}, Dale D. and {Koekemoer}, Anton M. and {Lemaux}, Brian C. and {Lubin}, Lori and {Mantha}, Kameswara Bharadwaj and {Marchesi}, Stefano and {Martig}, Marie and {Moreno}, Jorge and {Pampliega}, Belen Alcalde and {Patton}, David R. and {Salvato}, Mara and {Treister}, Ezequiel},
        title = "{Investigating the Effect of Galaxy Interactions on Star Formation at 0.5 < z < 3.0}",
      journal = {\apj},
         year = 2022,
        month = nov,
       volume = {940},
       number = {1},
          eid = {4},
        pages = {4},
          doi = {10.3847/1538-4357/ac96eb},
archivePrefix = {arXiv},
       eprint = {2209.15587},
 primaryClass = {astro-ph.GA},
       adsurl = {https://ui.adsabs.harvard.edu/abs/2022ApJ...940....4S}
}

@ARTICLE{Sok:2022,
       author = {{Sok}, Visal and {Muzzin}, Adam and {Jablonka}, Pascale and {Marsan}, Z. Cemile and {Tan}, Vivian Y.~Y. and {Alcorn}, Leo and {Marchesini}, Danilo and {Stefanon}, Mauro},
        title = "{Finite-resolution Deconvolution of Multiwavelength Imaging of 20,000 Galaxies in the COSMOS Field: The Evolution of Clumpy Galaxies over Cosmic Time}",
      journal = {\apj},
         year = 2022,
        month = jan,
       volume = {924},
       number = {1},
          eid = {7},
        pages = {7},
          doi = {10.3847/1538-4357/ac2f40},
archivePrefix = {arXiv},
       eprint = {2110.07612},
 primaryClass = {astro-ph.GA},
       adsurl = {https://ui.adsabs.harvard.edu/abs/2022ApJ...924....7S}
}

@ARTICLE{Tissera:2022,
       author = {{Tissera}, Patricia B. and {Rosas-Guevara}, Yetli and {Sillero}, Emanuel and {Pedrosa}, Susana E. and {Theuns}, Tom and {Bignone}, Lucas},
        title = "{The evolution of the oxygen abundance gradients in star-forming galaxies in the EAGLE simulations}",
      journal = {\mnras},
         year = 2022,
        month = apr,
       volume = {511},
       number = {2},
        pages = {1667-1684},
          doi = {10.1093/mnras/stab3644},
archivePrefix = {arXiv},
       eprint = {2112.06553},
 primaryClass = {astro-ph.GA},
       adsurl = {https://ui.adsabs.harvard.edu/abs/2022MNRAS.511.1667T}
}

@ARTICLE{Horstman:2021,
       author = {{Horstman}, Katelyn and {Shapley}, Alice E. and {Sanders}, Ryan L. and {Mobasher}, Bahram and {Reddy}, Naveen A. and {Kriek}, Mariska and {Coil}, Alison L. and {Siana}, Brian and {Shivaei}, Irene and {Freeman}, William R. and {Azadi}, Mojegan and {Price}, Sedona H. and {Leung}, Gene C.~K. and {Fetherolf}, Tara and {de Groot}, Laura and {Zick}, Tom and {Fornasini}, Francesca M. and {Barro}, Guillermo},
        title = "{The MOSDEF survey: differences in SFR and metallicity for morphologically selected mergers at z {\ensuremath{\sim}} 2}",
      journal = {\mnras},
         year = 2021,
        month = feb,
       volume = {501},
       number = {1},
        pages = {137-145},
          doi = {10.1093/mnras/staa3502},
archivePrefix = {arXiv},
       eprint = {2008.04327},
 primaryClass = {astro-ph.GA},
       adsurl = {https://ui.adsabs.harvard.edu/abs/2021MNRAS.501..137H}
}

@ARTICLE{McLeod:2021,
       author = {{McLeod}, D.~J. and {McLure}, R.~J. and {Dunlop}, J.~S. and {Cullen}, F. and {Carnall}, A.~C. and {Duncan}, K.},
        title = "{The evolution of the galaxy stellar-mass function over the last 12 billion years from a combination of ground-based and HST surveys}",
      journal = {\mnras},
         year = 2021,
        month = may,
       volume = {503},
       number = {3},
        pages = {4413-4435},
          doi = {10.1093/mnras/stab731},
archivePrefix = {arXiv},
       eprint = {2009.03176},
 primaryClass = {astro-ph.GA},
       adsurl = {https://ui.adsabs.harvard.edu/abs/2021MNRAS.503.4413M}
}

@ARTICLE{Simons:2021,
       author = {{Simons}, Raymond C. and {Papovich}, Casey and {Momcheva}, Ivelina and {Trump}, Jonathan R. and {Brammer}, Gabriel and {Estrada-Carpenter}, Vicente and {Backhaus}, Bren E. and {Cleri}, Nikko J. and {Finkelstein}, Steven L. and {Giavalisco}, Mauro and {Ji}, Zhiyuan and {Jung}, Intae and {Matharu}, Jasleen and {Weiner}, Benjamin},
        title = "{CLEAR: The Gas-phase Metallicity Gradients of Star-forming Galaxies at 0.6 < z < 2.6}",
      journal = {\apj},
         year = 2021,
        month = dec,
       volume = {923},
       number = {2},
          eid = {203},
        pages = {203},
          doi = {10.3847/1538-4357/ac28f4},
archivePrefix = {arXiv},
       eprint = {2011.03553},
 primaryClass = {astro-ph.GA},
       adsurl = {https://ui.adsabs.harvard.edu/abs/2021ApJ...923..203S}
}

@ARTICLE{Roberts-Borsani:2021,
       author = {{Roberts-Borsani}, Guido and {Treu}, Tommaso and {Mason}, Charlotte and {Schmidt}, Kasper B. and {Jones}, Tucker and {Fontana}, Adriano},
        title = "{Improving z {\ensuremath{\sim}} 7-11 Galaxy Property Estimates with JWST/NIRCam Medium-band Photometry}",
      journal = {\apj},
         year = 2021,
        month = apr,
       volume = {910},
       number = {2},
          eid = {86},
        pages = {86},
          doi = {10.3847/1538-4357/abe45b},
archivePrefix = {arXiv},
       eprint = {2102.04469},
 primaryClass = {astro-ph.GA},
       adsurl = {https://ui.adsabs.harvard.edu/abs/2021ApJ...910...86R}
}

@ARTICLE{Bluck:2020,
       author = {{Bluck}, Asa F.~L. and {Maiolino}, Roberto and {Piotrowska}, Joanna M. and {Trussler}, James and {Ellison}, Sara L. and {S{\'a}nchez}, Sebastian F. and {Thorp}, Mallory D. and {Teimoorinia}, Hossen and {Moreno}, Jorge and {Conselice}, Christopher J.},
        title = "{How do central and satellite galaxies quench? - Insights from spatially resolved spectroscopy in the MaNGA survey}",
      journal = {\mnras},
         year = 2020,
        month = nov,
       volume = {499},
       number = {1},
        pages = {230-268},
          doi = {10.1093/mnras/staa2806},
archivePrefix = {arXiv},
       eprint = {2009.05341},
 primaryClass = {astro-ph.GA},
       adsurl = {https://ui.adsabs.harvard.edu/abs/2020MNRAS.499..230B}
}

@ARTICLE{Boardman:2020a,
       author = {{Boardman}, N. and {Zasowski}, G. and {Seth}, A. and {Newman}, J. and {Andrews}, B. and {Bershady}, M. and {Bird}, J. and {Chiappini}, C. and {Fielder}, C. and {Fraser-McKelvie}, A. and {Jones}, A. and {Licquia}, T. and {Masters}, K.~L. and {Minchev}, I. and {Schiavon}, R.~P. and {Brownstein}, J.~R. and {Drory}, N. and {Lane}, R.~R.},
        title = "{Milky Way analogues in MaNGA: multiparameter homogeneity and comparison to the Milky Way}",
      journal = {\mnras},
         year = 2020,
        month = jan,
       volume = {491},
       number = {3},
        pages = {3672-3701},
          doi = {10.1093/mnras/stz3126},
archivePrefix = {arXiv},
       eprint = {1910.12896},
 primaryClass = {astro-ph.GA},
       adsurl = {https://ui.adsabs.harvard.edu/abs/2020MNRAS.491.3672B}
}

@ARTICLE{Bustamante:2020,
       author = {{Bustamante}, Sebasti{\'a}n and {Ellison}, Sara L. and {Patton}, David R. and {Sparre}, Martin},
        title = "{Galaxy pairs in the Sloan Digital Sky Survey - XIV. Galaxy mergers do not lie on the fundamental metallicity relation}",
      journal = {\mnras},
         year = 2020,
        month = may,
       volume = {494},
       number = {3},
        pages = {3469-3480},
          doi = {10.1093/mnras/staa1025},
archivePrefix = {arXiv},
       eprint = {2004.06121},
 primaryClass = {astro-ph.GA},
       adsurl = {https://ui.adsabs.harvard.edu/abs/2020MNRAS.494.3469B}
}

@ARTICLE{Curti:2020,
       author = {{Curti}, Mirko and {Maiolino}, Roberto and {Cirasuolo}, Michele and {Mannucci}, Filippo and {Williams}, Rebecca J. and {Auger}, Matt and {Mercurio}, Amata and {Hayden-Pawson}, Connor and {Cresci}, Giovanni and {Marconi}, Alessandro and {Belfiore}, Francesco and {Cappellari}, Michele and {Cicone}, Claudia and {Cullen}, Fergus and {Meneghetti}, Massimo and {Ota}, Kazuaki and {Peng}, Yingjie and {Pettini}, Max and {Swinbank}, Mark and {Troncoso}, Paulina},
        title = "{The KLEVER Survey: spatially resolved metallicity maps and gradients in a sample of 1.2 < z < 2.5 lensed galaxies}",
      journal = {\mnras},
         year = 2020,
        month = feb,
       volume = {492},
       number = {1},
        pages = {821-842},
          doi = {10.1093/mnras/stz3379},
archivePrefix = {arXiv},
       eprint = {1910.13451},
 primaryClass = {astro-ph.GA},
       adsurl = {https://ui.adsabs.harvard.edu/abs/2020MNRAS.492..821C}
}

@ARTICLE{Grossi:2020,
       author = {{Grossi}, M. and {Garc{\'\i}a-Benito}, R. and {Cortesi}, A. and {Gon{\c{c}}alves}, D.~R. and {Gon{\c{c}}alves}, T.~S. and {Lopes}, P.~A.~A. and {Men{\'e}ndez-Delmestre}, K. and {Telles}, E.},
        title = "{Inverted metallicity gradients in two Virgo cluster star-forming dwarf galaxies: evidence of recent merging?}",
      journal = {\mnras},
         year = 2020,
        month = oct,
       volume = {498},
       number = {2},
        pages = {1939-1950},
          doi = {10.1093/mnras/staa2382},
archivePrefix = {arXiv},
       eprint = {2008.02212},
 primaryClass = {astro-ph.GA},
       adsurl = {https://ui.adsabs.harvard.edu/abs/2020MNRAS.498.1939G}
}

@ARTICLE{Hani:2020,
       author = {{Hani}, Maan H. and {Gosain}, Hayman and {Ellison}, Sara L. and {Patton}, David R. and {Torrey}, Paul},
        title = "{Interacting galaxies in the IllustrisTNG simulations - II: star formation in the post-merger stage}",
      journal = {\mnras},
         year = 2020,
        month = apr,
       volume = {493},
       number = {3},
        pages = {3716-3731},
          doi = {10.1093/mnras/staa459},
archivePrefix = {arXiv},
       eprint = {2001.04472},
 primaryClass = {astro-ph.GA},
       adsurl = {https://ui.adsabs.harvard.edu/abs/2020MNRAS.493.3716H}
}

@ARTICLE{Cibinel:2019,
       author = {{Cibinel}, A. and {Daddi}, E. and {Sargent}, M.~T. and {Le Floc'h}, E. and {Liu}, D. and {Bournaud}, F. and {Oesch}, P.~A. and {Amram}, P. and {Calabr{\`o}}, A. and {Duc}, P.-A. and {Pannella}, M. and {Puglisi}, A. and {Perret}, V. and {Elbaz}, D. and {Kokorev}, V.},
        title = "{Early- and late-stage mergers among main sequence and starburst galaxies at 0.2 {\ensuremath{\leq}} z {\ensuremath{\leq}} 2}",
      journal = {\mnras},
         year = 2019,
        month = jun,
       volume = {485},
       number = {4},
        pages = {5631-5651},
          doi = {10.1093/mnras/stz690},
archivePrefix = {arXiv},
       eprint = {1809.00715},
 primaryClass = {astro-ph.GA},
       adsurl = {https://ui.adsabs.harvard.edu/abs/2019MNRAS.485.5631C}
}

@ARTICLE{Frankel:2019,
       author = {{Frankel}, Neige and {Sanders}, Jason and {Rix}, Hans-Walter and {Ting}, Yuan-Sen and {Ness}, Melissa},
        title = "{The Inside-out Growth of the Galactic Disk}",
      journal = {\apj},
         year = 2019,
        month = oct,
       volume = {884},
       number = {2},
          eid = {99},
        pages = {99},
          doi = {10.3847/1538-4357/ab4254},
archivePrefix = {arXiv},
       eprint = {1909.07118},
 primaryClass = {astro-ph.GA},
       adsurl = {https://ui.adsabs.harvard.edu/abs/2019ApJ...884...99F}
}

@ARTICLE{Iyer:2019,
       author = {{Iyer}, Kartheik G. and {Gawiser}, Eric and {Faber}, Sandra M. and {Ferguson}, Henry C. and {Kartaltepe}, Jeyhan and {Koekemoer}, Anton M. and {Pacifici}, Camilla and {Somerville}, Rachel S.},
        title = "{Nonparametric Star Formation History Reconstruction with Gaussian Processes. I. Counting Major Episodes of Star Formation}",
      journal = {\apj},
         year = 2019,
        month = jul,
       volume = {879},
       number = {2},
          eid = {116},
        pages = {116},
          doi = {10.3847/1538-4357/ab2052},
archivePrefix = {arXiv},
       eprint = {1901.02877},
 primaryClass = {astro-ph.GA},
       adsurl = {https://ui.adsabs.harvard.edu/abs/2019ApJ...879..116I}
}

@ARTICLE{Kriek:2019,
       author = {{Kriek}, Mariska and {Price}, Sedona H. and {Conroy}, Charlie and {Suess}, Katherine A. and {Mowla}, Lamiya and {Pasha}, Imad and {Bezanson}, Rachel and {van Dokkum}, Pieter and {Barro}, Guillermo},
        title = "{Stellar Metallicities and Elemental Abundance Ratios of z {\ensuremath{\sim}} 1.4 Massive Quiescent Galaxies}",
      journal = {\apjl},
         year = 2019,
        month = aug,
       volume = {880},
       number = {2},
          eid = {L31},
        pages = {L31},
          doi = {10.3847/2041-8213/ab2e75},
archivePrefix = {arXiv},
       eprint = {1907.04327},
 primaryClass = {astro-ph.GA},
       adsurl = {https://ui.adsabs.harvard.edu/abs/2019ApJ...880L..31K}
}

@ARTICLE{Pearson:2019,
       author = {{Pearson}, W.~J. and {Wang}, L. and {Alpaslan}, M. and {Baldry}, I. and {Bilicki}, M. and {Brown}, M.~J.~I. and {Grootes}, M.~W. and {Holwerda}, B.~W. and {Kitching}, T.~D. and {Kruk}, S. and {van der Tak}, F.~F.~S.},
        title = "{Effect of galaxy mergers on star-formation rates}",
      journal = {\aap},
         year = 2019,
        month = nov,
       volume = {631},
          eid = {A51},
        pages = {A51},
          doi = {10.1051/0004-6361/201936337},
archivePrefix = {arXiv},
       eprint = {1908.10115},
 primaryClass = {astro-ph.GA},
       adsurl = {https://ui.adsabs.harvard.edu/abs/2019A&A...631A..51P}
}

@ARTICLE{Pena:2019,
       author = {{Pe{\~n}a}, M. and {Flores-Dur{\'a}n}, S.~N.},
        title = "{Metallicity Gradients in M 31, M 33, NGC 300 and the Milky Way Using Abundances of Different Elements}",
      journal = {\rmxaa},
         year = 2019,
        month = oct,
       volume = {55},
        pages = {255-271},
          doi = {10.22201/ia.01851101p.2019.55.02.13},
archivePrefix = {arXiv},
       eprint = {1907.05797},
 primaryClass = {astro-ph.GA},
       adsurl = {https://ui.adsabs.harvard.edu/abs/2019RMxAA..55..255P}
}

@ARTICLE{RodriguezMontero:2019,
       author = {{Rodr{\'\i}guez Montero}, Francisco and {Dav{\'e}}, Romeel and {Wild}, Vivienne and {Angl{\'e}s-Alc{\'a}zar}, Daniel and {Narayanan}, Desika},
        title = "{Mergers, starbursts, and quenching in the SIMBA simulation}",
      journal = {\mnras},
         year = 2019,
        month = dec,
       volume = {490},
       number = {2},
        pages = {2139-2154},
          doi = {10.1093/mnras/stz2580},
archivePrefix = {arXiv},
       eprint = {1907.12680},
 primaryClass = {astro-ph.GA},
       adsurl = {https://ui.adsabs.harvard.edu/abs/2019MNRAS.490.2139R}
}

@ARTICLE{Hayes:2018,
       author = {{Hayes}, Christian R. and {Majewski}, Steven R. and {Shetrone}, Matthew and {Fern{\'a}ndez-Alvar}, Emma and {Allende Prieto}, Carlos and {Schuster}, William J. and {Carigi}, Leticia and {Cunha}, Katia and {Smith}, Verne V. and {Sobeck}, Jennifer and {Almeida}, Andres and {Beers}, Timothy C. and {Carrera}, Ricardo and {Fern{\'a}ndez-Trincado}, J.~G. and {Garc{\'\i}a-Hern{\'a}ndez}, D.~A. and {Geisler}, Doug and {Lane}, Richard R. and {Lucatello}, Sara and {Matthews}, Allison M. and {Minniti}, Dante and {Nitschelm}, Christian and {Tang}, Baitian and {Tissera}, Patricia B. and {Zamora}, Olga},
        title = "{Disentangling the Galactic Halo with APOGEE. I. Chemical and Kinematical Investigation of Distinct Metal-poor Populations}",
      journal = {\apj},
         year = 2018,
        month = jan,
       volume = {852},
       number = {1},
          eid = {49},
        pages = {49},
          doi = {10.3847/1538-4357/aa9cec},
archivePrefix = {arXiv},
       eprint = {1711.05781},
 primaryClass = {astro-ph.GA},
       adsurl = {https://ui.adsabs.harvard.edu/abs/2018ApJ...852...49H}
}

@ARTICLE{Helmi:2018,
       author = {{Helmi}, Amina and {Babusiaux}, Carine and {Koppelman}, Helmer H. and {Massari}, Davide and {Veljanoski}, Jovan and {Brown}, Anthony G.~A.},
        title = "{The merger that led to the formation of the Milky Way's inner stellar halo and thick disk}",
      journal = {\nat},
         year = 2018,
        month = oct,
       volume = {563},
       number = {7729},
        pages = {85-88},
          doi = {10.1038/s41586-018-0625-x},
archivePrefix = {arXiv},
       eprint = {1806.06038},
 primaryClass = {astro-ph.GA},
       adsurl = {https://ui.adsabs.harvard.edu/abs/2018Natur.563...85H}
}

@ARTICLE{Kawata:2018,
       author = {{Kawata}, Daisuke and {Allende Prieto}, Carlos and {Brook}, Chris B. and {Casagrande}, Luca and {Ciuc{\u{a}}}, Ioana and {Gibson}, Brad K. and {Grand}, Robert J.~J. and {Hayden}, Michael R. and {Hunt}, Jason A.~S.},
        title = "{Metallicity gradient of the thick disc progenitor at high redshift}",
      journal = {\mnras},
         year = 2018,
        month = jan,
       volume = {473},
       number = {1},
        pages = {867-878},
          doi = {10.1093/mnras/stx2464},
archivePrefix = {arXiv},
       eprint = {1706.01474},
 primaryClass = {astro-ph.GA},
       adsurl = {https://ui.adsabs.harvard.edu/abs/2018MNRAS.473..867K}
}

@ARTICLE{Lemasle:2018,
       author = {{Lemasle}, B. and {Hajdu}, G. and {Kovtyukh}, V. and {Inno}, L. and {Grebel}, E.~K. and {Catelan}, M. and {Bono}, G. and {Fran{\c{c}}ois}, P. and {Kniazev}, A. and {da Silva}, R. and {Storm}, J.},
        title = "{Milky Way metallicity gradient from Gaia DR2 F/1O double-mode Cepheids}",
      journal = {\aap},
         year = 2018,
        month = oct,
       volume = {618},
          eid = {A160},
        pages = {A160},
          doi = {10.1051/0004-6361/201834050},
archivePrefix = {arXiv},
       eprint = {1809.07352},
 primaryClass = {astro-ph.SR},
       adsurl = {https://ui.adsabs.harvard.edu/abs/2018A&A...618A.160L}
}

@ARTICLE{Sanchez-Menguiano:2018,
       author = {{S{\'a}nchez-Menguiano}, L. and {S{\'a}nchez}, S.~F. and {P{\'e}rez}, I. and {Ruiz-Lara}, T. and {Galbany}, L. and {Anderson}, J.~P. and {Kr{\"u}hler}, T. and {Kuncarayakti}, H. and {Lyman}, J.~D.},
        title = "{The shape of oxygen abundance profiles explored with MUSE: evidence for widespread deviations from single gradients}",
      journal = {\aap},
         year = 2018,
        month = feb,
       volume = {609},
          eid = {A119},
        pages = {A119},
          doi = {10.1051/0004-6361/201731486},
archivePrefix = {arXiv},
       eprint = {1710.01188},
 primaryClass = {astro-ph.GA},
       adsurl = {https://ui.adsabs.harvard.edu/abs/2018A&A...609A.119S}
}

@ARTICLE{Silva:2018,
       author = {{Silva}, Andrea and {Marchesini}, Danilo and {Silverman}, John D. and {Skelton}, Rosalind and {Iono}, Daisuke and {Martis}, Nicholas and {Marsan}, Z. Cemile and {Tadaki}, Ken-ichi and {Brammer}, Gabriel and {kartaltepe}, Jeyhan},
        title = "{Galaxy Mergers up to Z < 2.5. I. The Star Formation Properties of Merging Galaxies at Separations of 3-15 kpc}",
      journal = {\apj},
         year = 2018,
        month = nov,
       volume = {868},
       number = {1},
          eid = {46},
        pages = {46},
          doi = {10.3847/1538-4357/aae847},
archivePrefix = {arXiv},
       eprint = {1809.09796},
 primaryClass = {astro-ph.GA},
       adsurl = {https://ui.adsabs.harvard.edu/abs/2018ApJ...868...46S}
}

@ARTICLE{Stanghellini:2018,
       author = {{Stanghellini}, Letizia and {Haywood}, Misha},
        title = "{Galactic Planetary Nebulae as Probes of Radial Metallicity Gradients and Other Abundance Patterns}",
      journal = {\apj},
         year = 2018,
        month = jul,
       volume = {862},
       number = {1},
          eid = {45},
        pages = {45},
          doi = {10.3847/1538-4357/aacaf8},
archivePrefix = {arXiv},
       eprint = {1806.02276},
 primaryClass = {astro-ph.GA},
       adsurl = {https://ui.adsabs.harvard.edu/abs/2018ApJ...862...45S}
}

@ARTICLE{Belfiore:2017,
       author = {{Belfiore}, Francesco and {Maiolino}, Roberto and {Tremonti}, Christy and {S{\'a}nchez}, Sebastian F. and {Bundy}, Kevin and {Bershady}, Matthew and {Westfall}, Kyle and {Lin}, Lihwai and {Drory}, Niv and {Boquien}, M{\'e}d{\'e}ric and {Thomas}, Daniel and {Brinkmann}, Jonathan},
        title = "{SDSS IV MaNGA - metallicity and nitrogen abundance gradients in local galaxies}",
      journal = {\mnras},
         year = 2017,
        month = jul,
       volume = {469},
       number = {1},
        pages = {151-170},
          doi = {10.1093/mnras/stx789},
archivePrefix = {arXiv},
       eprint = {1703.03813},
 primaryClass = {astro-ph.GA},
       adsurl = {https://ui.adsabs.harvard.edu/abs/2017MNRAS.469..151B}
}

@ARTICLE{Goddard:2017,
       author = {{Goddard}, D. and {Thomas}, D. and {Maraston}, C. and {Westfall}, K. and {Etherington}, J. and {Riffel}, R. and {Mallmann}, N.~D. and {Zheng}, Z. and {Argudo-Fern{\'a}ndez}, M. and {Lian}, J. and {Bershady}, M. and {Bundy}, K. and {Drory}, N. and {Law}, D. and {Yan}, R. and {Wake}, D. and {Weijmans}, A. and {Bizyaev}, D. and {Brownstein}, J. and {Lane}, R.~R. and {Maiolino}, R. and {Masters}, K. and {Merrifield}, M. and {Nitschelm}, C. and {Pan}, K. and {Roman-Lopes}, A. and {Storchi-Bergmann}, T. and {Schneider}, D.~P.},
        title = "{SDSS-IV MaNGA: Spatially resolved star formation histories in galaxies as a function of galaxy mass and type}",
      journal = {\mnras},
         year = 2017,
        month = apr,
       volume = {466},
       number = {4},
        pages = {4731-4758},
          doi = {10.1093/mnras/stw3371},
archivePrefix = {arXiv},
       eprint = {1612.01546},
 primaryClass = {astro-ph.GA},
       adsurl = {https://ui.adsabs.harvard.edu/abs/2017MNRAS.466.4731G}
}

@ARTICLE{Iyer:2017,
       author = {{Iyer}, Kartheik and {Gawiser}, Eric},
        title = "{Reconstruction of Galaxy Star Formation Histories through SED Fitting:The Dense Basis Approach}",
      journal = {\apj},
         year = 2017,
        month = apr,
       volume = {838},
       number = {2},
          eid = {127},
        pages = {127},
          doi = {10.3847/1538-4357/aa63f0},
archivePrefix = {arXiv},
       eprint = {1702.04371},
 primaryClass = {astro-ph.GA},
       adsurl = {https://ui.adsabs.harvard.edu/abs/2017ApJ...838..127I}
}

@ARTICLE{Kilic:2017,
       author = {{Kilic}, Mukremin and {Munn}, Jeffrey A. and {Harris}, Hugh C. and {von Hippel}, Ted and {Liebert}, James W. and {Williams}, Kurtis A. and {Jeffery}, Elizabeth and {DeGennaro}, Steven},
        title = "{The Ages of the Thin Disk, Thick Disk, and the Halo from Nearby White Dwarfs}",
      journal = {\apj},
         year = 2017,
        month = mar,
       volume = {837},
       number = {2},
          eid = {162},
        pages = {162},
          doi = {10.3847/1538-4357/aa62a5},
archivePrefix = {arXiv},
       eprint = {1702.06984},
 primaryClass = {astro-ph.SR},
       adsurl = {https://ui.adsabs.harvard.edu/abs/2017ApJ...837..162K}
}

@ARTICLE{Ma:2017,
       author = {{Ma}, Xiangcheng and {Hopkins}, Philip F. and {Feldmann}, Robert and {Torrey}, Paul and {Faucher-Gigu{\`e}re}, Claude-Andr{\'e} and {Kere{\v{s}}}, Du{\v{s}}an},
        title = "{Why do high-redshift galaxies show diverse gas-phase metallicity gradients?}",
      journal = {\mnras},
         year = 2017,
        month = apr,
       volume = {466},
       number = {4},
        pages = {4780-4794},
          doi = {10.1093/mnras/stx034},
archivePrefix = {arXiv},
       eprint = {1610.03498},
 primaryClass = {astro-ph.GA},
       adsurl = {https://ui.adsabs.harvard.edu/abs/2017MNRAS.466.4780M}
}

@ARTICLE{Schonrich:2017,
       author = {{Sch{\"o}nrich}, Ralph and {McMillan}, Paul J.},
        title = "{Understanding inverse metallicity gradients in galactic discs as a consequence of inside-out formation}",
      journal = {\mnras},
         year = 2017,
        month = may,
       volume = {467},
       number = {1},
        pages = {1154-1174},
          doi = {10.1093/mnras/stx093},
archivePrefix = {arXiv},
       eprint = {1605.02338},
 primaryClass = {astro-ph.GA},
       adsurl = {https://ui.adsabs.harvard.edu/abs/2017MNRAS.467.1154S}
}

@ARTICLE{Zheng:2017,
       author = {{Zheng}, Zheng and {Wang}, Huiyuan and {Ge}, Junqiang and {Mao}, Shude and {Li}, Cheng and {Li}, Ran and {Mo}, Houjun and {Goddard}, Daniel and {Bundy}, Kevin and {Li}, Hongyu and {Nair}, Preethi and {Lin}, Lihwai and {Long}, R.~J. and {Riffel}, Rog{\'e}rio and {Thomas}, Daniel and {Masters}, Karen and {Bizyaev}, Dmitry and {Brownstein}, Joel R. and {Zhang}, Kai and {Law}, David R. and {Drory}, Niv and {Roman Lopes}, Alexandre and {Malanushenko}, Olena},
        title = "{SDSS-IV MaNGA: environmental dependence of stellar age and metallicity gradients in nearby galaxies}",
      journal = {\mnras},
         year = 2017,
        month = mar,
       volume = {465},
       number = {4},
        pages = {4572-4588},
          doi = {10.1093/mnras/stw3030},
archivePrefix = {arXiv},
       eprint = {1612.01523},
 primaryClass = {astro-ph.GA},
       adsurl = {https://ui.adsabs.harvard.edu/abs/2017MNRAS.465.4572Z}
}

@ARTICLE{Cunha:2016,
       author = {{Cunha}, K. and {Frinchaboy}, P.~M. and {Souto}, D. and {Thompson}, B. and {Zasowski}, G. and {Allende Prieto}, C. and {Carrera}, R. and {Chiappini}, C. and {Donor}, J. and {Garc{\'\i}a-Hern{\'a}ndez}, D.~A. and {Garc{\'\i}a P{\'e}rez}, A.~E. and {Hayden}, M.~R. and {Holtzman}, J. and {Jackson}, K.~M. and {Johnson}, J.~A. and {Majewski}, S.~R. and {M{\'e}sz{\'a}ros}, S. and {Meyer}, B. and {Nidever}, D.~L. and {O'Connell}, J. and {Schiavon}, R.~P. and {Schultheis}, M. and {Shetrone}, M. and {Simmons}, A. and {Smith}, V.~V. and {et al.}},
        title = "{Chemical abundance gradients from open clusters in the Milky Way disk: Results from the APOGEE survey}",
      journal = {Astronomische Nachrichten},
         year = 2016,
        month = sep,
       volume = {337},
       number = {8-9},
        pages = {922},
          doi = {10.1002/asna.201612398},
archivePrefix = {arXiv},
       eprint = {1601.03099},
 primaryClass = {astro-ph.GA},
       adsurl = {https://ui.adsabs.harvard.edu/abs/2016AN....337..922C}
}

@ARTICLE{Nelson:2016,
       author = {{Nelson}, Erica June and {van Dokkum}, Pieter G. and {F{\"o}rster Schreiber}, Natascha M. and {Franx}, Marijn and {Brammer}, Gabriel B. and {Momcheva}, Ivelina G. and {Wuyts}, Stijn and {Whitaker}, Katherine E. and {Skelton}, Rosalind E. and {Fumagalli}, Mattia and {Hayward}, Christopher C. and {Kriek}, Mariska and {Labb{\'e}}, Ivo and {Leja}, Joel and {Rix}, Hans-Walter and {Tacconi}, Linda J. and {van der Wel}, Arjen and {van den Bosch}, Frank C. and {Oesch}, Pascal A. and {Dickey}, Claire and {Ulf Lange}, Johannes},
        title = "{Where Stars Form: Inside-out Growth and Coherent Star Formation from HST H{\ensuremath{\alpha}} Maps of 3200 Galaxies across the Main Sequence at 0.7 < z < 1.5}",
      journal = {\apj},
         year = 2016,
        month = sep,
       volume = {828},
       number = {1},
          eid = {27},
        pages = {27},
          doi = {10.3847/0004-637X/828/1/27},
archivePrefix = {arXiv},
       eprint = {1507.03999},
 primaryClass = {astro-ph.GA},
       adsurl = {https://ui.adsabs.harvard.edu/abs/2016ApJ...828...27N}
}

@ARTICLE{Wuyts:2016,
       author = {{Wuyts}, Eva and {Wisnioski}, Emily and {Fossati}, Matteo and {F{\"o}rster Schreiber}, Natascha M. and {Genzel}, Reinhard and {Davies}, Ric and {Mendel}, J. Trevor and {Naab}, Thorsten and {R{\"o}ttgers}, Bernhard and {Wilman}, David J. and {Wuyts}, Stijn and {Bandara}, Kaushala and {Beifiori}, Alessandra and {Belli}, Sirio and {Bender}, Ralf and {Brammer}, Gabriel B. and {Burkert}, Andreas and {Chan}, Jeffrey and {Galametz}, Audrey and {Kulkarni}, Sandesh K. and {Lang}, Philipp and {Lutz}, Dieter and {Momcheva}, Ivelina G. and {Nelson}, Erica J. and {Rosario}, David and {Saglia}, Roberto P. and {Seitz}, Stella and {Tacconi}, Linda J. and {Tadaki}, Ken-ichi and {{\"U}bler}, Hannah and {van Dokkum}, Pieter},
        title = "{The Evolution of Metallicity and Metallicity Gradients from z = 2.7 to 0.6 with KMOS$^{3D}$}",
      journal = {\apj},
         year = 2016,
        month = aug,
       volume = {827},
       number = {1},
          eid = {74},
        pages = {74},
          doi = {10.3847/0004-637X/827/1/74},
archivePrefix = {arXiv},
       eprint = {1603.01139},
 primaryClass = {astro-ph.GA},
       adsurl = {https://ui.adsabs.harvard.edu/abs/2016ApJ...827...74W}
}

@ARTICLE{GonzalezDelgado:2015,
       author = {{Gonz{\'a}lez Delgado}, R.~M. and {Garc{\'\i}a-Benito}, R. and {P{\'e}rez}, E. and {Cid Fernandes}, R. and {de Amorim}, A.~L. and {Cortijo-Ferrero}, C. and {Lacerda}, E.~A.~D. and {L{\'o}pez Fern{\'a}ndez}, R. and {Vale-Asari}, N. and {S{\'a}nchez}, S.~F. and {Moll{\'a}}, M. and {Ruiz-Lara}, T. and {S{\'a}nchez-Bl{\'a}zquez}, P. and {Walcher}, C.~J. and {Alves}, J. and {Aguerri}, J.~A.~L. and {Bekerait{\'e}}, S. and {Bland-Hawthorn}, J. and {Galbany}, L. and {Gallazzi}, A. and {Husemann}, B. and {Iglesias-P{\'a}ramo}, J. and {Kalinova}, V. and {L{\'o}pez-S{\'a}nchez}, A.~R. and {Marino}, R.~A. and {M{\'a}rquez}, I. and {Masegosa}, J. and {Mast}, D. and {M{\'e}ndez-Abreu}, J. and {Mendoza}, A. and {del Olmo}, A. and {P{\'e}rez}, I. and {Quirrenbach}, A. and {Zibetti}, S.},
        title = "{The CALIFA survey across the Hubble sequence. Spatially resolved stellar population properties in galaxies}",
      journal = {\aap},
         year = 2015,
        month = sep,
       volume = {581},
          eid = {A103},
        pages = {A103},
          doi = {10.1051/0004-6361/201525938},
archivePrefix = {arXiv},
       eprint = {1506.04157},
 primaryClass = {astro-ph.GA},
       adsurl = {https://ui.adsabs.harvard.edu/abs/2015A&A...581A.103G}
}

@ARTICLE{Grazian:2015,
       author = {{Grazian}, A. and {Fontana}, A. and {Santini}, P. and {Dunlop}, J.~S. and {Ferguson}, H.~C. and {Castellano}, M. and {Amorin}, R. and {Ashby}, M.~L.~N. and {Barro}, G. and {Behroozi}, P. and {Boutsia}, K. and {Caputi}, K.~I. and {Chary}, R.~R. and {Dekel}, A. and {Dickinson}, M.~E. and {Faber}, S.~M. and {Fazio}, G.~G. and {Finkelstein}, S.~L. and {Galametz}, A. and {Giallongo}, E. and {Giavalisco}, M. and {Grogin}, N.~A. and {Guo}, Y. and {Kocevski}, D. and {Koekemoer}, A.~M. and {Koo}, D.~C. and {Lee}, K. -S. and {Lu}, Y. and {Merlin}, E. and {Mobasher}, B. and {Nonino}, M. and {Papovich}, C. and {Paris}, D. and {Pentericci}, L. and {Reddy}, N. and {Renzini}, A. and {Salmon}, B. and {Salvato}, M. and {Sommariva}, V. and {Song}, M. and {Vanzella}, E.},
        title = "{The galaxy stellar mass function at 3.5 {\ensuremath{\leq}}z {\ensuremath{\leq}} 7.5 in the CANDELS/UDS, GOODS-South, and HUDF fields}",
      journal = {\aap},
         year = 2015,
        month = mar,
       volume = {575},
          eid = {A96},
        pages = {A96},
          doi = {10.1051/0004-6361/201424750},
archivePrefix = {arXiv},
       eprint = {1412.0532},
 primaryClass = {astro-ph.GA},
       adsurl = {https://ui.adsabs.harvard.edu/abs/2015A&A...575A..96G}
}

@ARTICLE{Licquia:2015b,
       author = {{Licquia}, Timothy C. and {Newman}, Jeffrey A. and {Brinchmann}, Jarle},
        title = "{Unveiling the Milky Way: A New Technique for Determining the Optical Color and Luminosity of Our Galaxy}",
      journal = {\apj},
         year = 2015,
        month = aug,
       volume = {809},
       number = {1},
          eid = {96},
        pages = {96},
          doi = {10.1088/0004-637X/809/1/96},
archivePrefix = {arXiv},
       eprint = {1508.04446},
 primaryClass = {astro-ph.GA},
       adsurl = {https://ui.adsabs.harvard.edu/abs/2015ApJ...809...96L}
}

@ARTICLE{Maciel:2015,
       author = {{Maciel}, W.~J. and {Costa}, R.~D.~D. and {Cavichia}, O.},
        title = "{Radial abundance gradients from planetary nebulae at different distances from the galactic plane}",
      journal = {\rmxaa},
         year = 2015,
        month = oct,
       volume = {51},
        pages = {165},
          doi = {10.48550/arXiv.1505.07640},
archivePrefix = {arXiv},
       eprint = {1505.07640},
 primaryClass = {astro-ph.GA},
       adsurl = {https://ui.adsabs.harvard.edu/abs/2015RMxAA..51..165M}
}

@ARTICLE{Santini:2015,
       author = {{Santini}, P. and {Ferguson}, H.~C. and {Fontana}, A. and {Mobasher}, B. and {Barro}, G. and {Castellano}, M. and {Finkelstein}, S.~L. and {Grazian}, A. and {Hsu}, L.~T. and {Lee}, B. and {Lee}, S. -K. and {Pforr}, J. and {Salvato}, M. and {Wiklind}, T. and {Wuyts}, S. and {Almaini}, O. and {Cooper}, M.~C. and {Galametz}, A. and {Weiner}, B. and {Amorin}, R. and {Boutsia}, K. and {Conselice}, C.~J. and {Dahlen}, T. and {Dickinson}, M.~E. and {Giavalisco}, M. and {Grogin}, N.~A. and {Guo}, Y. and {Hathi}, N.~P. and {Kocevski}, D. and {Koekemoer}, A.~M. and {Kurczynski}, P. and {Merlin}, E. and {Mortlock}, A. and {Newman}, J.~A. and {Paris}, D. and {Pentericci}, L. and {Simons}, R. and {Willner}, S.~P.},
        title = "{Stellar Masses from the CANDELS Survey: The GOODS-South and UDS Fields}",
      journal = {\apj},
         year = 2015,
        month = mar,
       volume = {801},
       number = {2},
          eid = {97},
        pages = {97},
          doi = {10.1088/0004-637X/801/2/97},
archivePrefix = {arXiv},
       eprint = {1412.5180},
 primaryClass = {astro-ph.GA},
       adsurl = {https://ui.adsabs.harvard.edu/abs/2015ApJ...801...97S}
}

@ARTICLE{Genovali:2014,
       author = {{Genovali}, K. and {Lemasle}, B. and {Bono}, G. and {Romaniello}, M. and {Fabrizio}, M. and {Ferraro}, I. and {Iannicola}, G. and {Laney}, C.~D. and {Nonino}, M. and {Bergemann}, M. and {Buonanno}, R. and {Fran{\c{c}}ois}, P. and {Inno}, L. and {Kudritzki}, R. -P. and {Matsunaga}, N. and {Pedicelli}, S. and {Primas}, F. and {Th{\'e}venin}, F.},
        title = "{On the fine structure of the Cepheid metallicity gradient in the Galactic thin disk}",
      journal = {\aap},
         year = 2014,
        month = jun,
       volume = {566},
          eid = {A37},
        pages = {A37},
          doi = {10.1051/0004-6361/201323198},
archivePrefix = {arXiv},
       eprint = {1403.6128},
 primaryClass = {astro-ph.GA},
       adsurl = {https://ui.adsabs.harvard.edu/abs/2014A&A...566A..37G}
}

@ARTICLE{Stott:2014,
       author = {{Stott}, John P. and {Sobral}, David and {Swinbank}, A.~M. and {Smail}, Ian and {Bower}, Richard and {Best}, Philip N. and {Sharples}, Ray M. and {Geach}, James E. and {Matthee}, Jorryt},
        title = "{A relationship between specific star formation rate and metallicity gradient within z {\ensuremath{\sim}} 1 galaxies from KMOS-HiZELS}",
      journal = {\mnras},
         year = 2014,
        month = sep,
       volume = {443},
       number = {3},
        pages = {2695-2704},
          doi = {10.1093/mnras/stu1343},
archivePrefix = {arXiv},
       eprint = {1407.1047},
 primaryClass = {astro-ph.GA},
       adsurl = {https://ui.adsabs.harvard.edu/abs/2014MNRAS.443.2695S}
}
\bibliographystyle{aasjournalv7}

\appendix
\section{Radial profiles as a function of effective radius}\label{app:reff}

For this study, we have done the bulk of the analysis with gradients scaled to dex/kpc. This is because our aim is to compare progenitor samples at different redshifts to show growth and evolution in the resolved properties, which is better served with physical units. However, many studies of resolved properties such as age and metallicity gradients present their results scaled to dex/$R_{eff}$, and the effective radius varies for each galaxy. In order to also draw proper comparisons to the literature, we plot the average (stacked and normalized) property gradients at each redshift bin as a function of $R_{eff}$, shown in Figure \ref{fig:all-grads-reff}. Each gradient is divided into elliptical annuli of width $0.1 R_{eff}$ and the gradients are plotted out to 3 $R_{eff}$. The effective radius for each galaxy is the circularized half-light radius from the CANUCS DR1 catalogs measured in Photutils on the F150W 20mas resolution image.

In Figure \ref{fig:all-slopes-reff}, we plot the slope evolution over redshift with the slopes in dex/$R_{eff}$, in a similar fashion to Figure \ref{fig:all-slopes}. Overall, the average slopes for each redshift bin follow similar trends whether they were measured in kpc or in $R_{eff}$. However, the amplitude of the slope is larger in $R_{eff}$, and increases with decreasing redshift, which is the result of galaxies physically increasing in size. Interestingly, the property whose slopes exhibited the most change from scaling in dex/kpc to dex/$R_{eff}$ are the sSFR slopes. The non-merger slopes for sSFR and age as a function of redshift seem to be more inversely related when scaled to $R_{eff}$ than scaled to kpc. This potentially strengthens the interpretation that MWAs switch to lockstep growth at $z\sim 2$ rather than continue growing inside-out.

\begin{figure*}
\centering
\gridline{\fig{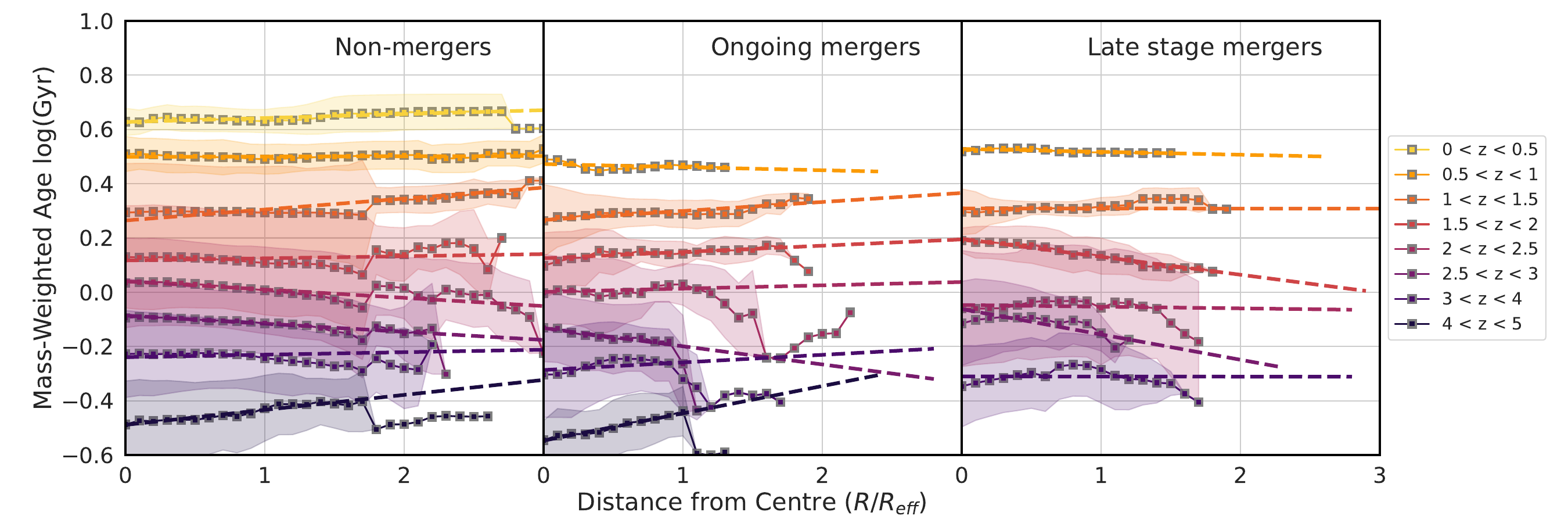}{\textwidth}{}}
\gridline{\fig{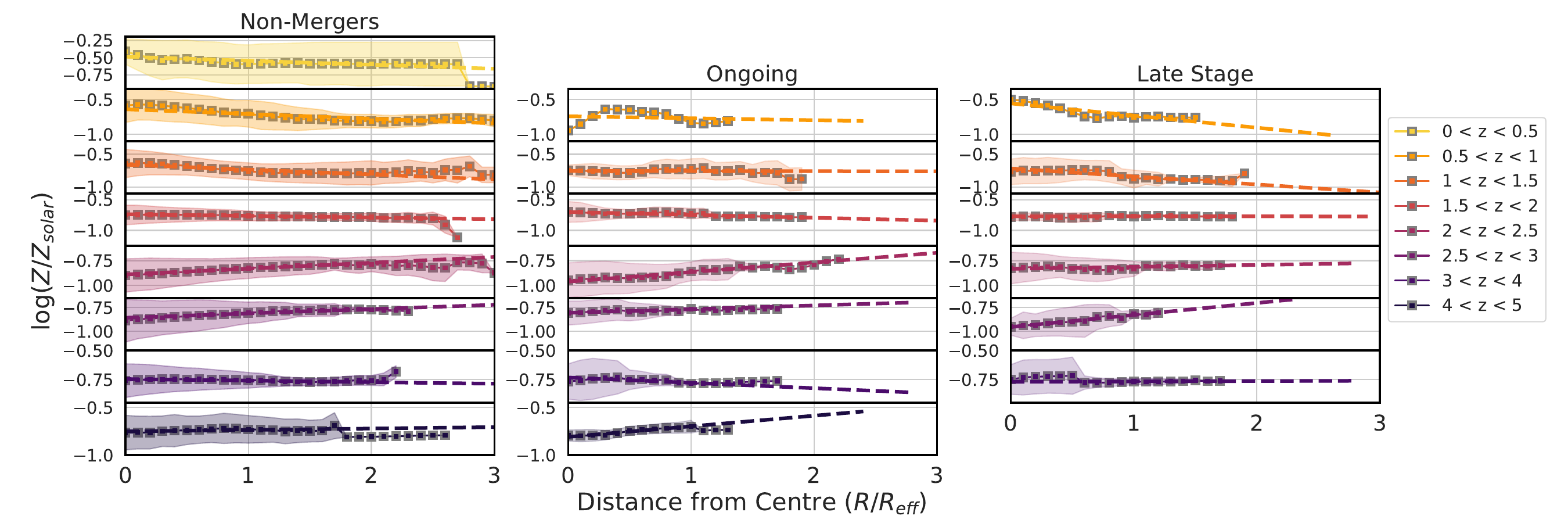}{\textwidth}{}}
\gridline{\fig{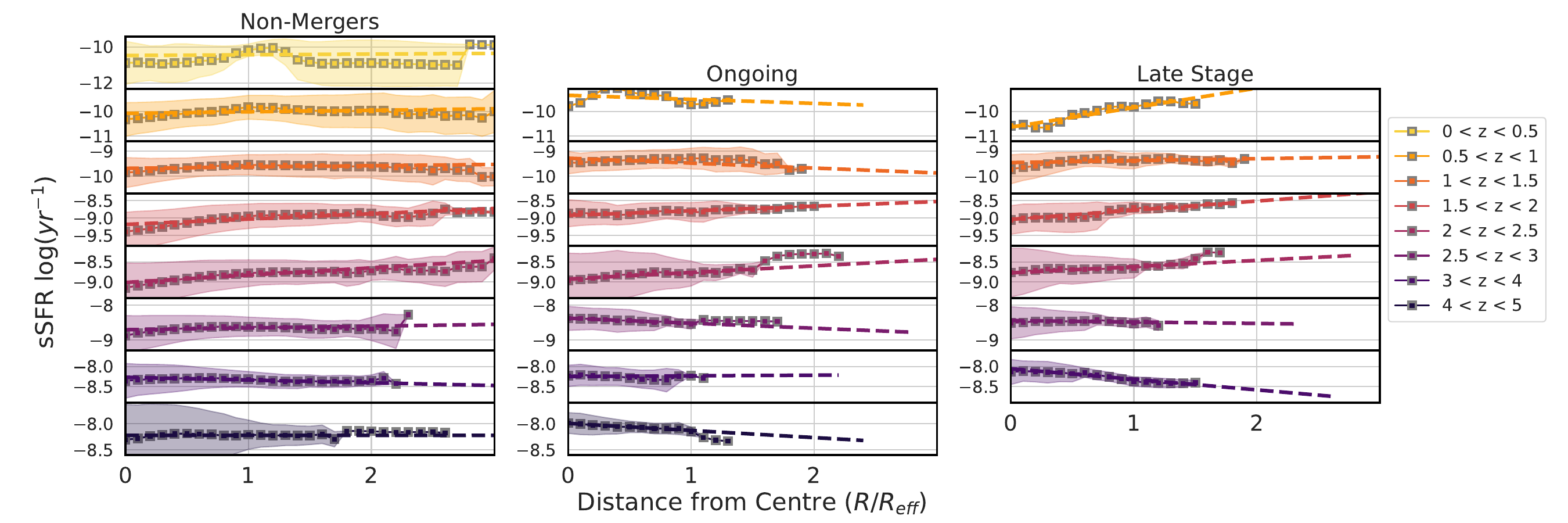}{\textwidth}{}}
\caption{Stacked and normalized property gradients at each redshift bin, similar to Figures \ref{fig:mw_ages_stack}, \ref{fig:met_stack} and \ref{fig:ssfr_stack}, but as a function of the galaxy's effective radius ($R_{eff}$) instead of kpc.}\label{fig:all-grads-reff}
\end{figure*}

\begin{figure}
    \centering
    \includegraphics[width=\columnwidth]{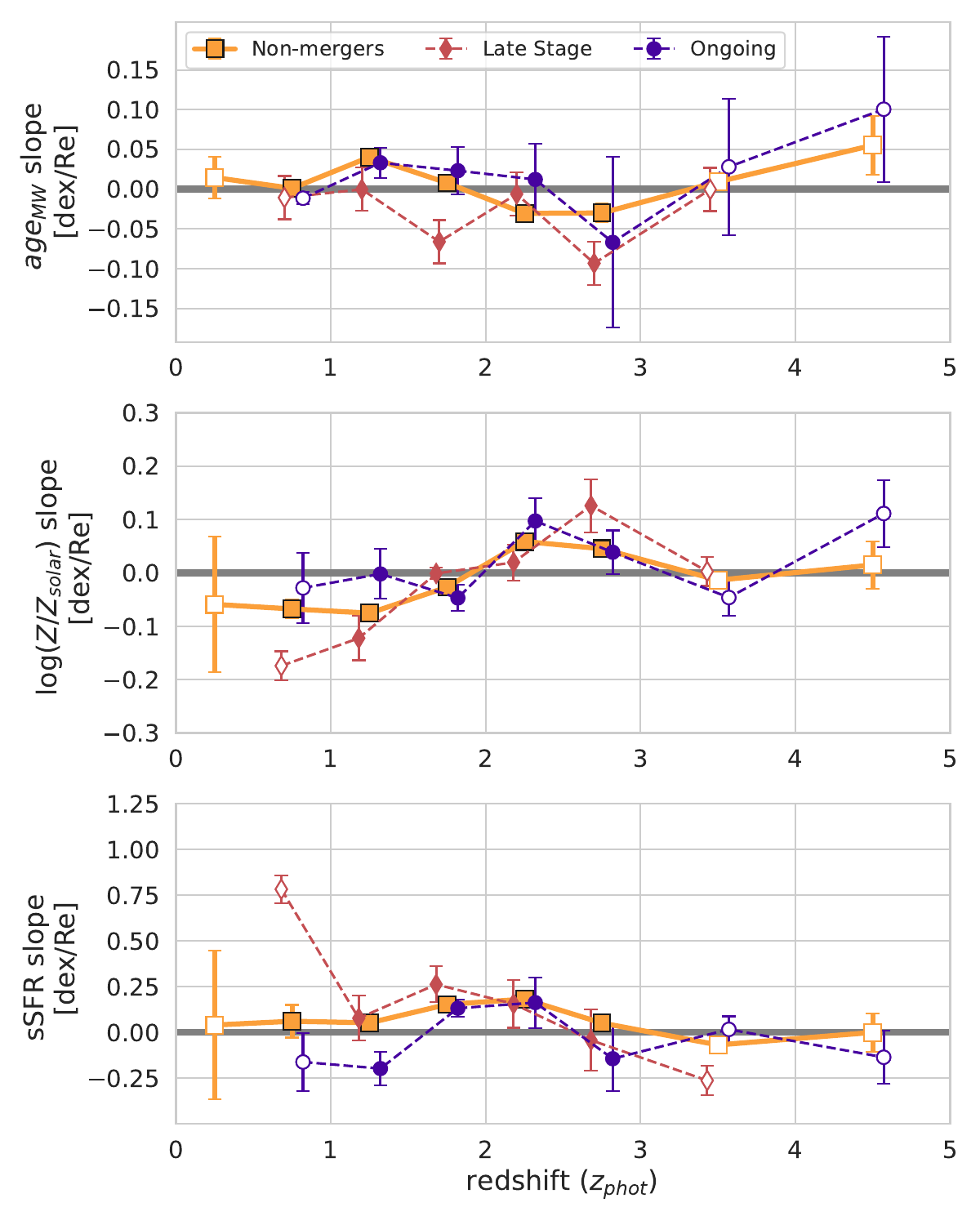}
    \caption{Slopes of property gradients as a function of redshift in units of dex/$R_{eff}$. Overall, the slopes have similar evolution with redshift, but the magnitude of the slope is larger when measured in $R_{eff}$ than in kpc.}
    \label{fig:all-slopes-reff}
\end{figure}


\section{Dust Gradients and Age-Dust-Metallicity Degeneracy}\label{appendix-b}

Age, metallicity, and dust are three factors which can redden the spectrum of a galaxy.  This is known as the age-dust-metallicity degeneracy. For degeneracy between age and metallicity, with modern multiband SED-fitting, it is possible for separation of age and metallicity without spectroscopic data (e.g. \citealt{Lee:2007, Eminian:2008}). This is because the bluer filters can probe the main-sequence turn off point to obtain age, while red and NIR filters are sensitive to the giant branch stars, which probe stellar metallicity \citep{Bell:2000, MacArthur:2004}. However, the degeneracy between dust and metallicity remains, as well as age and dust for star-forming galaxies (The UVJ diagram can effectively separate quiescent and highly dusty star-forming galaxies, see \citealt{Wuyts:2007} and \citealt{Williams:2009}, and \citealt{Antwi-Danso:2023} for an implementation similar to the method used in this paper). Modern SED-fitting codes use an energy balance model between UV and IR emission. The portion of IR emission coming from dust is balanced against the energy lost to attenuation from the UV (e.g. \citealt{Burgarella:2005, daCunha:2015, Salim:2018, Jones:2023}). Although this energy balance is better constrained with far-IR data that is most sensitive to dust temperature and mass \citep{Iyer:2025}, and we are limited to NIRCam and NIRISS data, it is still possible to show that the age-dust-metallicity degeneracy does not strongly affect our results. 

In Figure \ref{fig:corner-plots} we show four corner plots that present the distributions of dust-age, dust-metallicity, and age-metallicity distributions for every spatial bin of every galaxy in our sample. The dust, age, and metallicity for each bin are the best fit results from Dense Basis SED-fitting as described in \S\ref{sec:sed-fitting}. The top row has the distributions for spatial bins in the CLU fields on the left, versus the NCF fields on the right. This division is because the NCF fields include more medium band filters than the CLU fields. Medium bands are better at limiting the range certain emission lines appear (e.g. \citealt{Roberts-Borsani:2021,Withers:2023, Sarrouh:2024}), and thus can more precisely determine galaxy properties, such as photometric redshift. Based on the two CLU and NCF corner plots however, there does not appear to be a noticeable difference in the distributions of the spatial bins and in their 2D contours for dust, age, and metallicity. Another factor that can affect galaxy properties from SED-fitting is signal to noise, so the bottom row are two corner plots, one of spatial bins with size $\leq5$ pixels on the left, and the other of spatial bins with size $> 5$ on the right. These also correspond to bins which are of central regions of galaxies, or bins on the outskirts of galaxies. There is also no noticeable difference in the distributions or the 2D contour plots for smaller/central bins versus larger/outskirts bins. All four corner plots show that there are no major degenerate features in dust-age or metallicity-age distributions. There are some bins in the dust-metallicity distribution which show signs of degeneracy between the two properties, however they typically lie outside of the $2\sigma$ confidence contour.

Additionally, if this degeneracy has a noticeable effect on the properties of galaxies, we would expect the radial profile of dust to be the \textit{inverse} of either the age or the metallicity gradient. We test this by comparing the dust gradients against the gradients of the other degenerate properties. In the top panel of Figure \ref{fig:dust-grad}, we plot each individual radial dust gradient scaled by kpc for our sample of MWA progenitors, separated into eight subplots by their redshift bin. The thin grey lines are non-mergers while the thin blue lines are mergers. The stacked and normalized average dust gradient plotted as the thicker colored line. As a comparison, the middle and bottom panels of Figure \ref{fig:dust-grad} is a similar plot for each individual \textit{metallicity}  and \textit{mass-weighted age} gradient respectively, as well as their stacked and normalized average gradient at each redshift bin.  Note that these average gradients is slightly different from the ones presented in Figures \ref{fig:mw_ages_stack} and \ref{fig:met_stack} because it is an average of both merger and non-merger samples, but it is very similar to the non-merger profiles because 730 out of the 872 objects in our sample are non-mergers).  A typical $L_\star$ galaxy, which is a star-forming spiral at $z=0$ and what the Milky Way is classified as, would have $A_V \sim 1$. The evolution of the $A_V$ gradient with redshift for our sample is in line with what is expected for $L_\star$ galaxies. Outliers with $A_V \geq 1.25$, which is roughly $1\sigma$ above the average $A_V$ consists of only 33 out of the total 872 galaxies in our MWA progenitor sample.

The average dust gradient at each redshift shows that as a whole, dust gradients have slopes with amplitudes that are larger than the age and metallicity slopes. The gradients in each redshift bin also have slopes which \textit{follow} the trend of the non-merger sample's metallicity slopes, instead of an inverse relationship, which one would expect if the dust and metallicity were degenerate. The dust slopes also do not have an inverse relationship with the age gradients. Merging galaxies tend to have lower $A_V$ at all radii compared to non-merging galaxies (Only 2 of the 33 ``dusty" outliers are mergers). Notably, this is not seen in the metallicity gradients, as mergers do not have particularly high or low metallicity compared to the non-merger sample (which was also shown in \S\ref{sec:met-grads} and Figure \ref{fig:merger-nonmerger-compare}  in \S\ref{sec:merger-diff}. Mergers also tend to have younger ages than non-mergers, which would also not make sense if dust and age were degenerate, since if they were degenerate, we would expect mergers to either be younger and dustier, or older and less dusty. Since dust attenuation in $A_V$ is generally not an issue for merging galaxies, this also means that the enhanced sSFR for ongoing mergers is also not affected by dust attenuation.

\begin{figure*}
\gridline{\fig{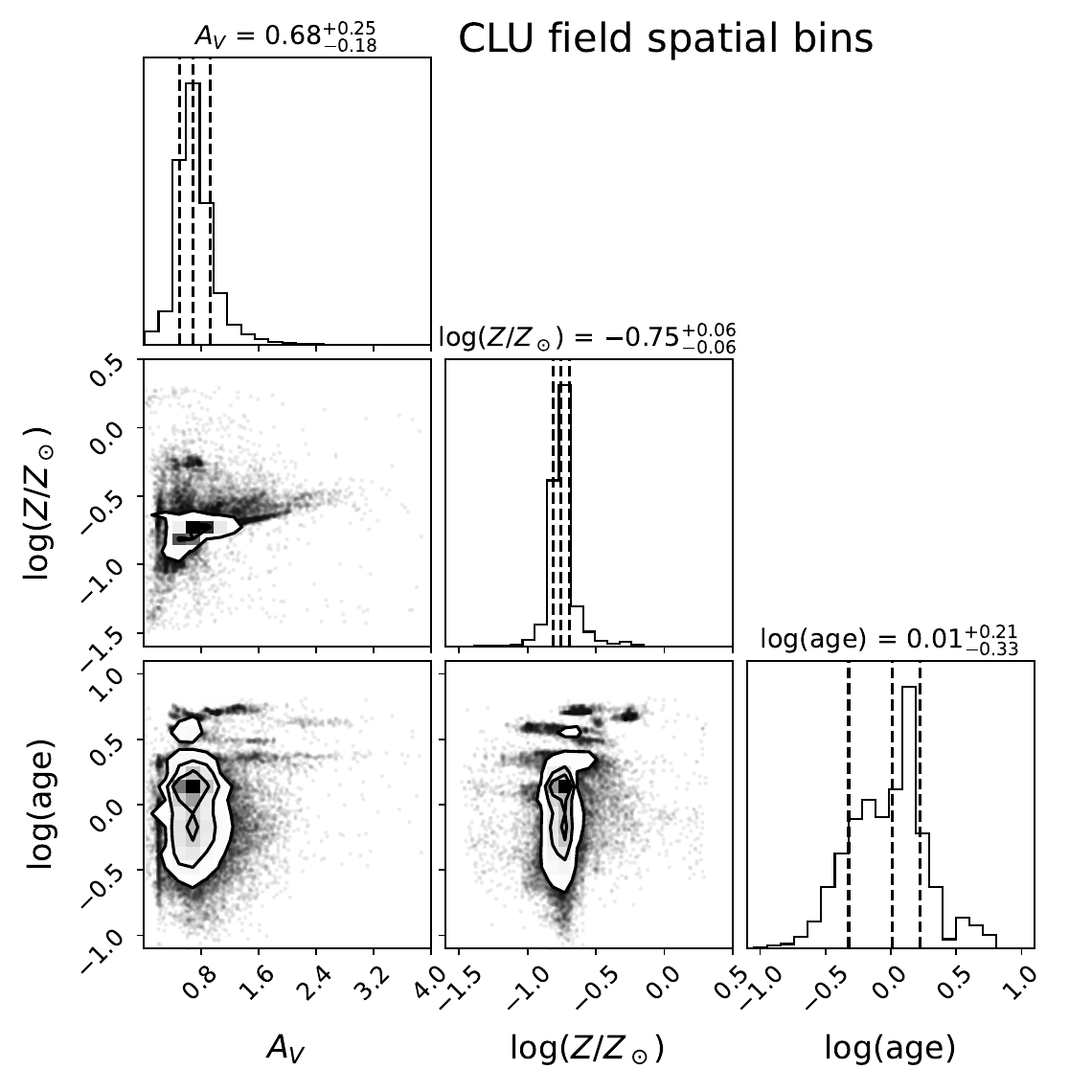}{0.5\textwidth}{}\fig{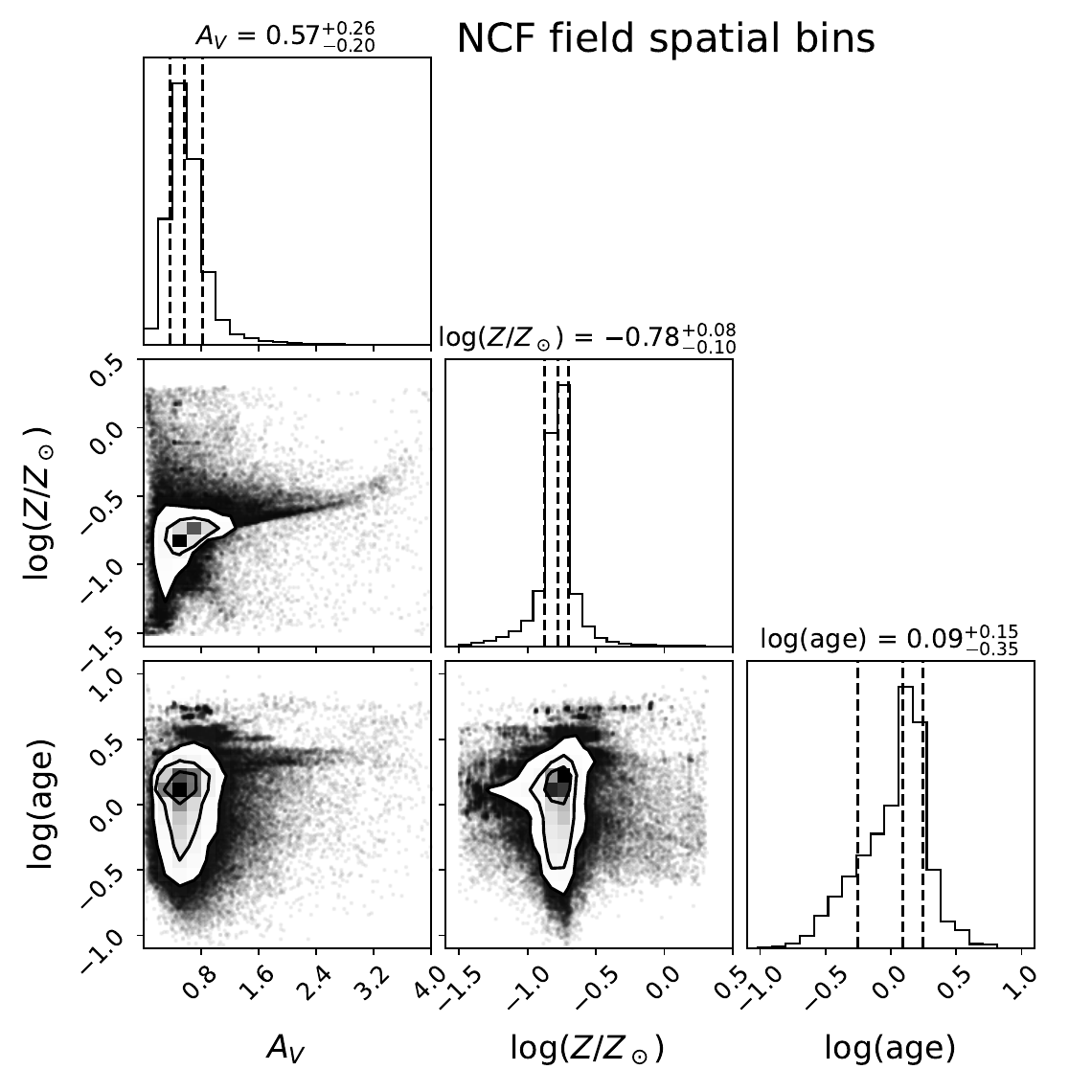}{0.5\textwidth}{}}
\vspace{-2em}
\gridline{\fig{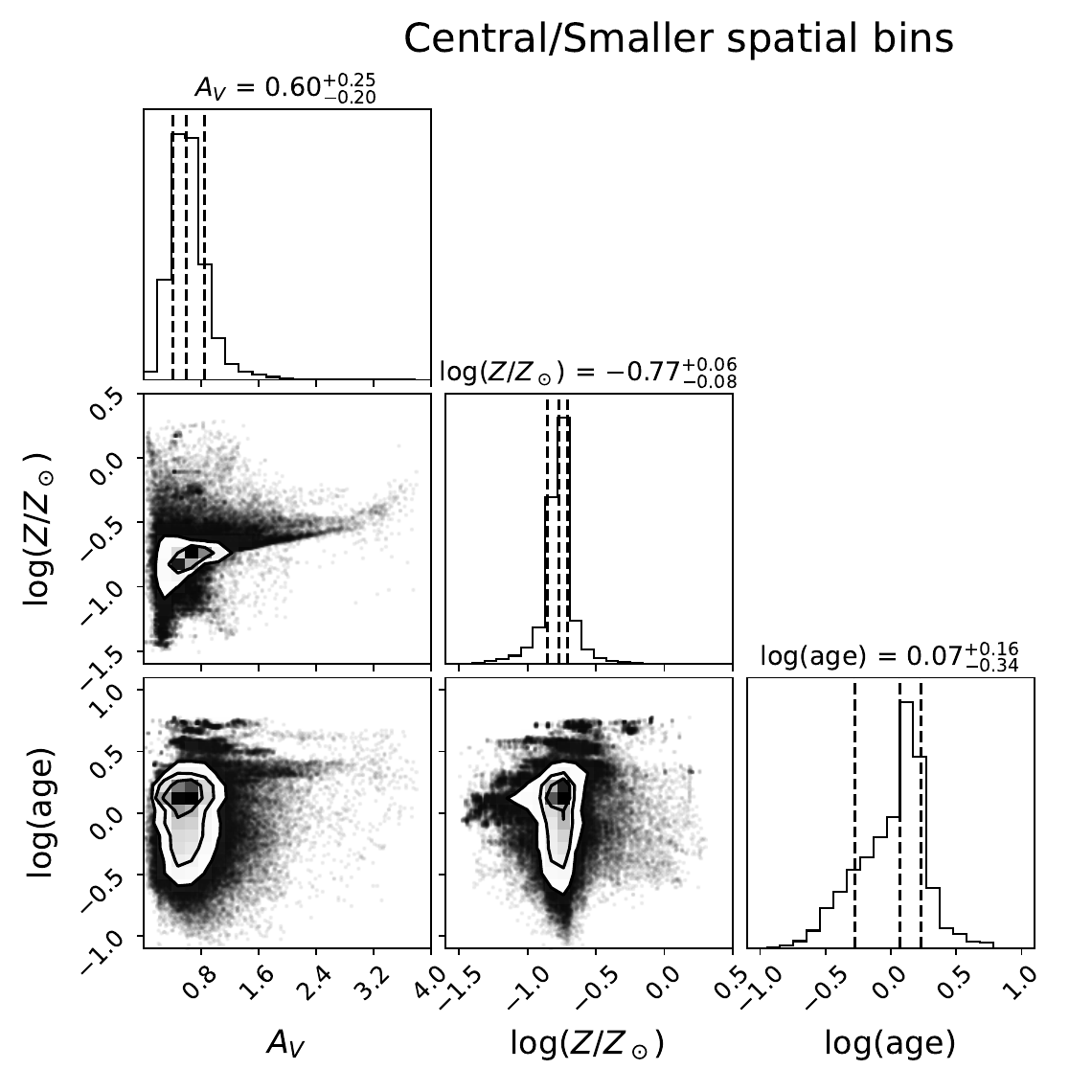}{0.5\textwidth}{}\fig{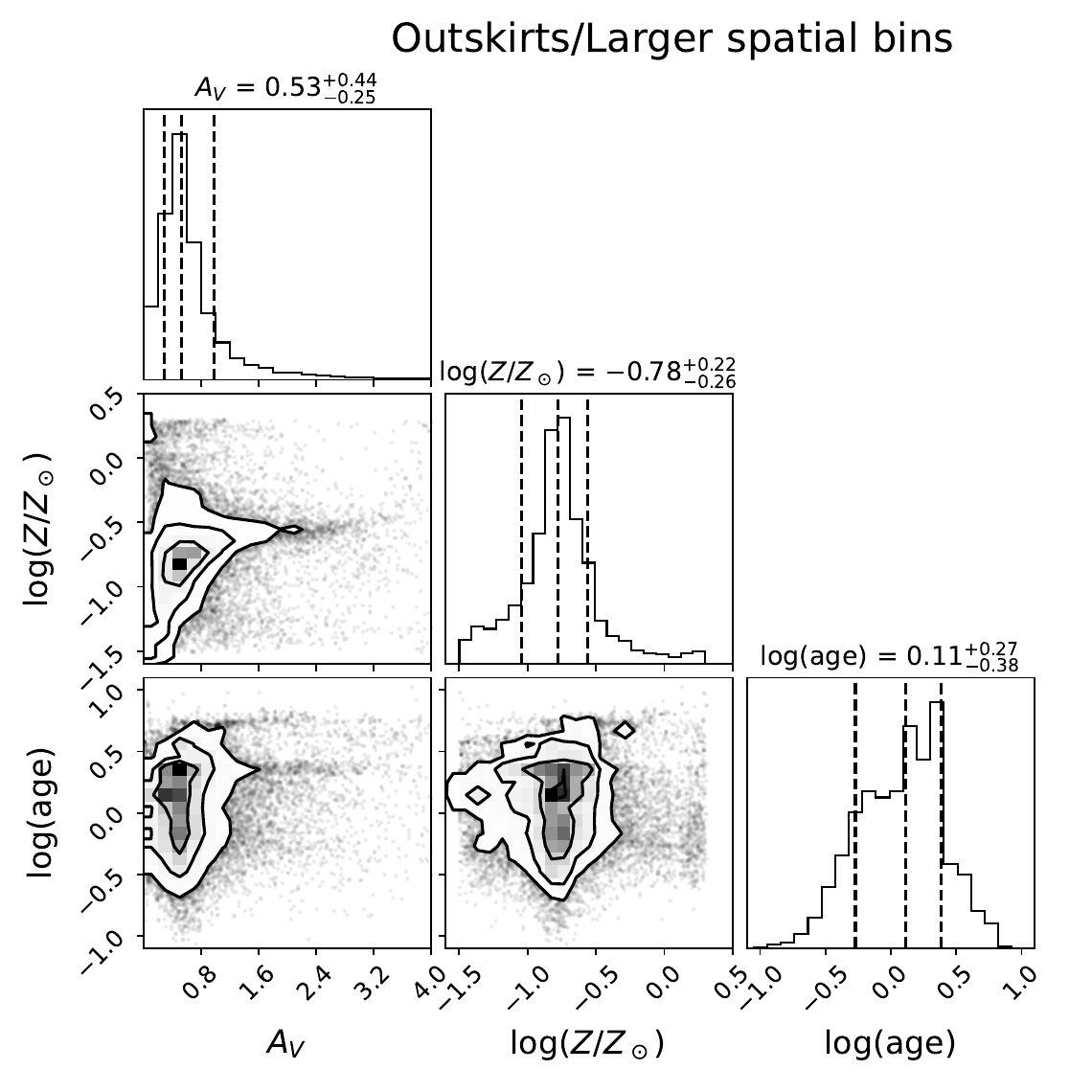}{0.5\textwidth}{}}
\caption{Top row: Corner plots of dust, metallicity, and stellar age for spatial bins of galaxies in the CLU fields versus the NCF fields. Bottom rows are similar corner plots but between smaller sized central bins (bin size $\leq 5$ pixels), versus larger, less central bins (bin size $> 5$ pixels). The 2D contours are for $1\sigma$, $1.5\sigma$, and $2\sigma$ confidence intervals. }\label{fig:corner-plots}
\end{figure*}
\begin{figure*}
    \centering
\gridline{\fig{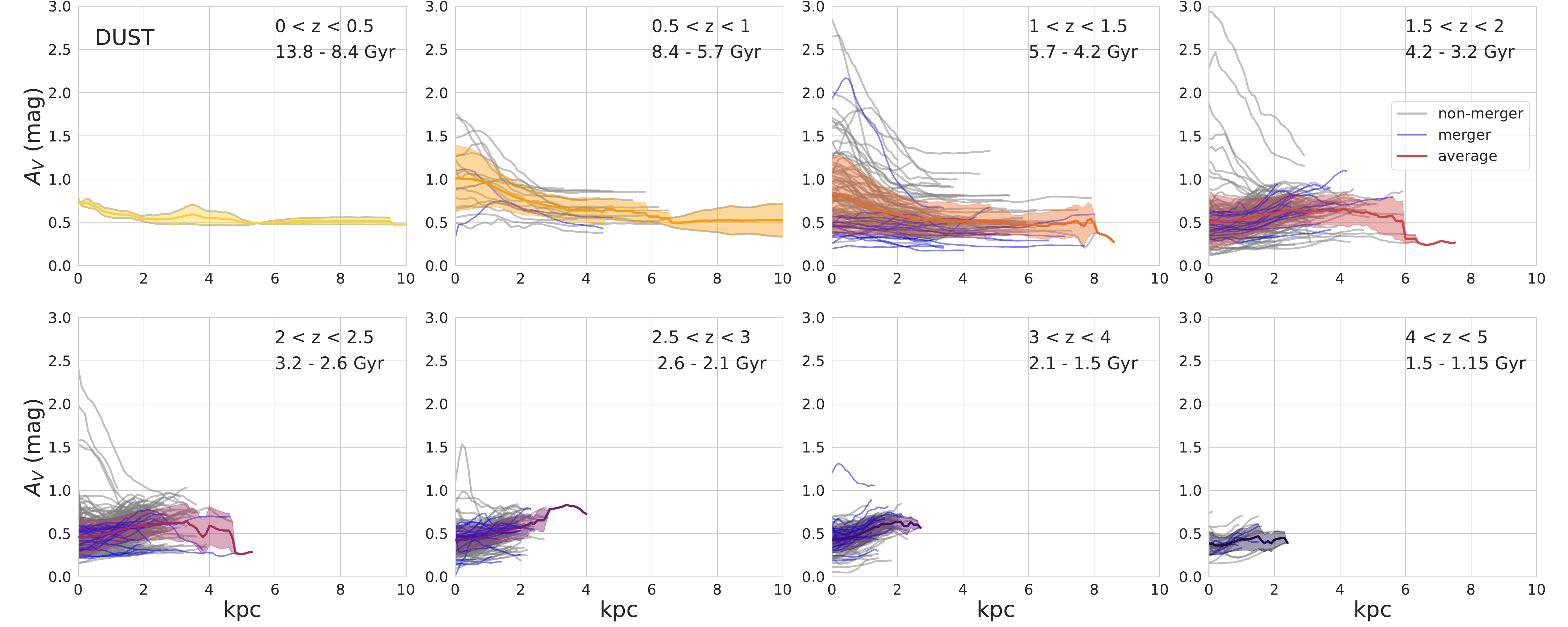}{\textwidth}{}}
\vspace{-2em}
\gridline{\fig{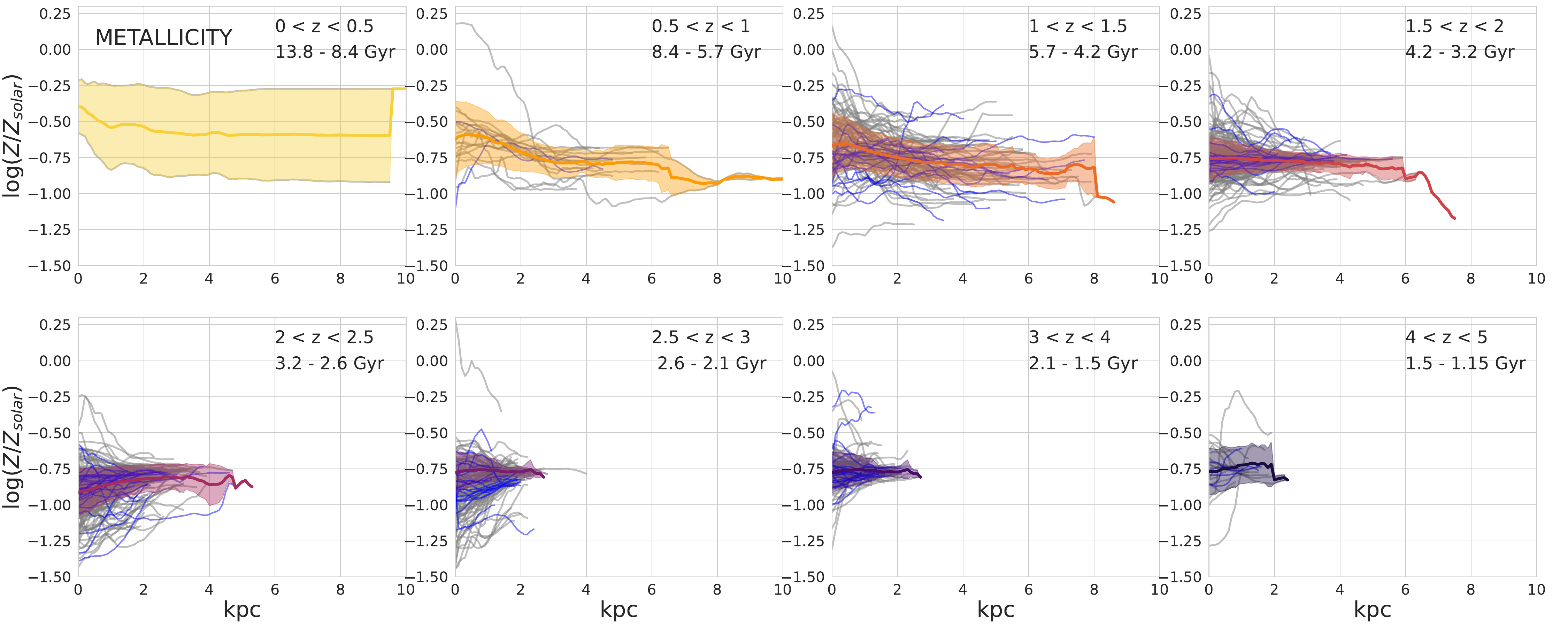}{\textwidth}{}}
\vspace{-2em}
\gridline{\fig{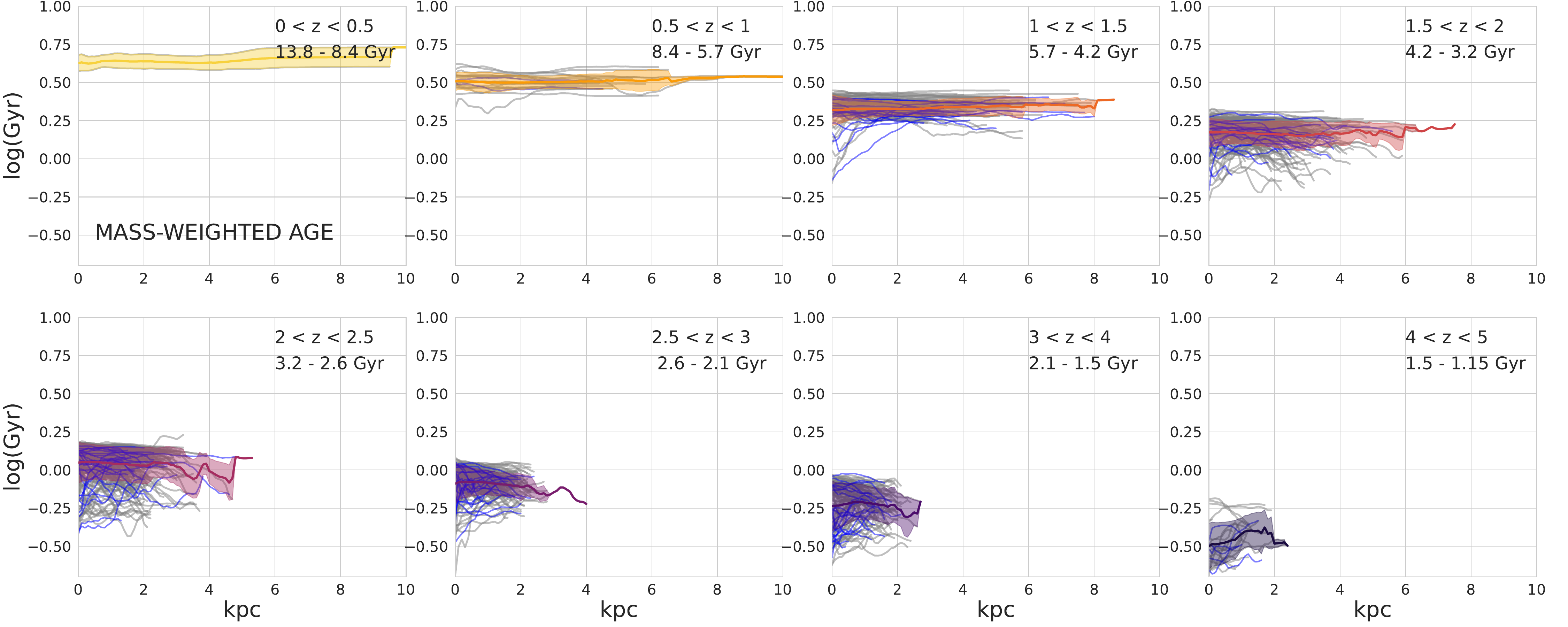}{\textwidth}{}}
\vspace{-2em}
    \caption{\textit{Top panel:} The radial dust attentuation ($A_V$ transmission) gradient of each galaxy in our sample, separated into eight redshift bins, as a function of kpc. Solid colored lines in each subplot represent the stacked and normalized average of that redshift bin. Shaded regions is the $1\sigma$ scatter, or the 18th-84th percentile. Thin grey lines are the individual dust gradients for non-mergers, thin blue lines are mergers. The values of $A_V$, accounting for the 18th-84th percentile scatter range from 0.25 to 1.25. \textit{Middle and bottom panels:} Similar to the top panel but for metallicity in $\log(Z/\Zsun)$ and mass-weighted age in log(Gyr).}
    \label{fig:dust-grad}
\end{figure*}
\end{document}